\documentclass[structabstract]{aa}
\usepackage{graphicx}
\usepackage{txfonts}
\usepackage{natbib}
\bibpunct{(}{)}{;}{a}{}{,} 

\def\gtrsim{\mathrel{\hbox{\rlap{\hbox{\lower4pt\hbox{$\sim$}}}\hbox{$>$}}}}
\def\lesssim{\mathrel{\hbox{\rlap{\hbox{\lower4pt\hbox{$\sim$}}}\hbox{$<$}}}}
\newcommand{\hi}{H{\sc i} }
\newcommand{\hii}{H{\sc ii} }

\newcommand{\oiii}{[O{\sc iii}]$\,$}
\newcommand{\oii}{[O{\sc ii}]$\,$}
\newcommand{\sii}{[S{\sc ii}]$\,$}

\newcommand{\msun}{M$_{\odot}$$\,$}

\newcommand{\arsec}{$^{\prime\prime}$}

\begin{document}

   \title{Young stellar clusters and associations in M33}

  \author{M. Grossi
          \inst{1,2}
          \and
          E. Corbelli\inst{2} \and C. Giovanardi\inst{2} \and L. Magrini\inst{2}
          }
\institute{CAAUL, Observat\'orio Astron\'omico de Lisboa, Universidade de Lisboa, Tapada de Ajuda, 1349-018, Lisboa, PT\\
\email{grossi@oal.ul.pt}\\
\and
INAF-Osservatorio Astrofisico di Arcetri, L.go E. Fermi 5, 50125 Firenze, IT}

   \date{}

\abstract{}
{We analyse multi-wavelength observations  of 32 young star clusters and associations in M33 with known oxygen abundance (8 $<$ 12 + log(O/H) $<$ 8.7).
The data set includes ultraviolet (UV), optical, mid-infrared (MIR), CO (1-0) and 21-cm line (H{\sc i}) observations. We derive the spectral energy distribution (SED) of these systems and the properties of their  gaseous environment to investigate the process of star formation and the interplay with the interstellar medium (ISM).}
{We determine age, bolometric luminosities, masses, and the extinction by comparing the multi-band integrated photometry to single-age stellar population models. The best-fit solutions have been obtained using  the Large Magellanic Cloud (LMC) extinction curves.}
{The stellar system ages range between 2 and 15 Myr, masses are between $3\times 10^2$ and $4\times 10^4$ \msun, and the intrinsic extinction, A$_V$, varies from 0.3 to 1 mag.  We find a correlation between age and extinction (young clusters being more reddened than older ones), and between the cluster mass and size. From the MIR emission we infer the presence of a dust component around the clusters whose fractional luminosity at 24 $\mu$m, L$_{24}$/L$_{Bol}$, decreases with the galactocentric distance.  However, the total infrared luminosity inferred from L$_{24}$ is smaller than what we derive from the extinction corrections.
We also find that the H$\alpha$ luminosity predicted by population synthesis models is higher than the observed one, especially for low-mass systems (M $< 10^4$ \msun). This difference is reduced, but not erased, when the incomplete sampling of the initial mass function (IMF) at the high-mass end is taken into account.}
{Our results suggest that a non-negligible fraction of UV ionising and non-ionising radiation is leaking into the ISM outside the \hii regions. This agrees with the large UV and H$\alpha$ diffuse fractions observed in M33, but it implies that stellar systems younger than 3 Myr retain, on average,  only 30\% of their Lyman-continuum photons. However, the uncertainties in cluster ages and the stochastic fluctuations of the IMF do not allow us to accurately quantify this issue. We also consider the possibility that this discrepancy is the consequence of a suppressed or delayed formation of the most massive stars, given that it mainly affects the young and less massive clusters. We do not find any clear correlation between the cluster mass and the gas surface density or metallicity.
}


\maketitle

%

\section{Introduction}

M33 is the nearest late-type spiral, its properties being in between earlier-type spirals like
the Milky Way and M31 and the star-forming dwarf galaxies in the Local Group.
With a relatively high star-formation rate per unit area (3.4 $\times 10^{-3}$ \msun
yr$^{-1}$ kpc$^{-2}$), seven times higher than M31 \citep{1998ApJ...498..541K} and a low inclination, M33 is an ideal environment to study the properties of young stellar systems and star-forming regions.

Young star clusters and associations are important for studying the ongoing star-formation activity and the chemical evolution of their parent galaxy.
The study of young stellar populations enables us to analyse the initial phase of
the cluster evolution, which is mainly driven by hot massive stars.
Despite their short lifetime, these stars play an important role in galaxy evolution,
because they are the main source  of ionisation, heating, and the chemical enrichment of the interstellar medium (ISM). Massive stars trace the current star-formation activity, they power the infrared (IR) radiation of dust and, at the end of their lives, they eventually trigger further star-formation events via supernova explosions. Moreover, from the properties of the gas and dust distribution of the cluster environment one can better understand the mutual relation between star formation and the ISM.

A number of studies have analysed the cluster population of M33 both from space \citep{1999ApJS..122..431C,2001A&A...366..498C} and from the ground \citep{2001AJ....122.1796M,2002AJ....123.3141M,2004ChJAA...4..125M},
although they have mainly concentrated on rather evolved stellar systems with ages between a few tens of Myr and few Gyrs. Nonetheless, the population of young stellar systems ($\lesssim$ 10 Myr)  in
M33 has not been thoroughly investigated, and only a few studies have been dedicated to such young star-forming systems.

A recent multi-wavelength analysis of radio-selected sources in M33 led to the identification of 11 young, optically-visible stellar clusters without finding any highly obscured systems \citep{2006ApJS..162..329B}. Extinction seems generally lower in M33 than in the Milky Way, and the absence of large molecular complexes \citep{2007ApJ...661..830R} may imply that even in an early star-formation phase stellar clusters may not be highly extincted.
\citet{2009A&A...495..479C} examined the properties of young stellar clusters in M33 selected at 24 $\mu$m with H$\alpha$ counterparts and stressed the importance of combining multiple wavelength observations (from the UV to MIR) to study star-forming sites and to improve the understanding of massive star formation. \citet{2007A&A...476.1161V,2009A&A...493..453V}
showed that MIR images of M33 obtained with the {\em Spitzer} telescope allow us to identify \hii
regions with 24 $\mu$m luminosities as low as $2 \times 10^{37}$ erg s$^{-1}$ and to study the star-formation properties of the M33 disc.

Here we present a multi-wavelength analysis of a sample of 32 young clusters and associations in M33 selected from the catalogues of emission-line objects of \citet{2007A&A...470..865M} and \citet{2008ApJ...675.1213R}.
Our aim was to define a list of young stellar systems selected on the basis of their H$\alpha$ emission, for which a measure of their metal abundance was available. In addition, our objects have a 24 $\mu$m counterpart in the catalogue of Verley et al. (2007).
The final sample includes well known giant \hii regions such as NGC588 and IC132, located in the outskirts of the galaxy, large OB associations such as C400 and B0222, and more compact \hii regions such as B0670 or LGCHII11 with sizes smaller than 10 pc.

We use UV (GALEX), $UBVRI$ (Local Group Survey),  MIR ({\em Spitzer} IRAC and MIPS) observations to constrain ages, masses, extinction, and dust properties of the sample from the comparison to model spectral energy distributions (SEDs). We search for possible correlations among the derived parameters,  we compare the properties of these stellar systems to their metallicity, we look for radial gradients in the observed emissions at different wavelenghts, and finally, we use archival CO and \hi observations to relate the total stellar masses to the local gas content.

The data are presented in Sect. 2, and in Sect. 3 we discuss the sample selection. The
method to derive the photometric measurements in the different bands is described in Sect.
4, while an overview of the SED-fitting technique is given in Sect. 5. The results are
discussed topic by topic in Sect. 6. The issue of the ionising photon deficiency within the clusters is analysed in Sect. 7. The properties of the gaseous environment of the
sample are presented in Sect. 8, and in Sect. 9 we summarise our results and
conclusions.

 \begin{figure*}
   \centering
   \includegraphics[width=11cm,bb=14 14 525 558]{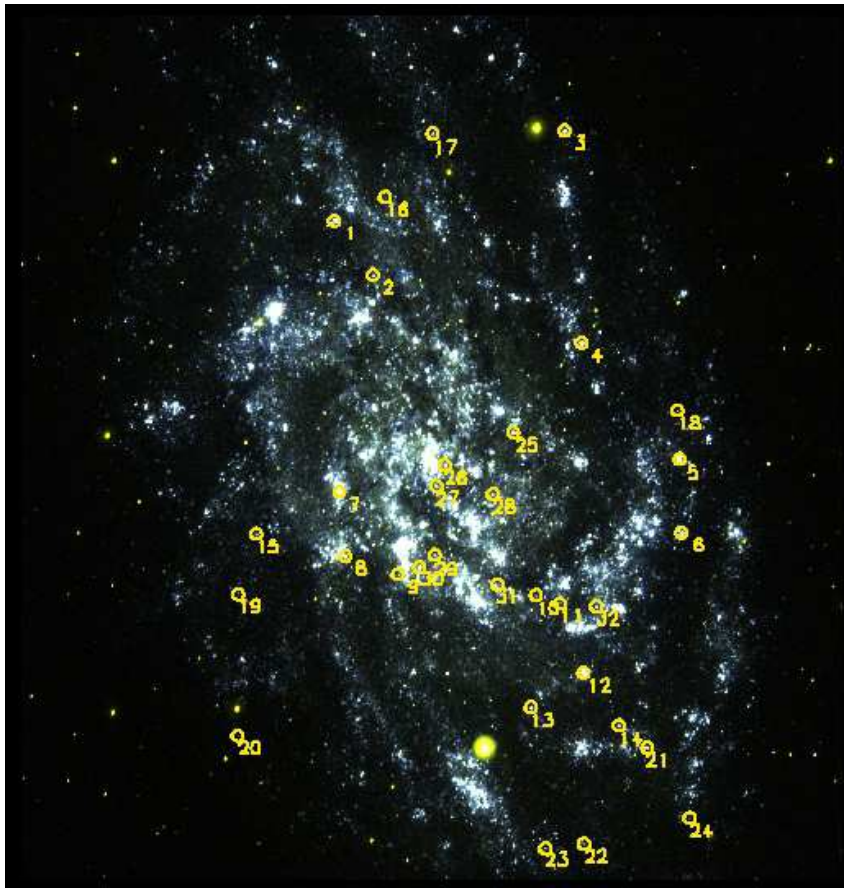}   

   \caption{Location of the 32 clusters selected in this study, overlayed on the GALEX
(FUV and NUV) images of M33.}  \label{Fig1}%
    \end{figure*}

\section{The data set: archival UV, optical and IR images}

\begin{table}
\caption{Sample of young clusters in M33 selected in this work.}
\begin{center}
\begin{minipage}{\textwidth}
\begin{tabular}{llccc}
\hline \hline
ID   & Name       &     RA         &   DEC    &  12 + log(O/H)     \\
     &    & \multicolumn{2}{c}{J2000.0}      &      \\
\hline \hline
1  &  B691 			& 01:34:16.6     & 30:51:54 &  8.26  $\pm$  0.02 $^a$\\  
2  &  VGHC~2-84 	& 01:34:06.7     & 30:48:56 &  8.35  $\pm$  0.04 $^a$\\  
3  &     IC132 		& 01:33:15.8     & 30:56:45 &  8.08  $\pm$  0.04 $^a$\\  
4  &  B290 			& 01:33:11.4     & 30:45:15 &  8.21  $\pm$  0.06 $^a$\\  
5  &    NGC588 	    & 01:32:45.9     & 30:38:54 &  8.31  $\pm$  0.06 $^a$\\  
6  &      C400 		& 01:32:44.4     & 30:35:17 &  7.98  $\pm$  0.08 $^b$\\  
7  &       MA2 		& 01:34:15.6     & 30:37:11 &  8.33  $\pm$  0.08 $^a$\\  
8  &       S14 		& 01:34:13.7     & 30:33:45 &  8.30  $\pm$  0.11 $^b$\\  
9  & GDK99~128 	    & 01:33:59.9     & 30:32:44 &  8.47  $\pm$  0.06 $^a$\\  
10 &      C129a 	& 01:33:23.4     & 30:31:35 &  8.45  $\pm$  0.08 $^b$\\  
11 &       C121 	& 01:33:17.1     & 30:31:11 &  8.35  $\pm$  0.06 $^b$\\  
12 &      B0221 	& 01:33:09.8     & 30:27:23 &  8.28  $\pm$  0.10 $^b$\\  
13 &      B0222 	& 01:33:25.0     & 30:25:31 &  8.10  $\pm$  0.18 $^b$\\  
14 &       C022 	& 01:33:01.6     & 30:24:31 &  8.31  $\pm$  0.08 $^b$\\  
15 &   B0745 		& 01:34:37.6     & 30:34:55 &  8.07  $\pm$  0.10 $^a$\\  
16 & B0670   		& 01:34:03.3     & 30:53:09 &  8.28  $\pm$  0.07 $^a$\\   
17 & B0637   		& 01:33:50.6     & 30:56:33 &  8.34  $\pm$  0.05 $^a$\\   
18 & LGC~HII~3  	& 01:32:45.9     & 30:41:35 &  8.24  $\pm$  0.05 $^a$\\   
19 & B0706   		& 01:34:42.2     & 30:31:42 &  8.32  $\pm$  0.12 $^a$\\   
20 &  LGC~HII~11    & 01:34:42.2     & 30:24:00 &  8.17  $\pm$  0.06 $^a$\\   
21 &  B0261      	& 01:32:54.6     & 30:23:22 &  8.36  $\pm$  0.08 $^b$\\   
22 &  MCM00Em24     & 01:33:10.8     & 30:18:08 &  8.18  $\pm$  0.25 $^a$\\   
23 &  B0254      	& 01:33:21.2     & 30:17:58 &  8.16  $\pm$  0.19 $^b$\\   
24 &  LGC~HII~2     & 01:32:43.0     & 30:19:31 &  8.25  $\pm$  0.06 $^a$\\   
25 &  B0045    		&  01:33:29.0    &  30:40:24  &   8.49  $\pm$   0.04 $^a$ \\ 
26 &  B0029      	&  01:33:47.8    &  30:38:38    &   8.21  $\pm$    0.10 $^b$ \\  
27 &  B0016      	&  01:33:50.1    &  30:37:30    &   7.98  $\pm$    0.19  $^b$\\  
28 &  B0033b     	&  01:33:34.9    &  30:37:05    &   8.40  $\pm$    0.08  $^b$\\  
29 &  B0015b     	&  01:33:50.3    &  30:33:42    &   8.73  $\pm$    0.23  $^b$\\  
30 &  B0013c     	&  01:33:54.1    &  30:33:10    &   8.26  $\pm$    0.12  $^b$\\  
31 &  B0208f     	&  01:33:34.1    &  30:32:13    &   8.07  $\pm$    0.06  $^b$\\  
32 &  C403       	&  01:33:07.6    &  30:31:01    &   8.27  $\pm$    0.07  $^b$\\  
\hline \hline
\label{list_clust}
\end{tabular}
\end{minipage}
\end{center}
\tablefoot{The identification names are
from: B--\citet{1974A&A....37...33B};
C-- \citet{1987A&A...174...28C};
GDK99--\citet{1999ApJS..120..247G};
LGC-HII-n--\citet{2007A&A...470..865M};
VGVC--\citet{1986A&AS...64..237V};
MCM00--\citet{2000A&A...355..713M};
MA--\citet{1942ApJ....95....5M};
S--\citet{1971ApJ...168..327S}.
Substructure designations within individual cluster
(indicated by the letters at the end of their name)
are taken from \citet{2002ASSL..221.....H}.
The oxygen abundance determinations are from: \tablefoottext{a}  \citet{2007A&A...470..865M};
\tablefoottext{b} \citet{2008ApJ...675.1213R}
}
\end{table}

\subsection{Galex images}

We used Galaxy Evolution Explorer \citep[GALEX;][]{2005ApJ...619L...1M} observations of M33
in the far-UV (FUV, $\lambda_{eff} = 1528$ \AA) and near-UV (NUV,
$\lambda_{eff}$ = 2271 \AA), distributed by \citet{2007ApJS..173..185G}. The angular
resolution (FWHM) is $\sim$ 4\arcsec \,in the FUV and $\sim$ 5$^{\prime\prime\!}$.5  in the NUV images, corresponding to approximately 20 pc at the M33 distance. The exposure time was 3400 sec in both filters. A more detailed description of GALEX observations of M33 and of the data reduction and calibration procedures can be found in
\citet{2005ApJ...619L..67T}.

\subsection{Ground-based optical data: the Local Group Survey}

$UBVRI$ images of M33 have been taken from the Local Group Survey (LGS) of galaxies
currently forming stars \citep{2006AJ....131.2478M}. M33 observations have been performed
with the Kitt Peak National Observatory 4-m telescope using the wide-field Mosaic cameras.
Narrowband images through the filters centred on \oiii $\lambda$5007, H$\alpha$ and \sii
$\lambda$6731 are also provided \citep{2007AJ....134.2474M}. The data are resampled to a
scale of 0.27 arcsec/pixel. The FWHM of the point spread function in the $V$ band is 1\farcs1,
corresponding to $\sim$4 pc.
More details about the
observations and data reduction can be found in these papers.

We also made use of an optical narrowband H$\alpha$ image of M33
obtained with the 0.6 meter Burrell-Schmidt
telescope at Kitt Peak National Observatory \citep{2000ApJ...541..597H}.
The 2048 $\times$ 2048 pixel CCD provides a field of view of 1.15 $\times$ 1.15 square
degrees, with a scale of
2.03 arcsec/pixel. The final mosaic covers an
area of about 1.75 $\times$ 1.75 square degrees.
More details on the observations and reduction procedure can be found in
Hoopes \& Walterbos (2000).

\subsection{{\em Spitzer} images}

{\em Spitzer} IRAC and MIPS images of M33 were obtained under the Guaranteed Time Observer (GTO)
programme PID 5 (PI R. Gehrz).
We used the IRAC (3.6-8 $\mu$m) and MIPS (24 $\mu$m) maps of M33 produced by \citet{2007A&A...476.1161V}.
The spatial resolution of the IRAC maps is between 1\farcs7  and 2\arcsec ($\sim$ 7 pc), while at 24 $\mu$m it is 6\arsec (24 pc).
The map scale is 1.2 arcsec/pixel in both IRAC and MIPS images.

\subsection{CO and HI maps}

We used the 3-mm CO map of M33 (J=1-0 rotational line) obtained from the combination of  the
Berkeley Illinois Maryland Association (BIMA) array
\citep{2003ApJS..149..343E,2007ApJ...661..830R}
and the Five College Radio Astronomy Observatory (FCRAO) single dish \citep{2004ApJ...602..723H}.
The final resolution of the merged map is 13\arcsec.
The median noise level of the channel maps is 240 mK, corresponding to a
mass sensitivity of 5 $\times 10^4$ \msun (4$\sigma$ detection in two adjacent channels; see
Rosolowsky et al. 2007) using a conversion factor X = 2
$\times 10^{20}$ cm$^{-2}$ (K km s$^{-1}$)$^{-1}$.

The 21-cm observations of M33 have been taken at the Westerbork
Radio Synthesis Telescope (WRST) array \citep{1987A&AS...67..509D}, with a spatial
resolution of 24\arcsec $\times$  48\arcsec.

 \begin{table*}
\caption{UV and optical photometric measurements. The
last column displays the aperture radius used for the photometry.}
\label{phottab}
\begin{minipage}{\textwidth}
\begin{center}
\begin{tabular}{lcccccccr}
\hline \hline
ID &       FUV      &      NUV       &        U       &        B       &       V        &        R       &         I      & $r_{ap}$ \\
   &     AB mag     &     AB mag     &     AB mag     &     AB mag     &     AB mag     &     AB mag     &     AB mag     & $\:\:(^{\prime\prime\!})$ \\
\hline \hline
01 & 15.22$\pm$0.00 & 15.31$\pm$0.00 & 14.80$\pm$0.08 & 14.63$\pm$0.06 & 14.66$\pm$0.04 & 14.76$\pm$0.04 & 14.76$\pm$0.04 & 11. \\
02 & 17.32$\pm$0.01 & 17.40$\pm$0.01 & 16.92$\pm$0.08 & 16.95$\pm$0.06 & 17.10$\pm$0.04 & 17.18$\pm$0.04 & 17.48$\pm$0.05 &  4. \\
03 & 15.76$\pm$0.01 & 15.69$\pm$0.00 & 15.08$\pm$0.08 & 14.87$\pm$0.06 & 14.82$\pm$0.04 & 15.24$\pm$0.04 & 15.43$\pm$0.04 &  9. \\
04 & 15.49$\pm$0.00 & 15.29$\pm$0.00 & 14.64$\pm$0.08 & 14.71$\pm$0.06 & 14.80$\pm$0.04 & 14.92$\pm$0.04 & 15.08$\pm$0.04 &  6. \\
05 & 14.84$\pm$0.00 & 14.66$\pm$0.00 & 14.59$\pm$0.08 & 14.40$\pm$0.06 & 14.44$\pm$0.04 & 14.89$\pm$0.04 & 15.16$\pm$0.04 & 12. \\
06 & 15.20$\pm$0.01 & 15.05$\pm$0.00 & 14.86$\pm$0.08 & 14.89$\pm$0.06 & 14.95$\pm$0.03 & 15.20$\pm$0.03 & 15.26$\pm$0.04 & 20. \\
07 & 14.42$\pm$0.01 & 14.34$\pm$0.00 & 14.09$\pm$0.08 & 14.16$\pm$0.06 & 14.30$\pm$0.03 & 14.41$\pm$0.03 & 14.64$\pm$0.06 &  8. \\
08 & 16.23$\pm$0.04 & 16.13$\pm$0.04 & 15.67$\pm$0.08 & 15.75$\pm$0.06 & 15.90$\pm$0.04 & 15.94$\pm$0.04 & 16.22$\pm$0.05 &  8. \\
09 & 18.67$\pm$0.11 & 18.85$\pm$0.19 & 17.61$\pm$0.08 & 17.82$\pm$0.06 & 17.97$\pm$0.04 & 17.83$\pm$0.04 & 18.34$\pm$0.06 &  3. \\
10 & 19.28$\pm$0.04 & 19.15$\pm$0.03 & 18.36$\pm$0.08 & 18.50$\pm$0.06 & 18.52$\pm$0.04 & 18.33$\pm$0.04 & 18.60$\pm$0.06 &  2. \\
11 & 18.69$\pm$0.03 & 18.68$\pm$0.03 & 17.51$\pm$0.08 & 17.68$\pm$0.06 & 17.80$\pm$0.05 & 17.70$\pm$0.04 & 17.94$\pm$0.06 &  5. \\
12 & 14.85$\pm$0.00 & 14.72$\pm$0.00 & 14.38$\pm$0.08 & 14.29$\pm$0.06 & 14.32$\pm$0.04 & 14.27$\pm$0.04 & 13.97$\pm$0.04 & 23. \\
13 & 16.74$\pm$0.01 & 16.71$\pm$0.01 & 16.13$\pm$0.08 & 16.14$\pm$0.06 & 16.24$\pm$0.04 & 16.19$\pm$0.04 & 16.20$\pm$0.04 &  9. \\
14 & 17.02$\pm$0.01 & 17.03$\pm$0.01 & 16.66$\pm$0.08 & 16.74$\pm$0.06 & 16.87$\pm$0.04 & 17.03$\pm$0.04 & 17.34$\pm$0.05 &  6. \\
15 & 18.07$\pm$0.02 & 18.13$\pm$0.02 & 17.20$\pm$0.08 & 17.36$\pm$0.06 & 17.50$\pm$0.04 & 17.47$\pm$0.04 & 17.81$\pm$0.05 &  5. \\
16 & 19.67$\pm$0.09 & 19.91$\pm$0.13 & 18.82$\pm$0.08 & 18.93$\pm$0.06 & 19.01$\pm$0.05 & 18.99$\pm$0.05 & 19.30$\pm$0.08 &  2. \\
17 & 17.29$\pm$0.02 & 17.47$\pm$0.01 & 17.16$\pm$0.08 & 17.01$\pm$0.06 & 17.05$\pm$0.04 & 17.52$\pm$0.05 & 17.65$\pm$0.06 &  8. \\
18 & 18.39$\pm$0.03 & 18.39$\pm$0.02 & 18.19$\pm$0.08 & 18.33$\pm$0.06 & 18.46$\pm$0.04 & 18.64$\pm$0.04 & 18.96$\pm$0.05 &  3. \\
19 & 19.28$\pm$0.04 & 19.42$\pm$0.04 & 17.90$\pm$0.08 & 17.81$\pm$0.06 & 17.76$\pm$0.04 & 17.70$\pm$0.04 & 17.94$\pm$0.05 &  6. \\
20 & 19.90$\pm$0.04 & 20.05$\pm$0.03 & 19.46$\pm$0.08 & 19.52$\pm$0.06 & 19.55$\pm$0.04 & 19.70$\pm$0.05 & 20.19$\pm$0.08 &  3. \\
21 & 16.89$\pm$0.01 & 16.81$\pm$0.01 & 16.60$\pm$0.08 & 16.59$\pm$0.06 & 16.72$\pm$0.04 & 16.94$\pm$0.04 & 17.17$\pm$0.05 &  7. \\
22 & 18.37$\pm$0.02 & 18.57$\pm$0.01 & 17.67$\pm$0.08 & 18.05$\pm$0.06 & 18.24$\pm$0.04 & 18.39$\pm$0.04 & 18.59$\pm$0.07 &  4. \\
23 & 18.09$\pm$0.03 & 18.13$\pm$0.02 & 17.27$\pm$0.08 & 17.62$\pm$0.06 & 17.77$\pm$0.04 & 17.97$\pm$0.05 & 18.15$\pm$0.08 &  7. \\
24 & 19.33$\pm$0.04 & 19.47$\pm$0.03 & 18.48$\pm$0.08 & 18.93$\pm$0.06 & 19.11$\pm$0.05 & 19.19$\pm$0.05 & 19.29$\pm$0.09 &  4. \\
25 & 16.25$\pm$0.01 & 16.25$\pm$0.01 & 15.58$\pm$0.08 & 15.55$\pm$0.06 & 15.66$\pm$0.04 & 15.72$\pm$0.04 & 15.97$\pm$0.05 &  9. \\
26 & 16.59$\pm$0.02 & 16.60$\pm$0.02 & 15.82$\pm$0.08 & 15.91$\pm$0.06 & 16.05$\pm$0.05 & 16.07$\pm$0.05 & 16.50$\pm$0.07 &  7. \\
27 & 17.40$\pm$0.04 & 17.44$\pm$0.01 & 16.75$\pm$0.08 & 16.94$\pm$0.07 & 17.17$\pm$0.07 & 17.22$\pm$0.06 & 17.54$\pm$0.09 &  5. \\
28 & 17.97$\pm$0.02 & 17.89$\pm$0.02 & 16.61$\pm$0.08 & 16.68$\pm$0.06 & 16.76$\pm$0.04 & 16.62$\pm$0.04 & 16.86$\pm$0.05 &  6. \\
29 & 17.63$\pm$0.03 & 17.65$\pm$0.03 & 16.57$\pm$0.08 & 16.75$\pm$0.07 & 16.84$\pm$0.05 & 16.69$\pm$0.05 & 16.84$\pm$0.09 &  9. \\
30 & 16.23$\pm$0.04 & 16.02$\pm$0.03 & 15.12$\pm$0.08 & 15.21$\pm$0.06 & 15.25$\pm$0.04 & 15.09$\pm$0.04 & 15.17$\pm$0.05 & 11. \\
31 & 16.78$\pm$0.01 & 16.63$\pm$0.01 & 15.99$\pm$0.08 & 16.09$\pm$0.06 & 16.30$\pm$0.04 & 16.36$\pm$0.04 & 16.74$\pm$0.06 &  7. \\
32 & 17.95$\pm$0.05 & 17.82$\pm$0.05 & 16.94$\pm$0.08 & 16.88$\pm$0.06 & 17.01$\pm$0.05 & 17.10$\pm$0.05 & 17.27$\pm$0.06 &  7. \\
\hline \hline
\end{tabular}
\end{center}
\end{minipage}
\end{table*}

\section{Sample selection}

The sample was selected  from the catalogues of emission line
objects of \citet{2007A&A...470..865M} and of \citet{2008ApJ...675.1213R}.
The two catalogues  contains 72 and 61  \hii regions respectively.
The oxygen abundances were obtained using the temperature-sensitive emission line \oiii$\lambda$4363,
with a direct measurement of the electron temperature.
The condition of a measured and not assumed electron temperature is essential, because the determination
of the ionic and total chemical abundances exponentially dependends on it.

The 32 \hii regions of the sample were selected on the basis of their shape and their position, preferentially
round and/or located in regions of the galaxy not too crowded in the H$\alpha$ map.
Then we searched for a 24 $\mu$m  counterpart in the {\em Spitzer}/MIPS map of Verley et al. (2007), and by cross-correlating the two lists of objects we ended up with 25 sources with metal abundances within the range 8 $<$ 12 + log(O/H) $<$ 8.7. We then added seven more regions
at large galactocentric radii, which have a clear detection in the 24 $\mu$m map even though they are not included in the Verley et al. (2007) catalogue.
The final sample includes well known giant \hii regions such as NGC588 and IC132, which are located in the outskirts of the galaxy, large OB association as C400 and B0222, and more compact
\hii regions such as B0670 or LGCHII11 with sizes smaller than 10 pc. Figure \ref{Fig1} shows
the position of the selected clusters and associations overlayed on the GALEX (FUV) image. In the rest of the paper we will refer to these young stellar systems by using the term  ``clusters'' even though, in many cases, their size is larger than that of typical bound clusters.

\begin{table*}
\caption{H$\alpha$ and {\em Spitzer} IRAC/MIPS photometry of the sample. Flux units are in ergs
s$^{-1}$ cm$^{-2}$.}
\label{MIRphot}
\begin{minipage}{\textwidth}
\begin{center}
\begin{tabular}{lrrrrrrr}
\hline \hline
    ID & $F^0_{H\alpha} \: \: \: \:$ & $\lambda F_{\lambda}^{3.6}\:\:\:$ &  $\lambda F_{\lambda}^{4.5}\:\:\:$ &
$\lambda F_{\lambda}^{5.8}\:\:\:$ &  $\lambda F_{\lambda}^{8.0}\:\:\:$ & $\lambda F_{\lambda}^{24}\:\:\:$ \\
       & $10^{-13}$ cgs$\:$ & $10^{-12}$ cgs$\:$ & $10^{-12}$ cgs$\:$ & $10^{-12}$ cgs$\:$ & $10^{-12}$ cgs$\:$
& $10^{-12}$ cgs$\:$ \\
\hline \hline
 1 & 25.29$\pm$0.13 &  6.10$\pm$0.74 &  4.23$\pm$0.53 &  9.73$\pm$1.06 & 24.18$\pm$2.51 & 23.78$\pm$2.40 \\
 2 &  4.82$\pm$0.15 &  0.22$\pm$0.10 &  0.17$\pm$0.07 &  0.50$\pm$0.11 &  1.65$\pm$0.23 &  0.75$\pm$0.09 \\
 3 & 52.85$\pm$1.62 &  1.20$\pm$0.68 &  1.00$\pm$0.37 &  1.54$\pm$0.27 &  7.76$\pm$0.88 &  6.71$\pm$0.69 \\
 4 & 26.33$\pm$0.56 &  2.65$\pm$0.38 &  2.16$\pm$0.31 &  6.43$\pm$0.73 & 16.40$\pm$1.71 & 15.77$\pm$1.60 \\
 5 & 61.11$\pm$0.36 &  1.57$\pm$0.26 &  1.37$\pm$0.22 &  2.31$\pm$0.31 &  9.63$\pm$1.03 &  9.85$\pm$1.01 \\
 6 &  4.80$\pm$0.24 &  1.92$\pm$0.32 &  1.06$\pm$0.20 &  3.07$\pm$0.39 &  7.04$\pm$0.77 &  3.14$\pm$0.33 \\
 7 & 32.64$\pm$0.38 &  1.92$\pm$0.31 &  1.42$\pm$0.23 &  3.90$\pm$0.47 & 26.65$\pm$2.75 & 18.56$\pm$1.88 \\
 8 & 11.48$\pm$0.34 &  4.08$\pm$0.55 &  2.51$\pm$0.35 & 10.67$\pm$1.16 & 26.15$\pm$2.69 & 16.17$\pm$1.64 \\
 9 &  4.18$\pm$0.09 &  0.42$\pm$0.12 &  0.30$\pm$0.09 &  1.26$\pm$0.20 &  5.10$\pm$0.64 &  2.20$\pm$0.24 \\
10 &  2.51$\pm$0.07 &  0.28$\pm$0.09 &  0.20$\pm$0.07 &  0.76$\pm$0.14 &  3.31$\pm$0.40 &  2.39$\pm$0.26 \\
11 &  2.58$\pm$0.06 &  0.32$\pm$0.10 &  0.22$\pm$0.07 &  0.93$\pm$0.16 &  2.76$\pm$0.35 &  0.75$\pm$0.10 \\
12 & 19.92$\pm$0.13 & 11.96$\pm$1.35 &  7.63$\pm$0.88 & 18.95$\pm$1.98 & 35.46$\pm$3.61 & 24.89$\pm$2.51 \\
13 &  7.29$\pm$0.05 &  1.62$\pm$0.28 &  1.20$\pm$0.21 &  3.12$\pm$0.39 &  7.98$\pm$0.87 &  4.32$\pm$0.45 \\
14 &  3.15$\pm$0.08 &  0.54$\pm$0.14 &  0.37$\pm$0.10 &  1.42$\pm$0.21 &  3.81$\pm$0.44 &  1.89$\pm$0.21 \\
15 &  3.66$\pm$0.07 &  0.42$\pm$0.12 &  0.37$\pm$0.10 &  1.18$\pm$0.19 &  3.29$\pm$0.39 &  5.69$\pm$0.59 \\
16 &  0.72$\pm$0.02 &  0.08$\pm$0.05 &  0.05$\pm$0.04 &  0.21$\pm$0.06 &  1.21$\pm$0.17 &  0.38$\pm$0.06 \\
17 &  5.94$\pm$0.09 &  0.12$\pm$0.06 &  0.10$\pm$0.05 &  0.16$\pm$0.06 &  2.11$\pm$0.28 &  0.58$\pm$0.08 \\
18 &  0.72$\pm$0.02 &  0.10$\pm$0.05 &  0.06$\pm$0.04 &  0.15$\pm$0.05 &  0.46$\pm$0.09 &  0.19$\pm$0.03 \\
19 &  4.13$\pm$0.08 &  0.16$\pm$0.07 &  0.12$\pm$0.05 &  0.29$\pm$0.08 &  2.14$\pm$0.27 &  0.46$\pm$0.06 \\
20 &  0.30$\pm$0.01 &  0.03$\pm$0.03 &  0.02$\pm$0.02 &  0.07$\pm$0.04 &  0.23$\pm$0.06 &  0.09$\pm$0.02 \\
21 &  4.88$\pm$0.15 &  0.17$\pm$0.08 &  0.12$\pm$0.06 &  0.18$\pm$0.06 &  1.30$\pm$0.21 &  0.51$\pm$0.08 \\
22 &  0.68$\pm$0.01 &  0.10$\pm$0.05 &  0.06$\pm$0.04 &  0.15$\pm$0.05 &  0.40$\pm$0.08 &  0.15$\pm$0.03 \\
23 &  1.78$\pm$0.04 &  0.09$\pm$0.05 &  0.07$\pm$0.04 &  0.08$\pm$0.04 &  0.48$\pm$0.10 &  0.13$\pm$0.03 \\
24 &  0.43$\pm$0.01 &  0.08$\pm$0.07 &  0.05$\pm$0.04 &  0.10$\pm$0.04 &  0.23$\pm$0.06 &  0.07$\pm$0.02 \\
25 & 25.56$\pm$0.87 &  4.76$\pm$0.60 &  3.96$\pm$0.50 & 14.34$\pm$1.52 & 39.06$\pm$3.99 & 72.34$\pm$7.26 \\
26 & 15.36$\pm$0.34 &  1.34$\pm$0.26 &  0.92$\pm$0.19 &  5.04$\pm$0.60 & 34.17$\pm$3.60 &  8.75$\pm$0.91 \\
27 &  4.14$\pm$0.14 &  1.54$\pm$0.26 &  1.09$\pm$0.19 &  5.15$\pm$0.60 & 16.57$\pm$1.73 &  8.75$\pm$0.90 \\
28 & 12.52$\pm$0.25 &  1.93$\pm$0.31 &  1.55$\pm$0.25 &  4.39$\pm$0.52 & 12.29$\pm$1.33 & 12.10$\pm$1.23 \\
29 &  6.74$\pm$0.23 &  0.91$\pm$0.21 &  0.63$\pm$0.15 &  3.11$\pm$0.40 &  4.64$\pm$0.66 &  2.70$\pm$0.30 \\
30 & 16.18$\pm$0.53 &  3.32$\pm$0.46 &  2.09$\pm$0.31 &  9.32$\pm$1.03 & 28.48$\pm$2.96 & 11.66$\pm$1.19 \\
31 & 18.67$\pm$0.47 &  1.78$\pm$0.29 &  1.11$\pm$0.20 &  6.18$\pm$0.70 & 45.39$\pm$4.63 & 12.18$\pm$1.24 \\
32 &  1.53$\pm$0.06 &  0.44$\pm$0.12 &  0.28$\pm$0.09 &  1.42$\pm$0.21 &  3.29$\pm$0.39 &  2.20$\pm$0.24 \\
\hline \hline
\end{tabular}
\end{center}
\end{minipage}
\end{table*}

   \begin{figure*}
   \centering
   \includegraphics[width=4.5cm,bb=25 30 430 230]{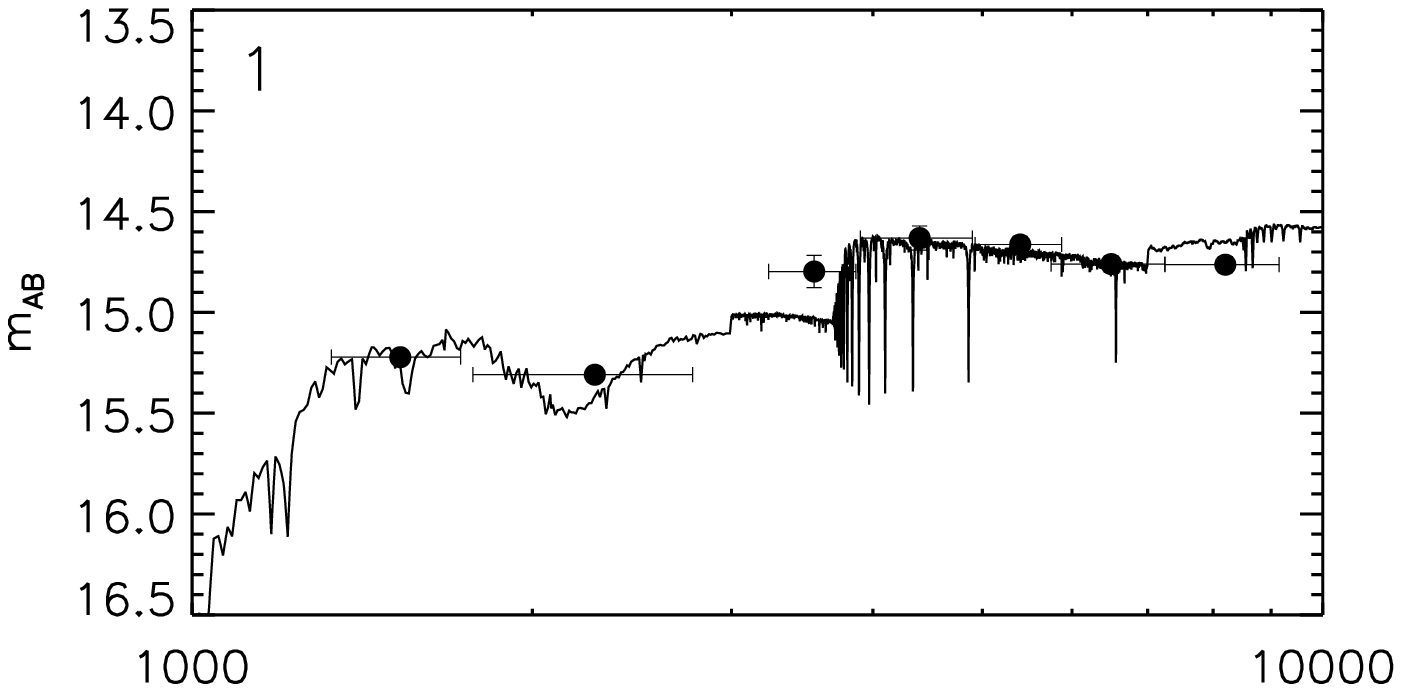}
   \includegraphics[width=4.5cm,bb=25 30 430 230]{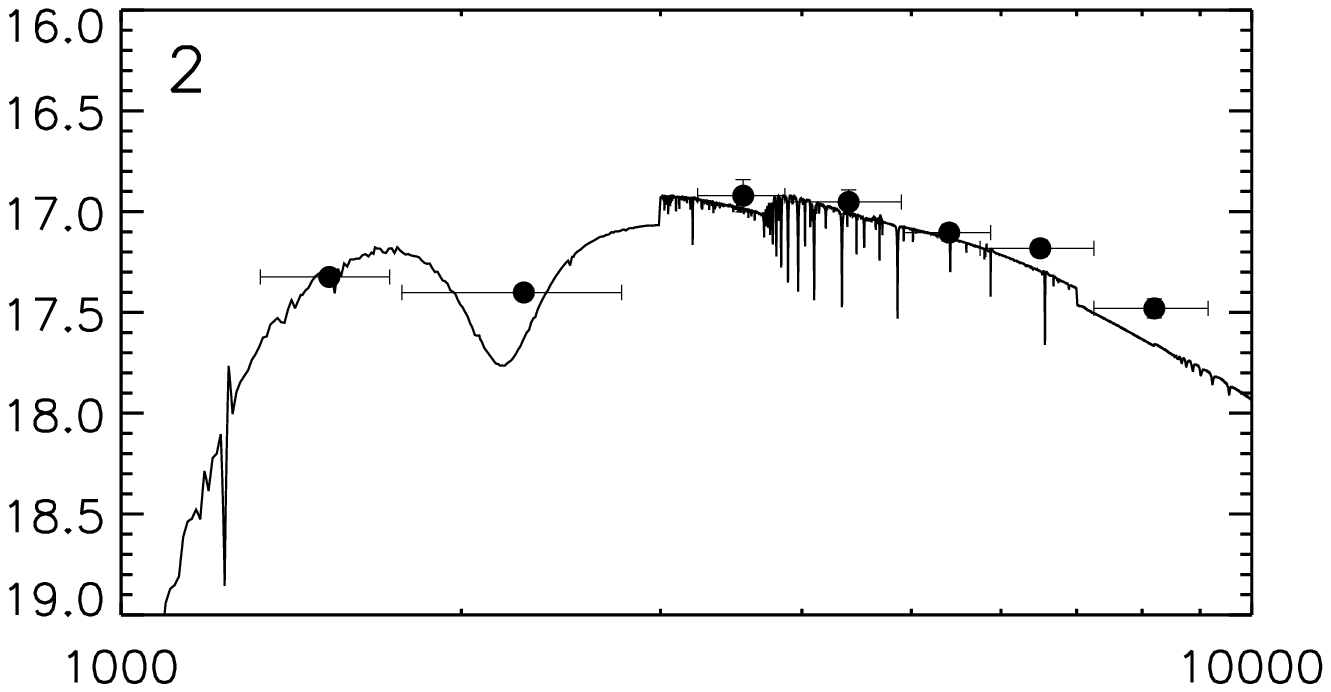}
   \includegraphics[width=4.5cm,bb=25 30 430 230]{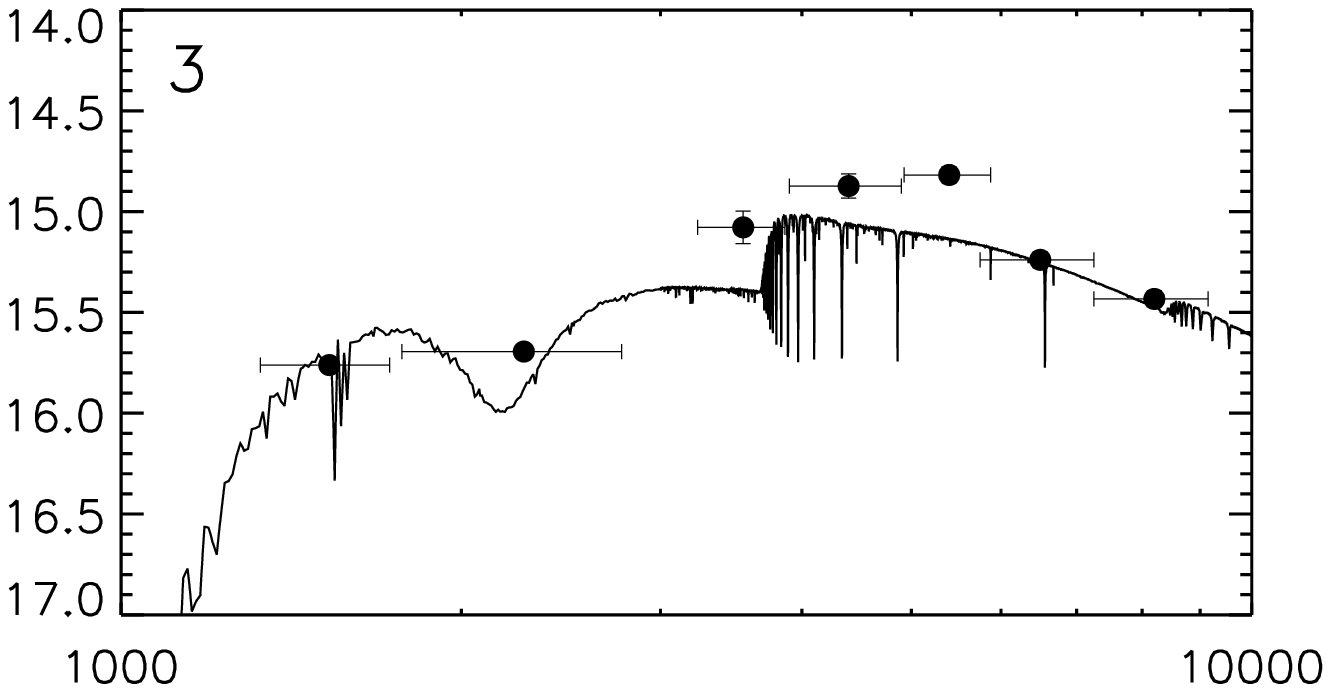}
   \includegraphics[width=4.5cm,bb=25 30 430 230]{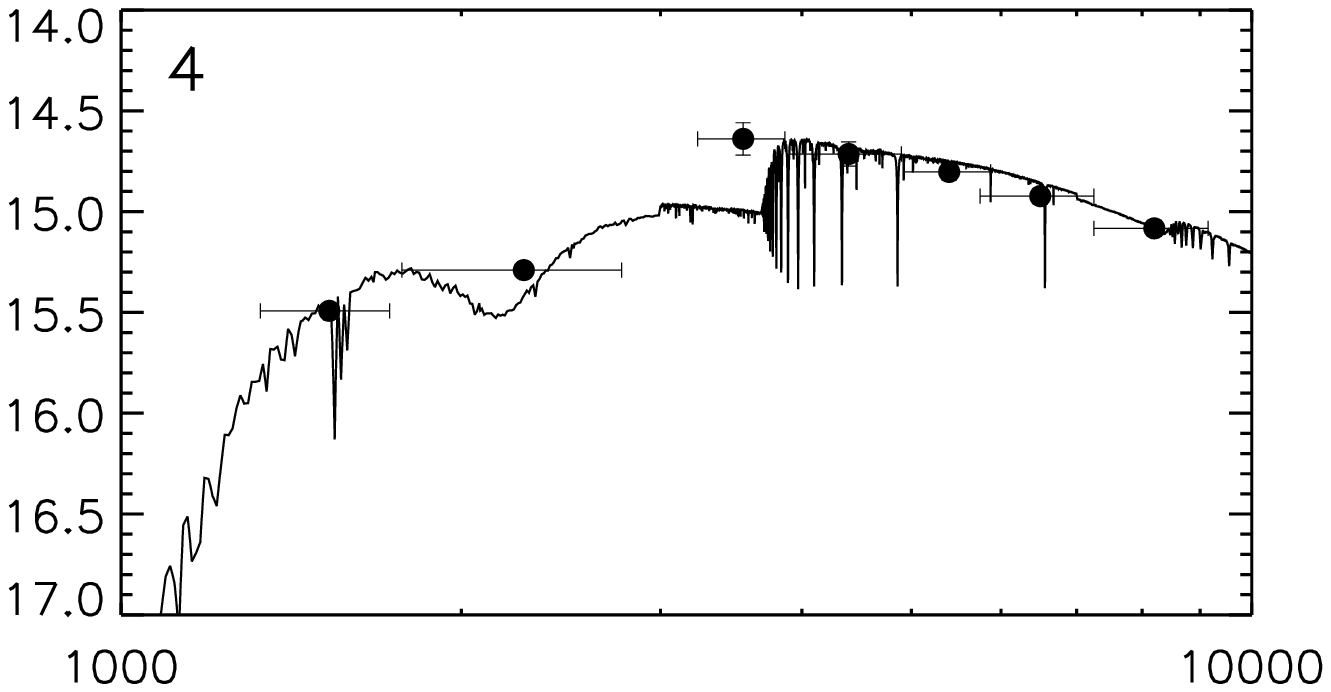}
   \includegraphics[width=4.5cm,bb=25 30 430 230]{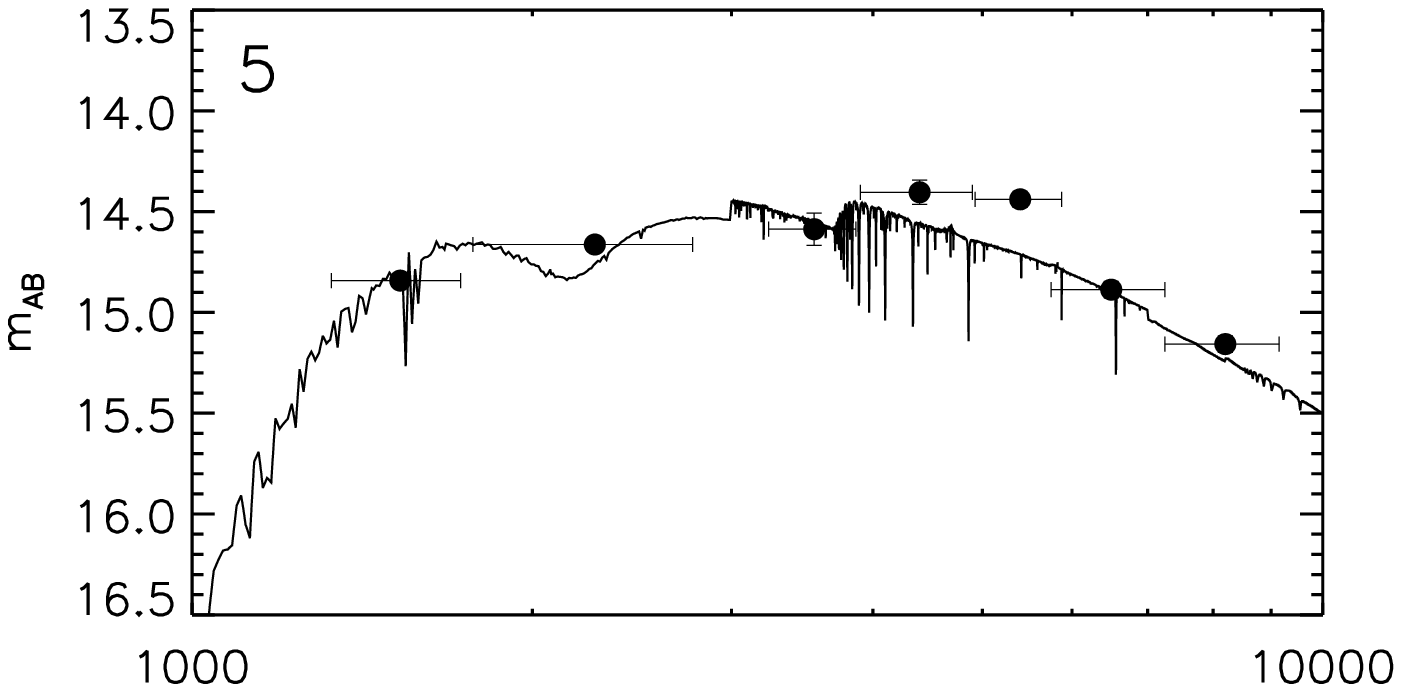}
   \includegraphics[width=4.5cm,bb=25 30 430 230]{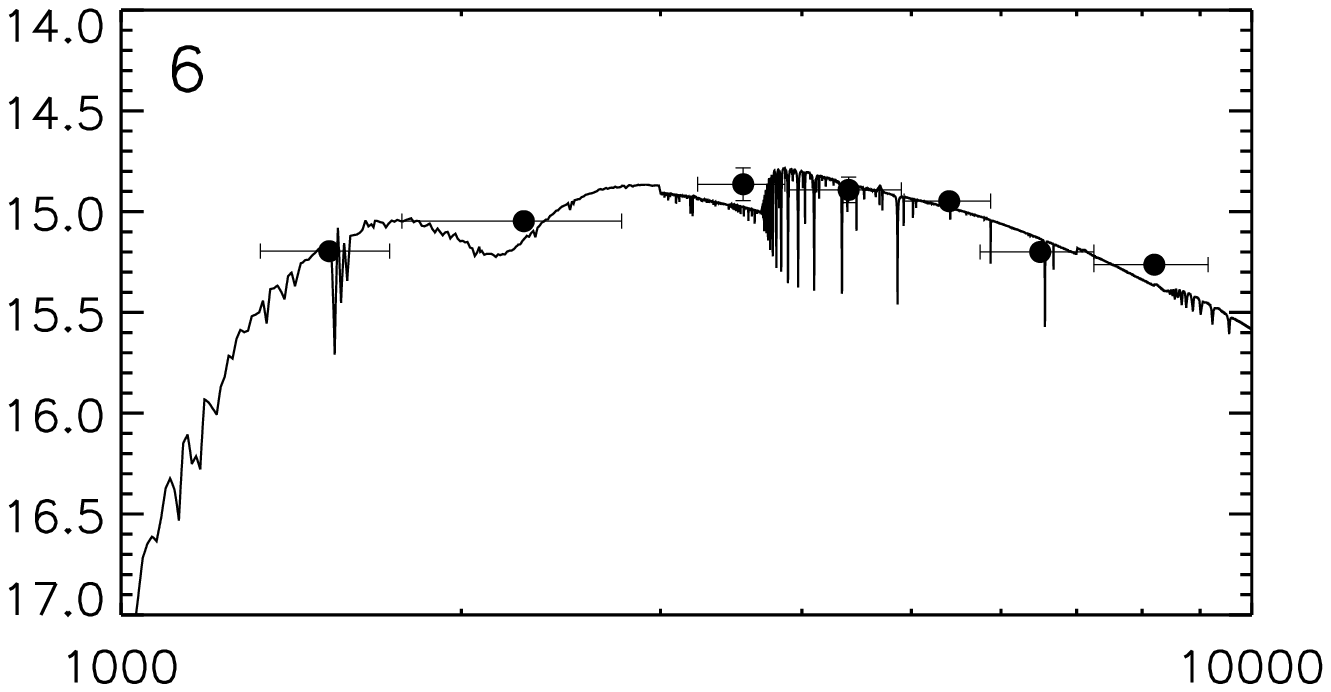}
   \includegraphics[width=4.5cm,bb=25 30 430 230]{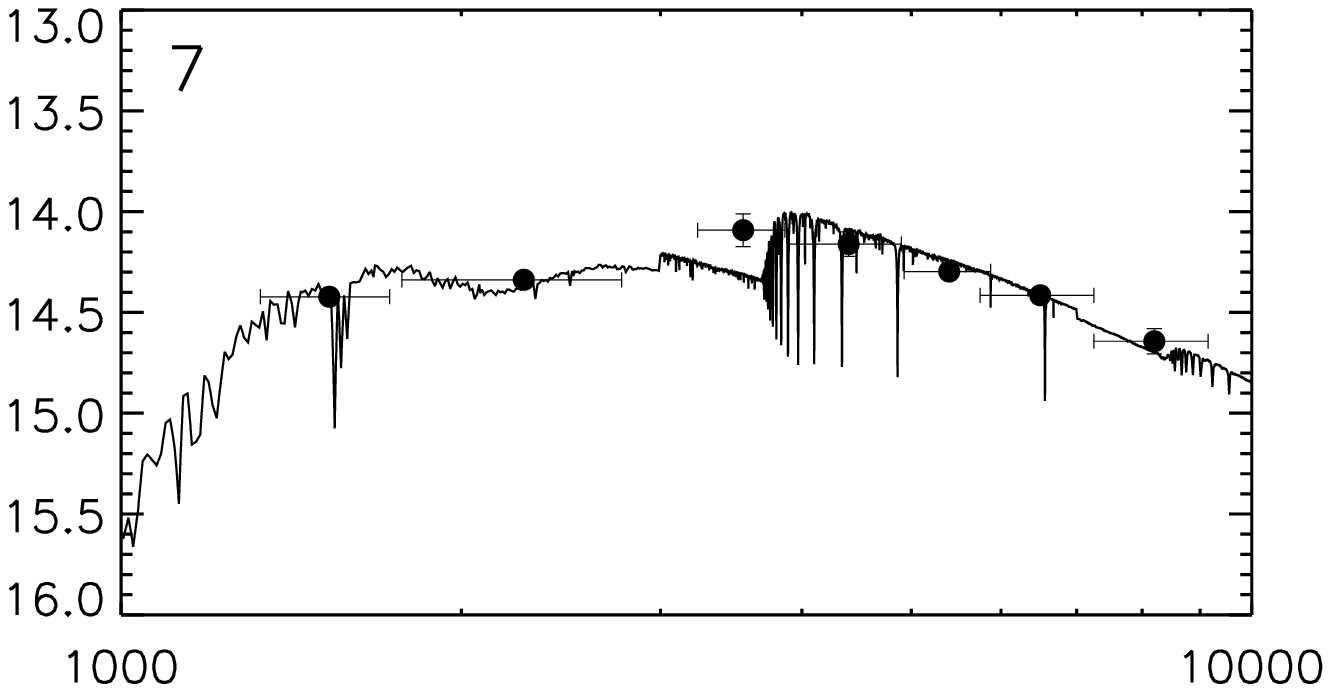}
   \includegraphics[width=4.5cm,bb=25 30 430 230]{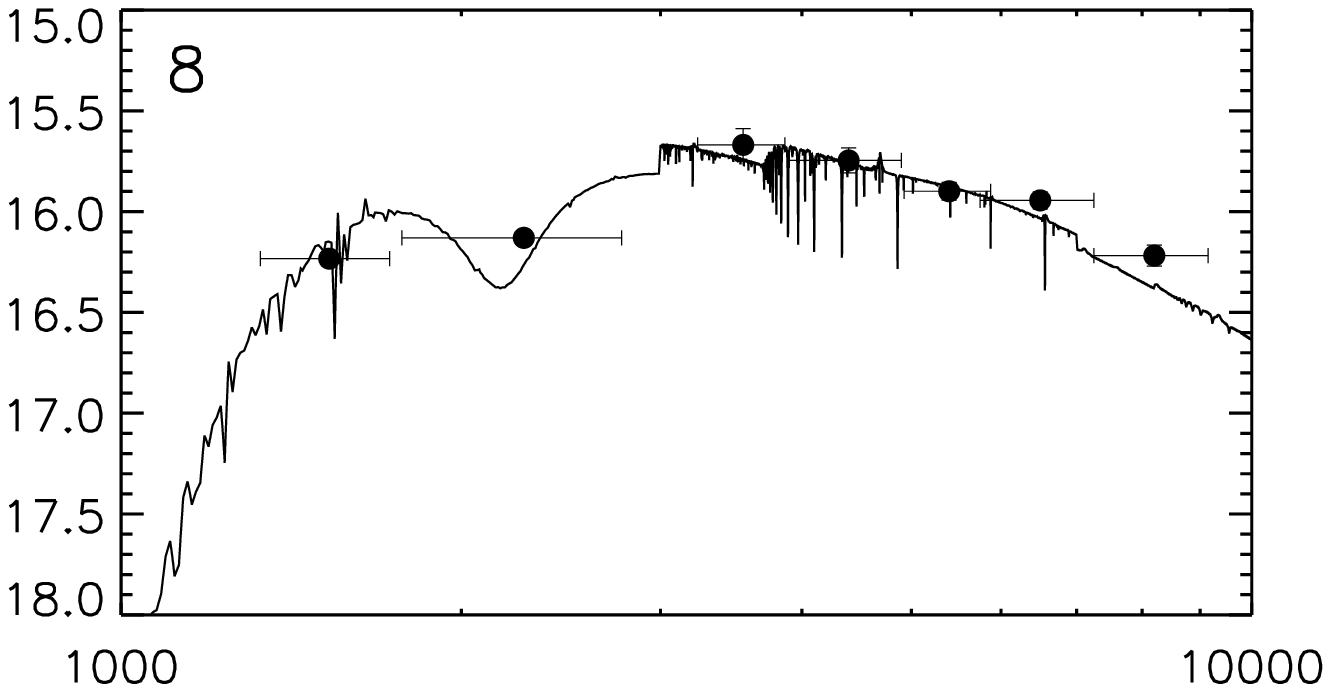}
   \includegraphics[width=4.5cm,bb=25 30 430 230]{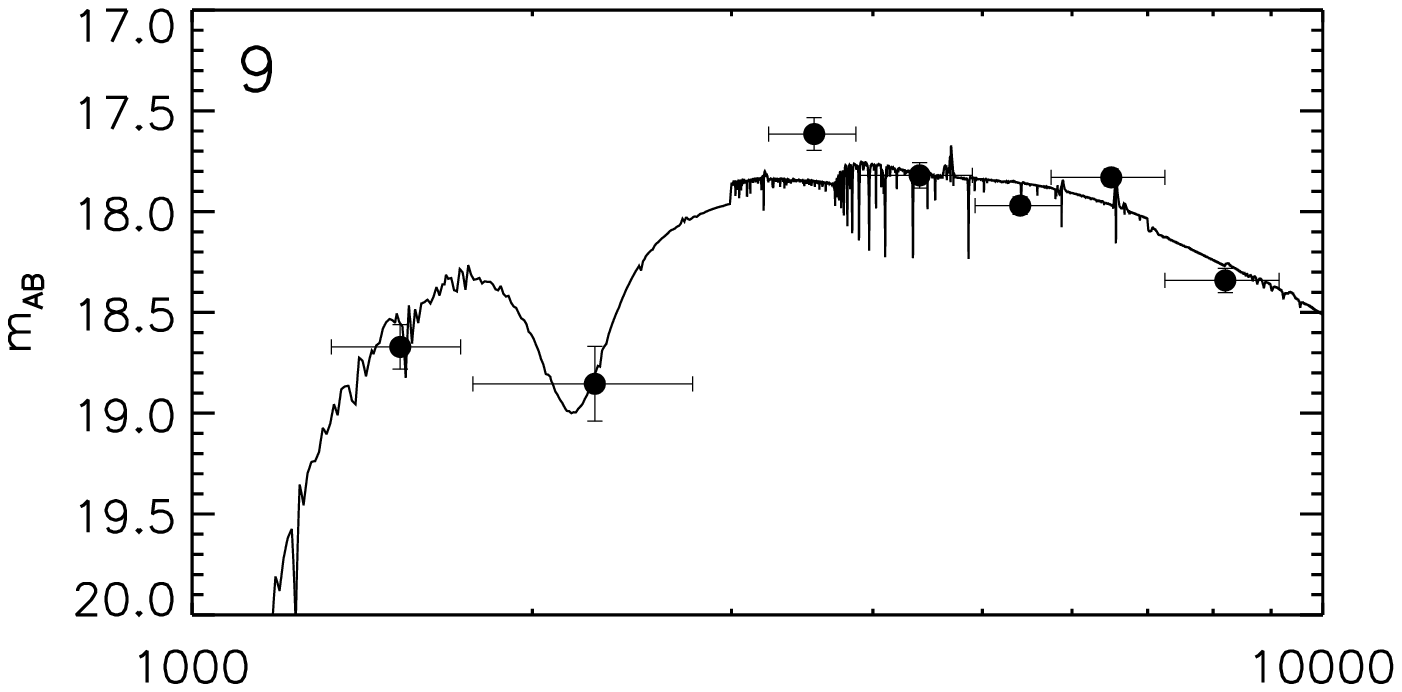}
   \includegraphics[width=4.5cm,bb=25 30 430 230]{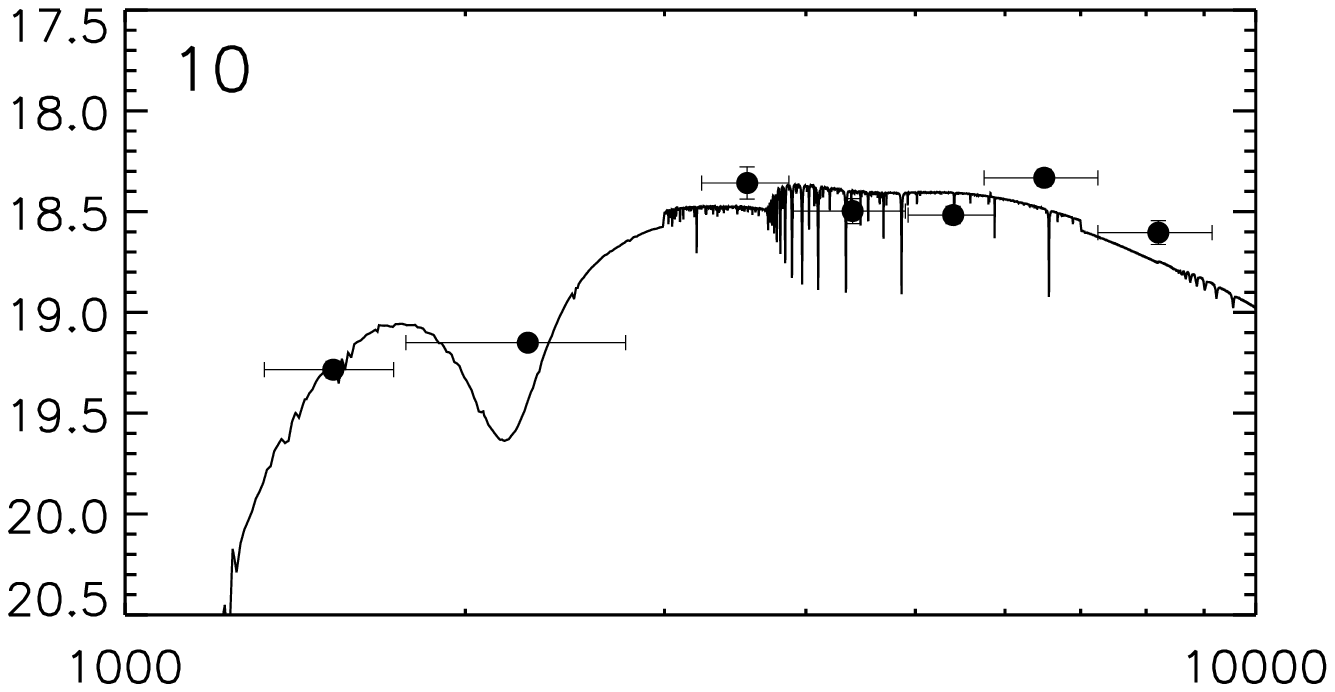}
   \includegraphics[width=4.5cm,bb=25 30 430 230]{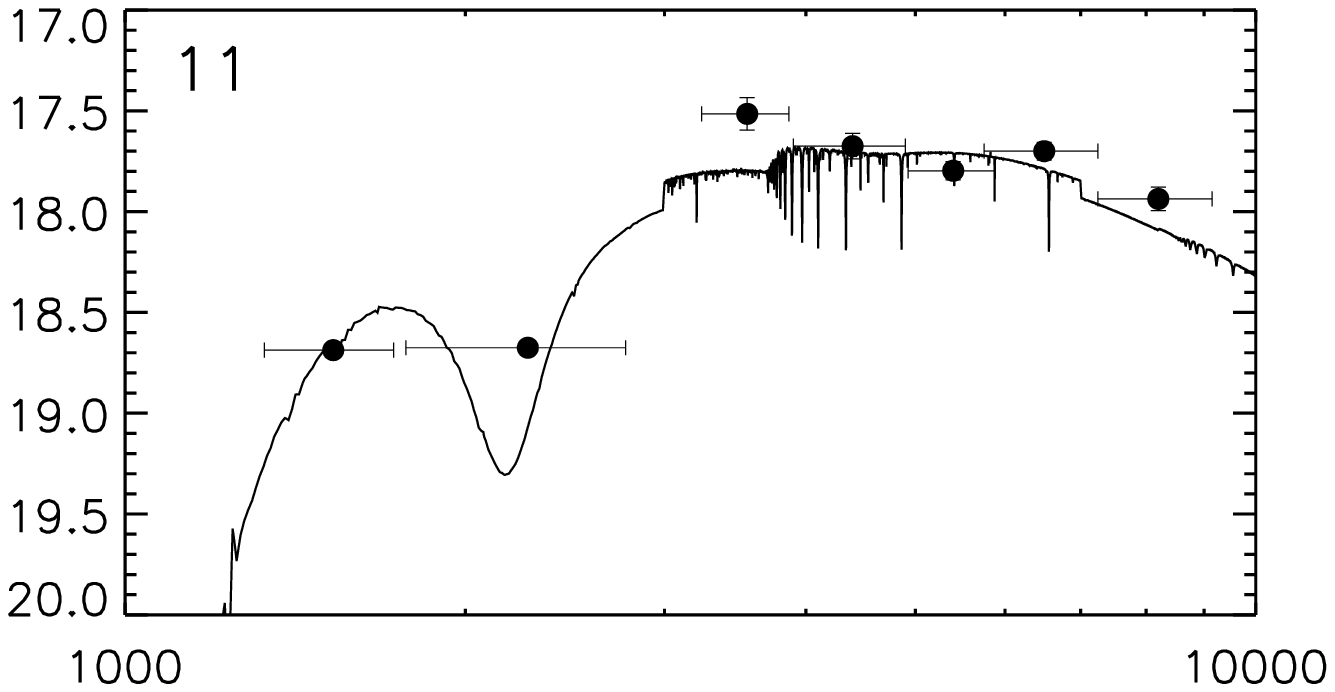}
   \includegraphics[width=4.5cm,bb=25 30 430 230]{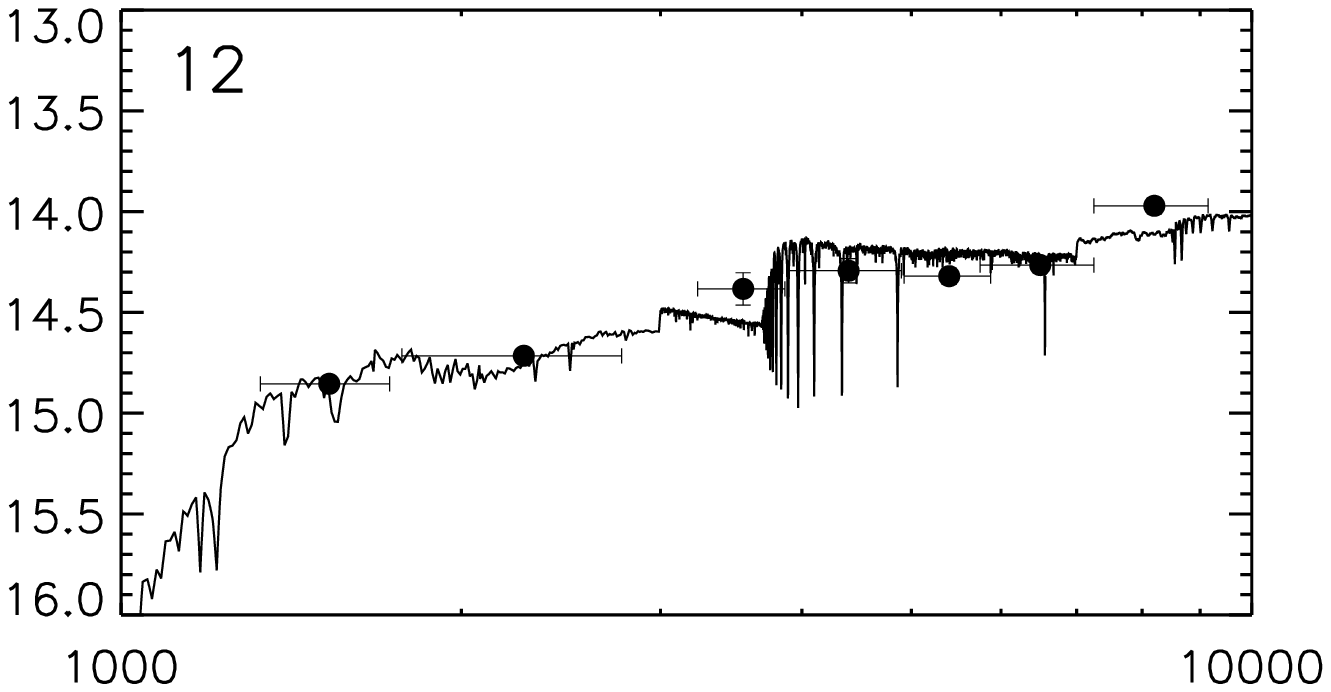}
   \includegraphics[width=4.5cm,bb=25 30 430 230]{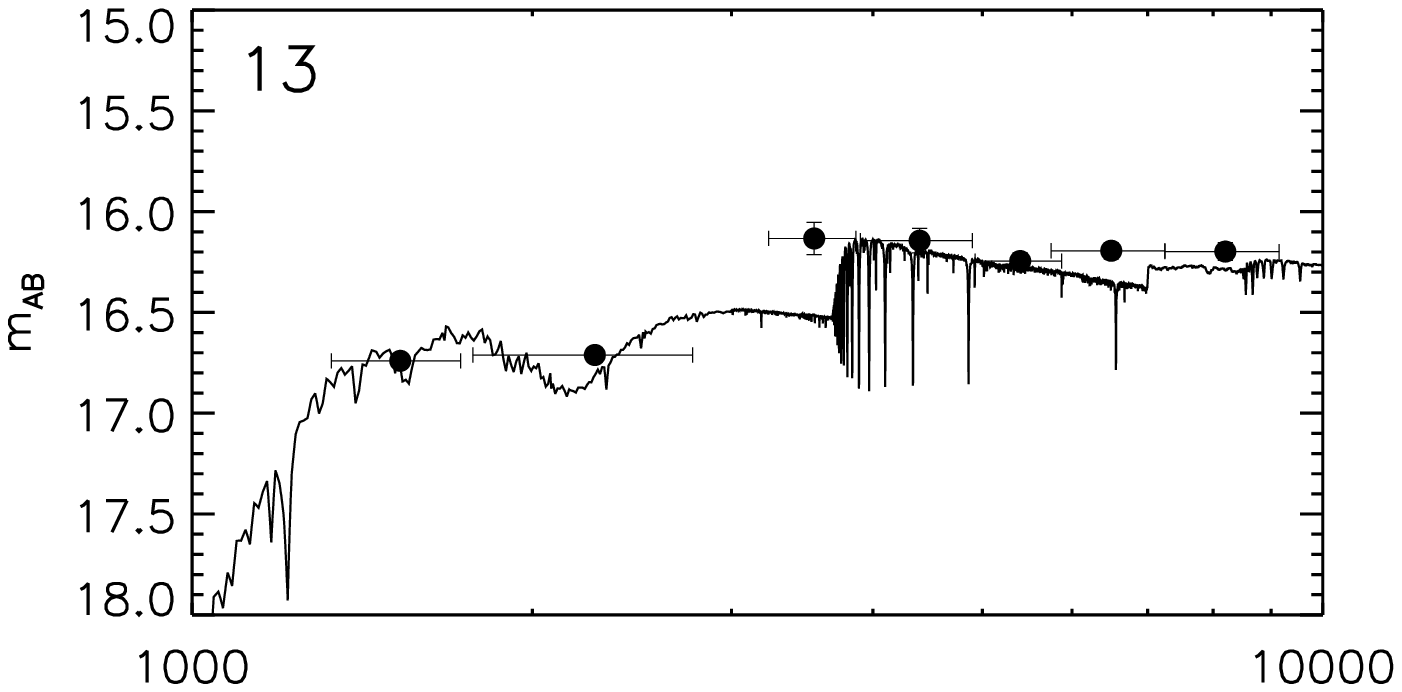}
   \includegraphics[width=4.5cm,bb=25 30 430 230]{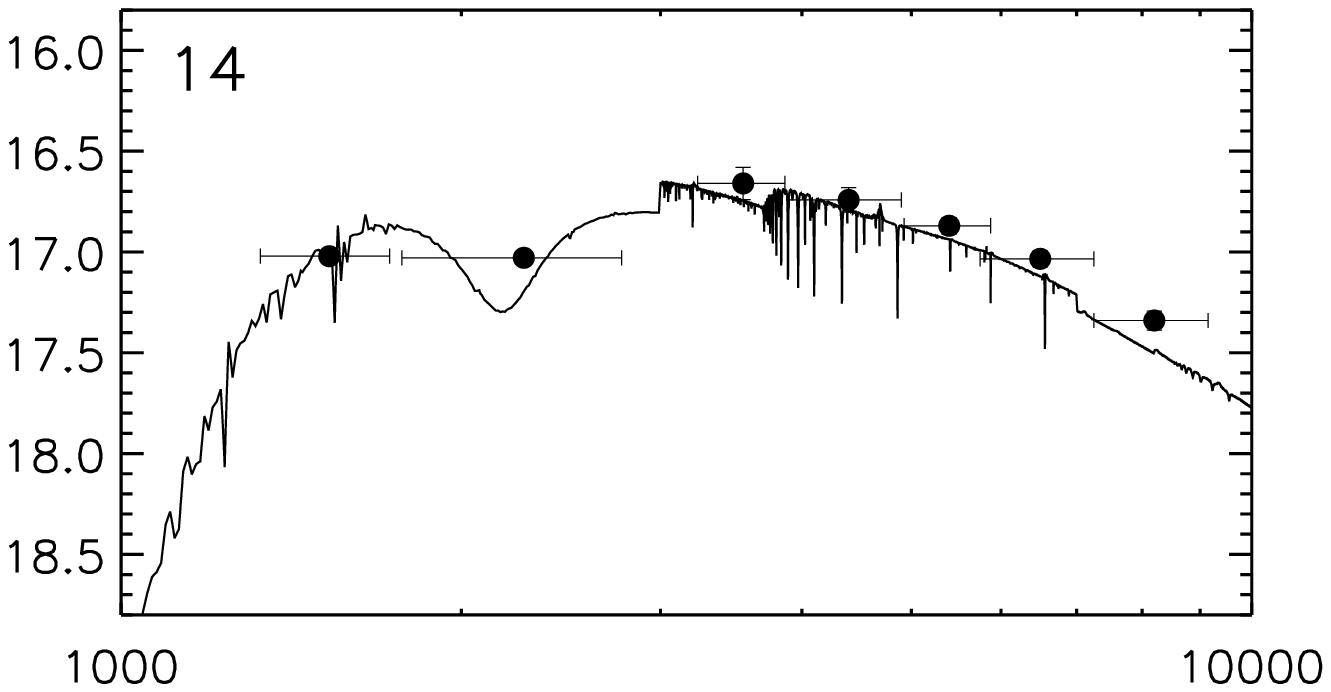}
   \includegraphics[width=4.5cm,bb=25 30 430 230]{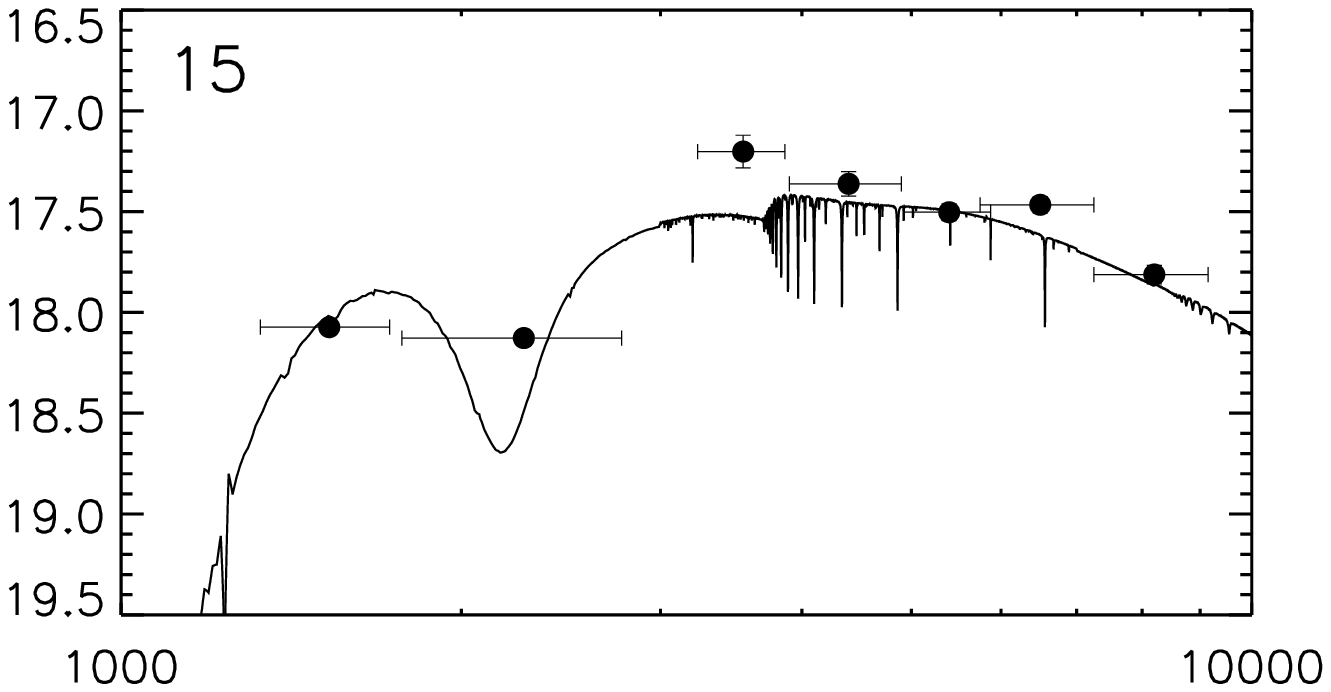}
   \includegraphics[width=4.5cm,bb=25 30 430 230]{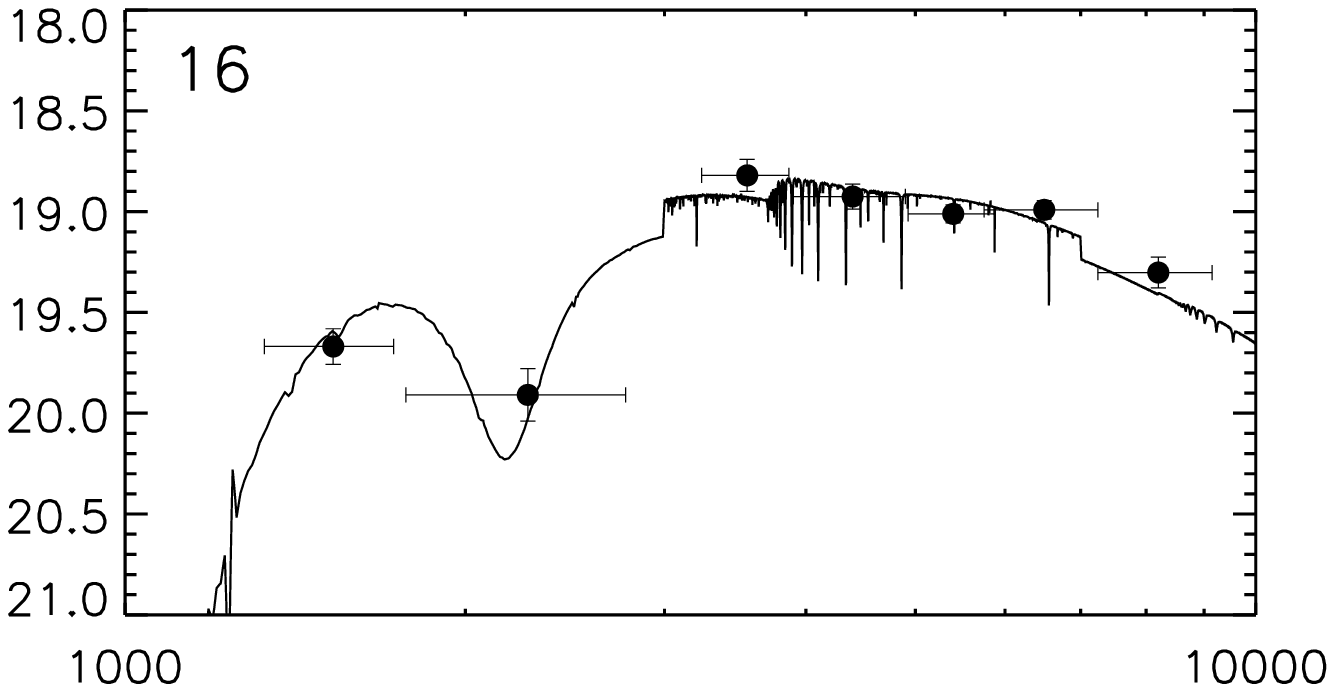}
   \includegraphics[width=4.5cm,bb=25 30 430 230]{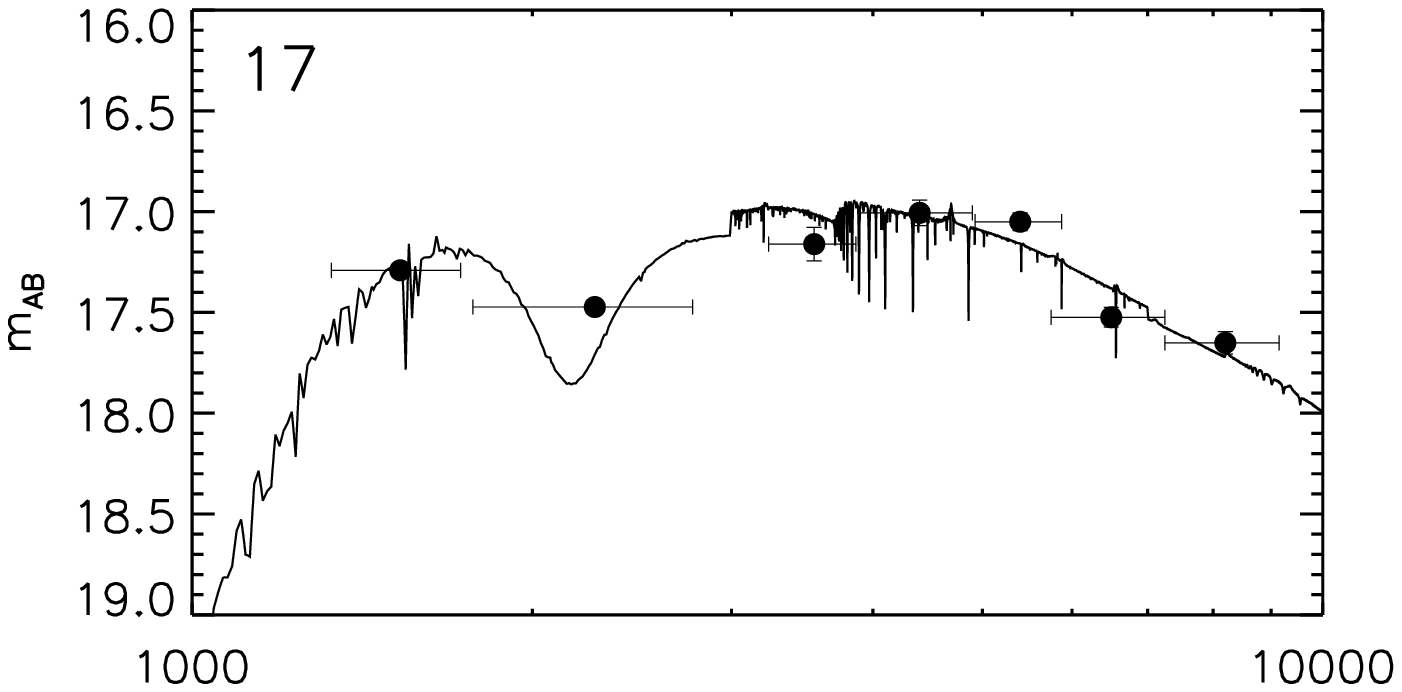}
   \includegraphics[width=4.5cm,bb=25 30 430 230]{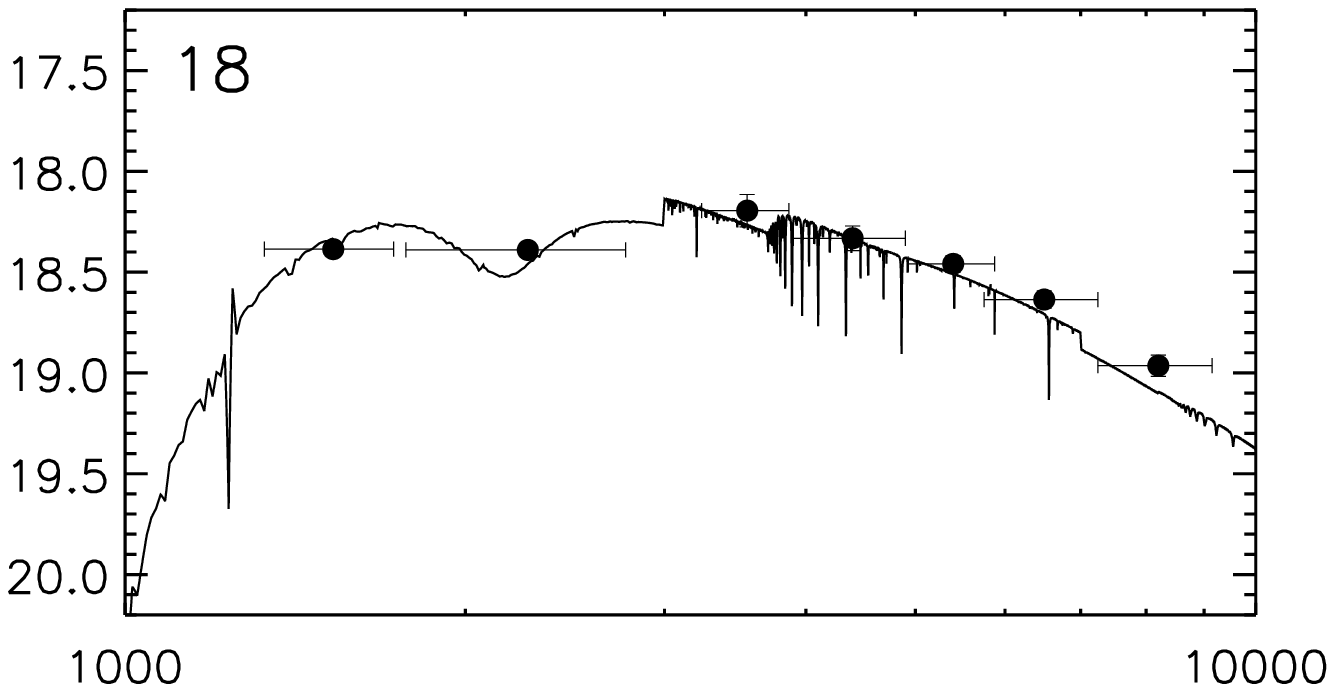}
   \includegraphics[width=4.5cm,bb=25 30 430 230]{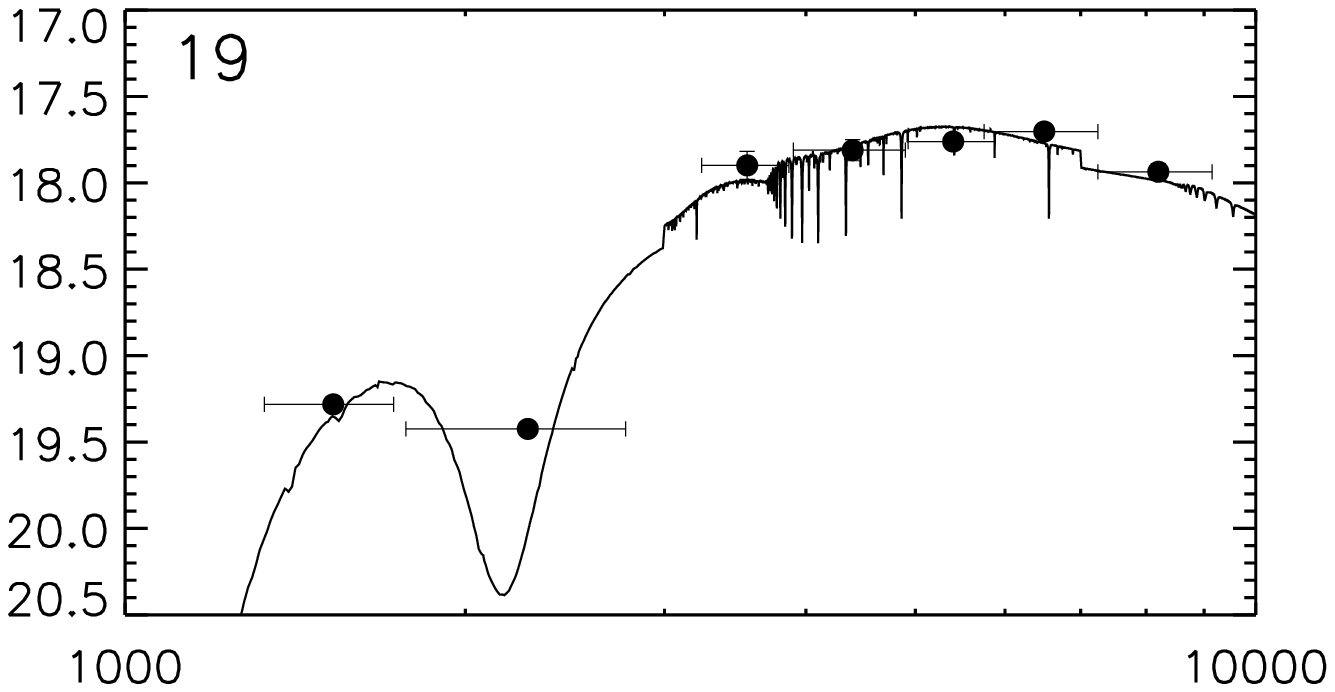}
   \includegraphics[width=4.5cm,bb=25 30 430 230]{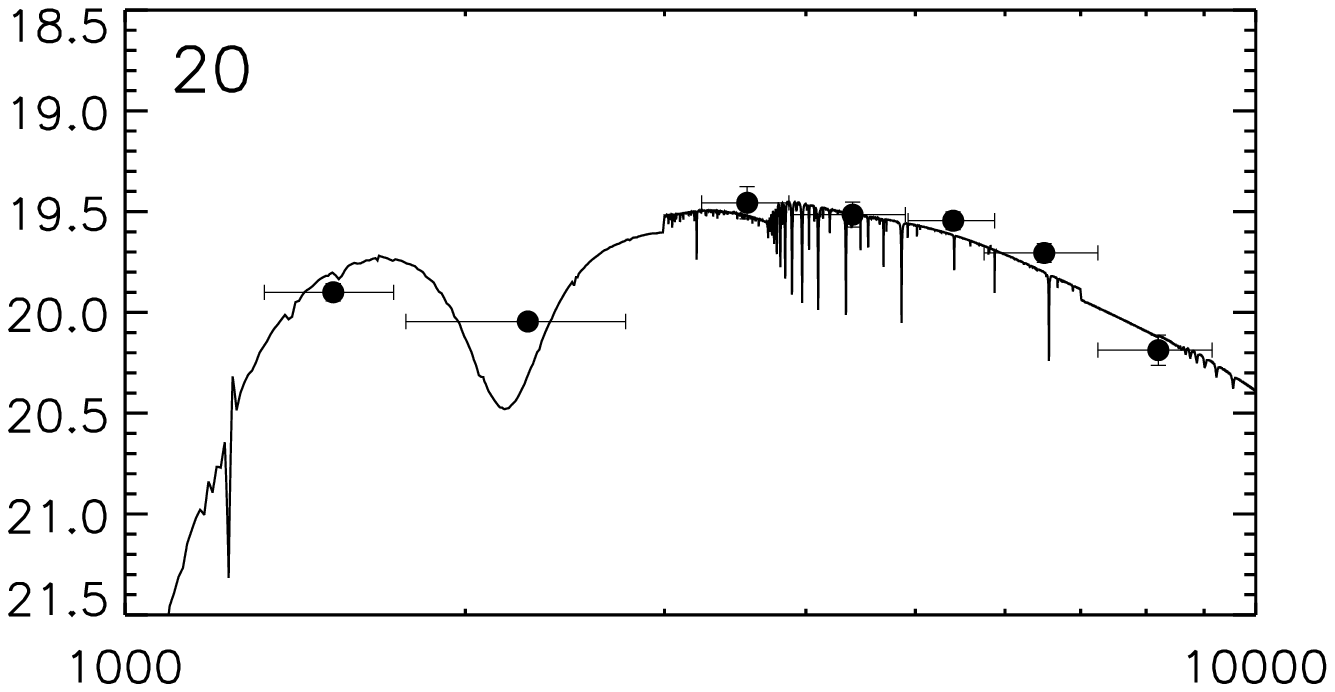}
   \includegraphics[width=4.5cm,bb=25 30 430 230]{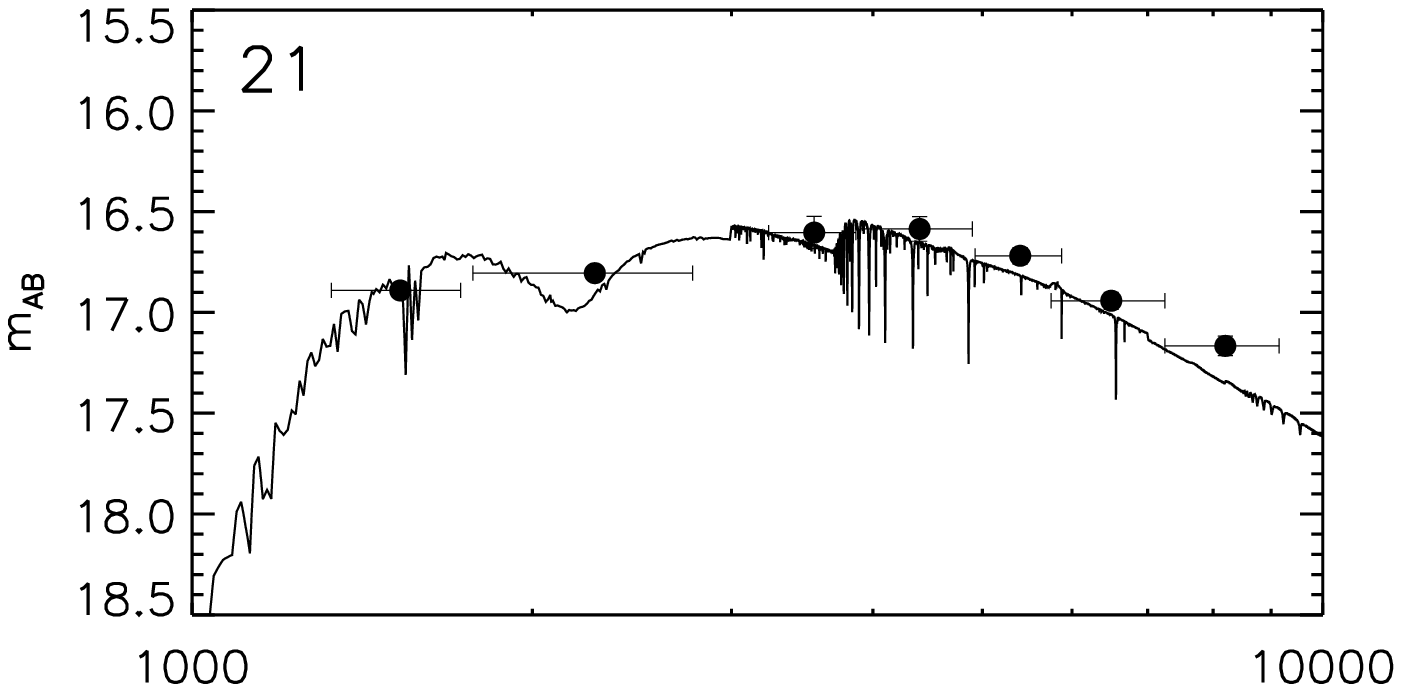}
   \includegraphics[width=4.5cm,bb=25 30 430 230]{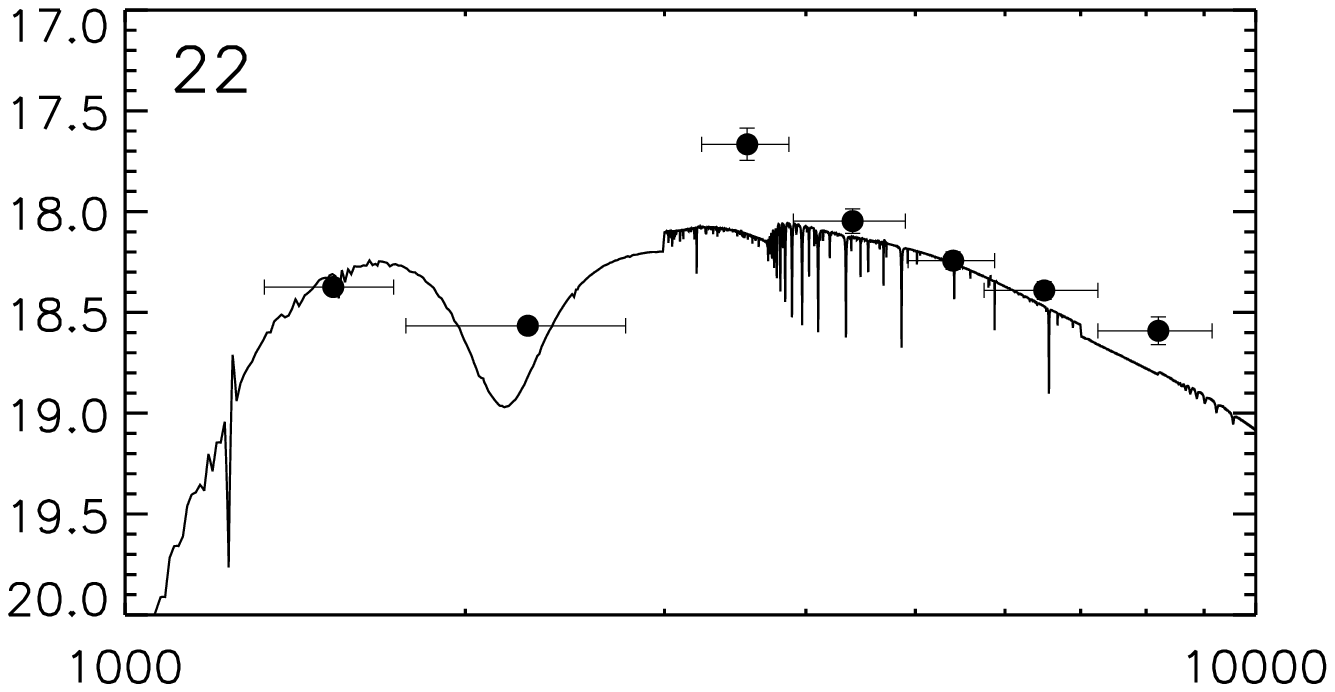}
   \includegraphics[width=4.5cm,bb=25 30 430 230]{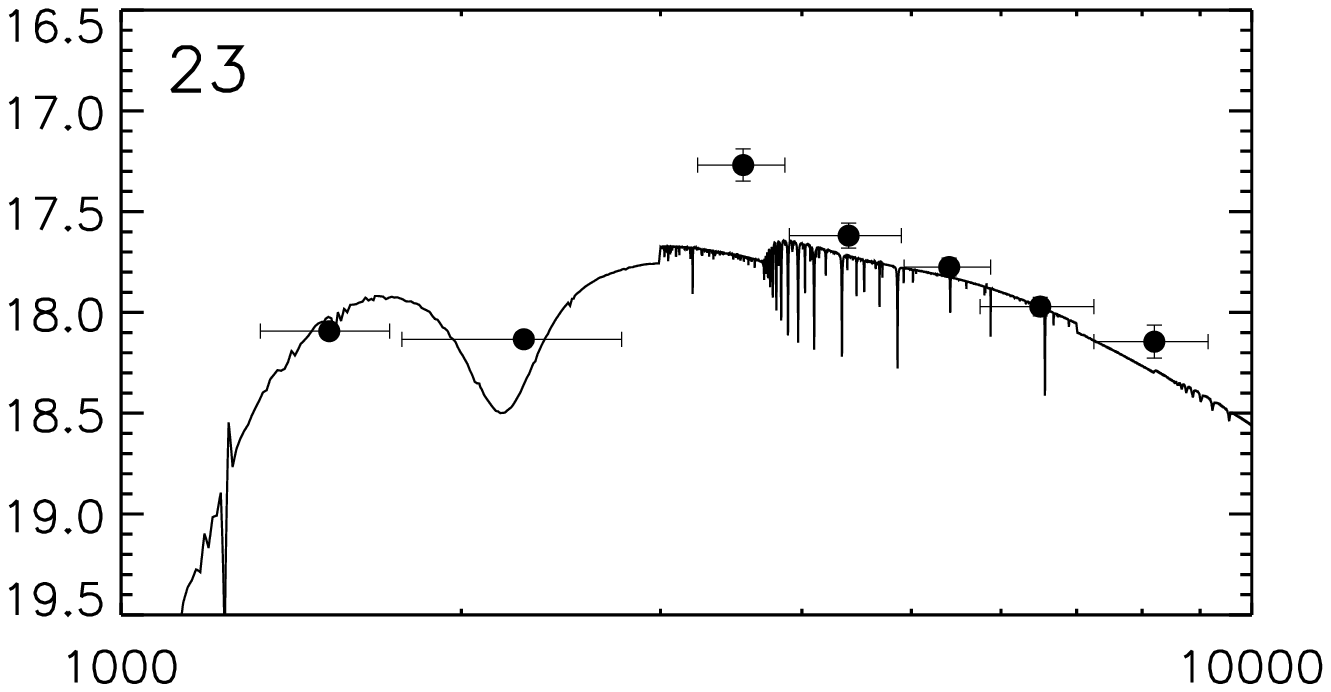}
   \includegraphics[width=4.5cm,bb=25 30 430 230]{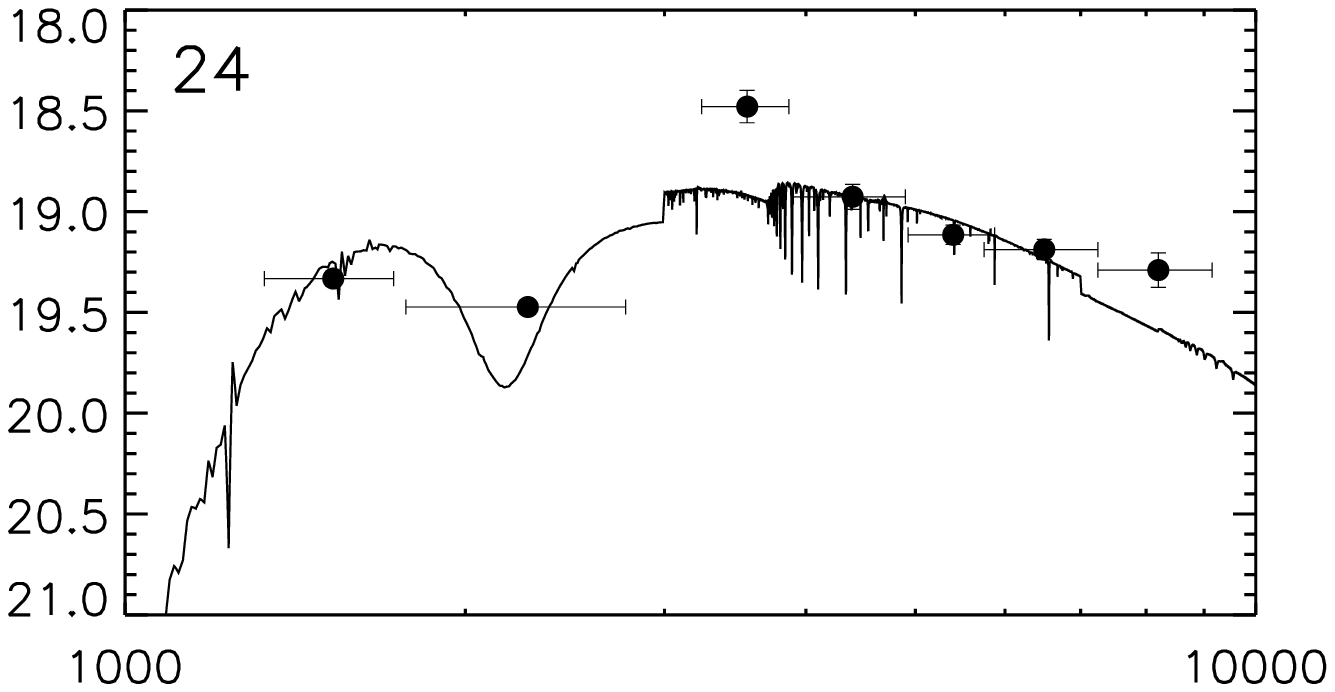}
   \includegraphics[width=4.5cm,bb=25 30 430 230]{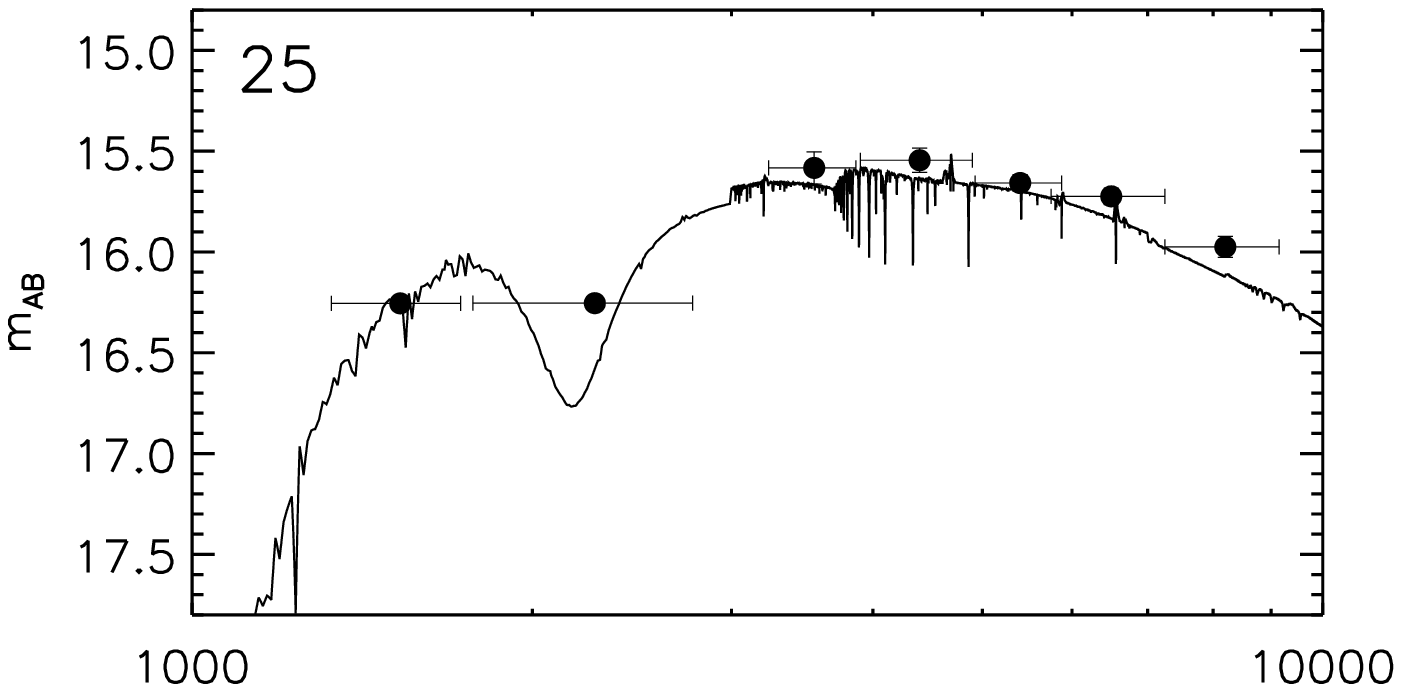}
   \includegraphics[width=4.5cm,bb=25 30 430 230]{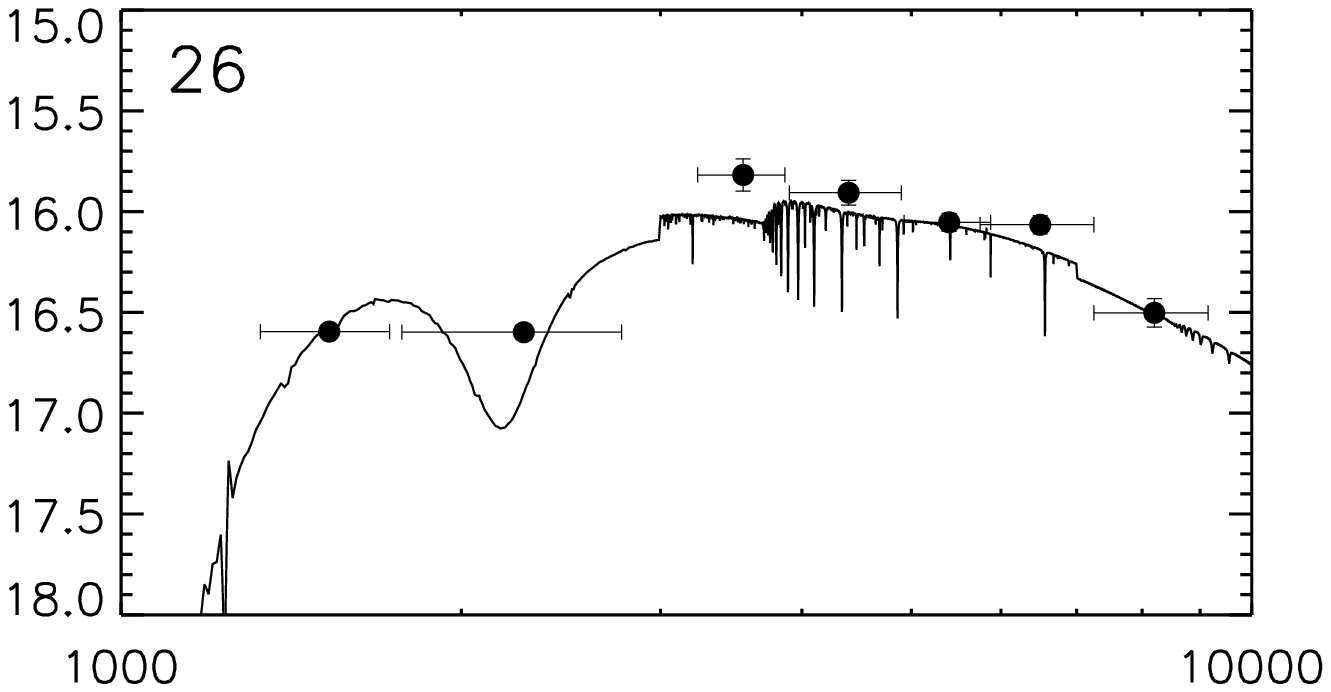}
   \includegraphics[width=4.5cm,bb=25 30 430 230]{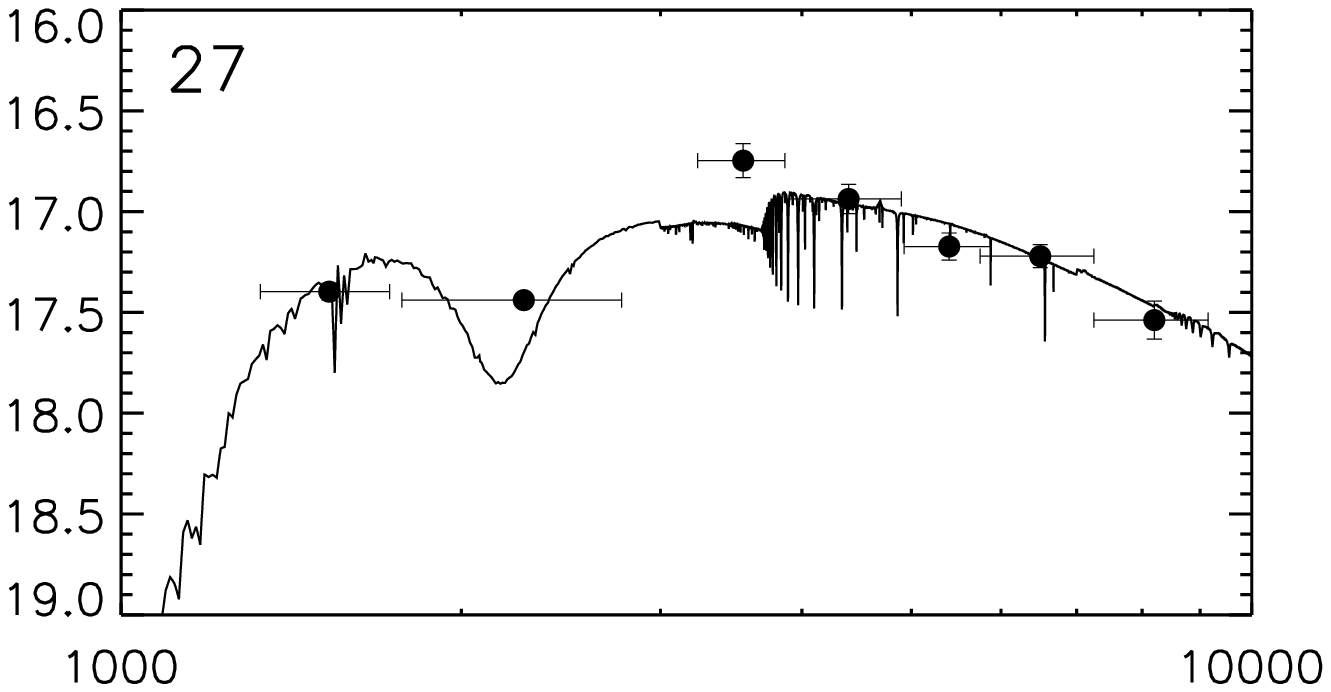}
   \includegraphics[width=4.5cm,bb=25 30 430 230]{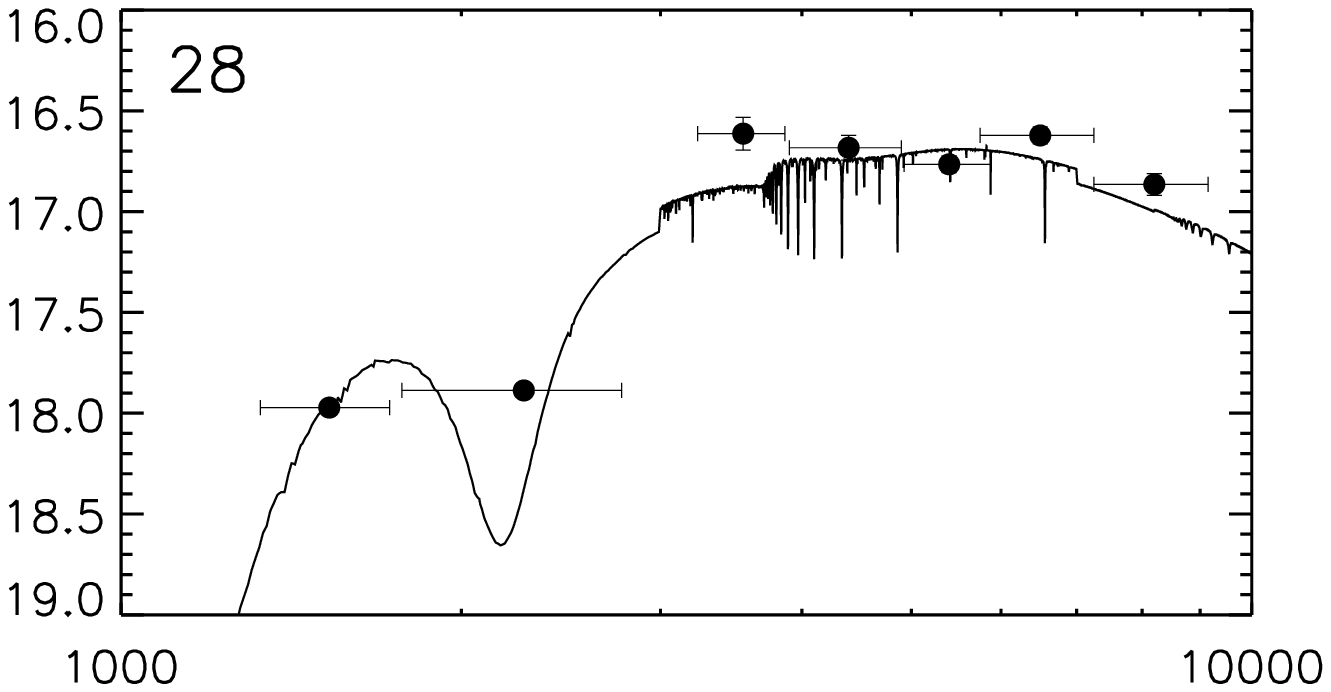}
   \includegraphics[width=4.5cm,bb=25 30 430 230]{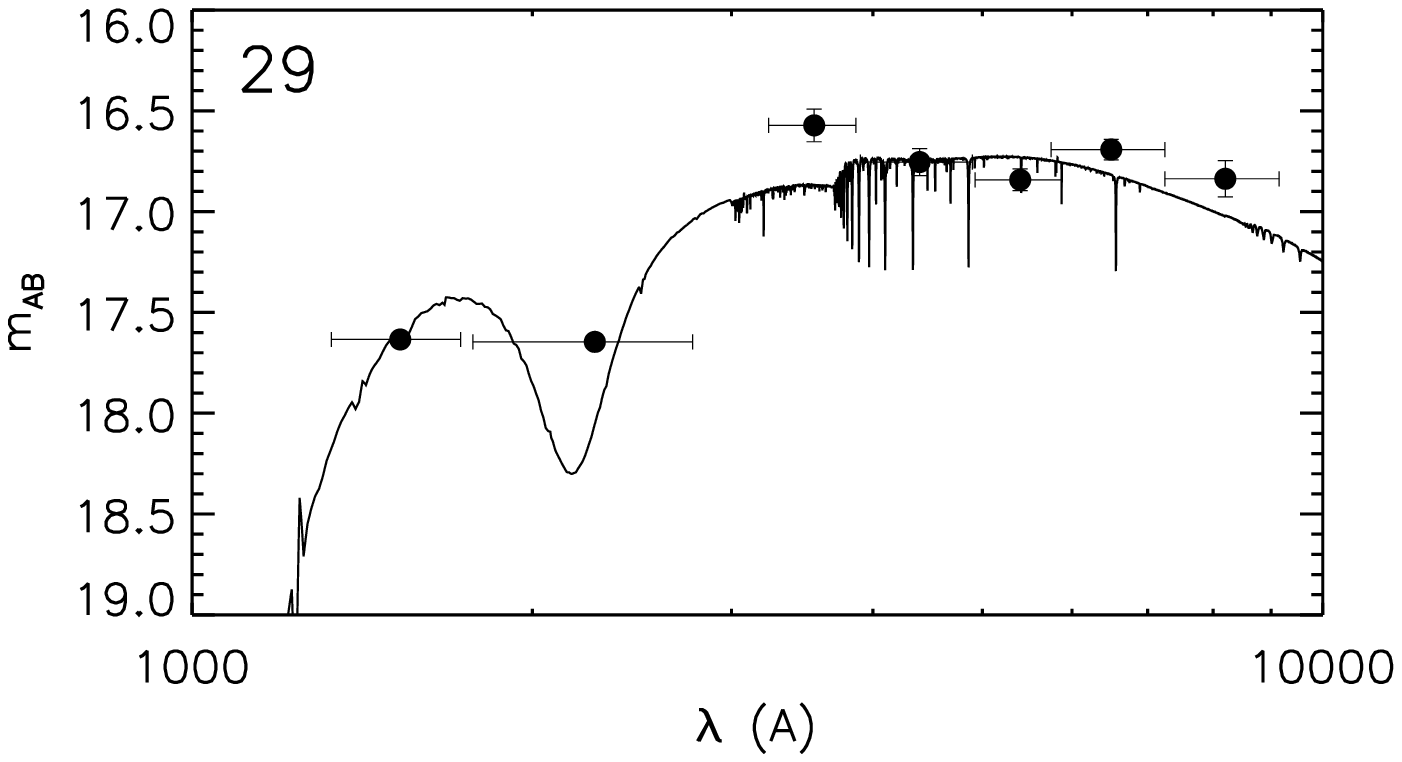}
   \includegraphics[width=4.5cm,bb=25 30 430 230]{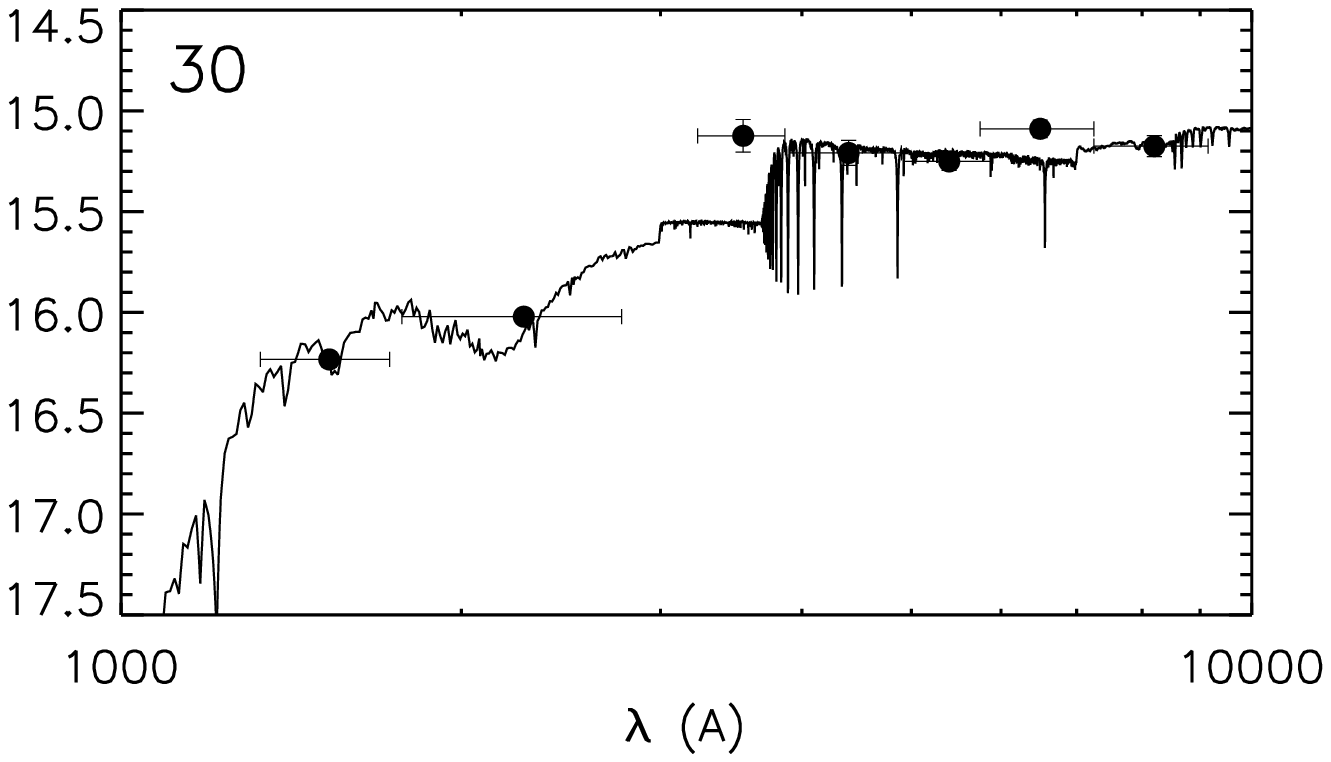}
   \includegraphics[width=4.5cm,bb=25 30 430 230]{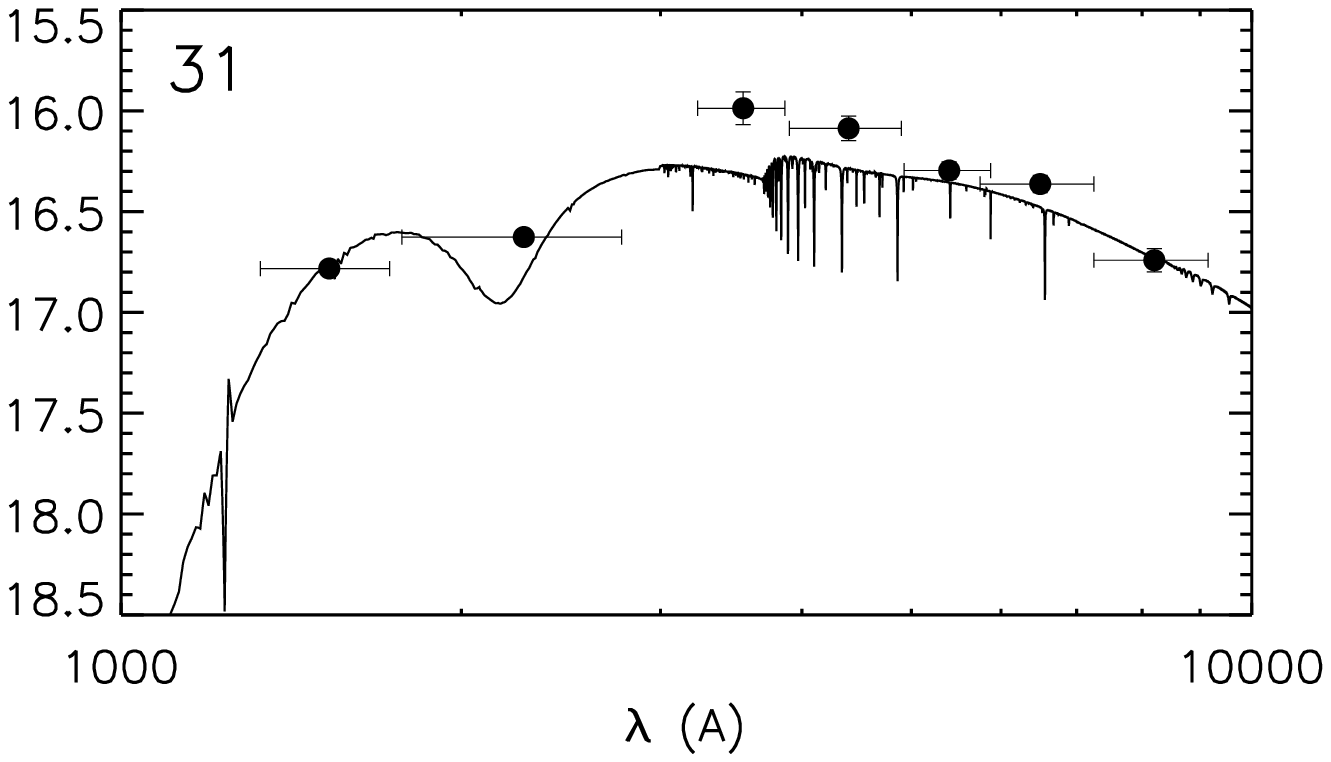}
   \includegraphics[width=4.5cm,bb=25 30 430 230]{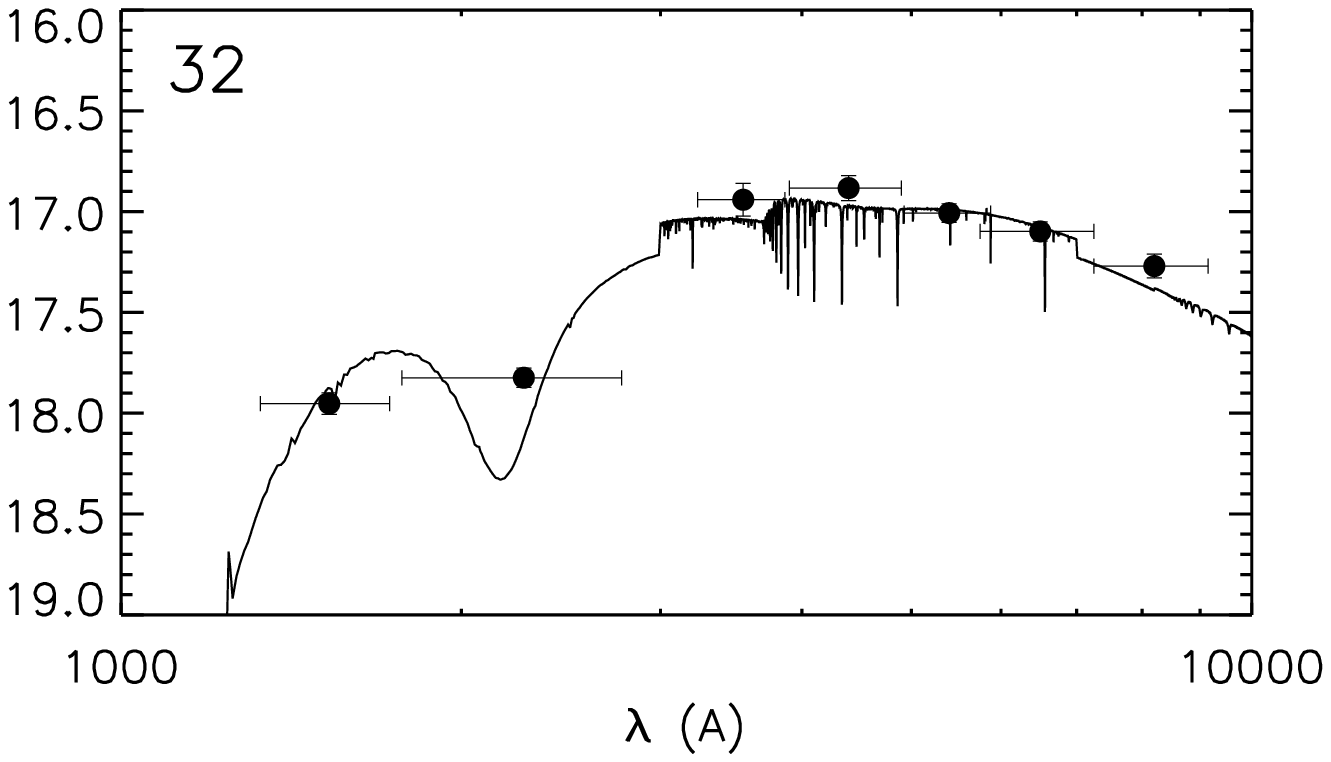}
    \caption{Output best-fit SEDs in units of AB magnitudes, overlayed on the measured integrated magnitudes
    from 1000 to 10000 $\AA$. Vertical bars indicate the uncertainty on the photometry, while the
horizontal ones show the approximate width of each filter.
The numbers at the top-left corners of the plots correspond to the cluster
ID given in Table \ref{list_clust}.}  \label{spec1}%
    \end{figure*}

   \begin{figure*}
   \centering
   \includegraphics[width=4.7cm,bb=20 30 410 230]{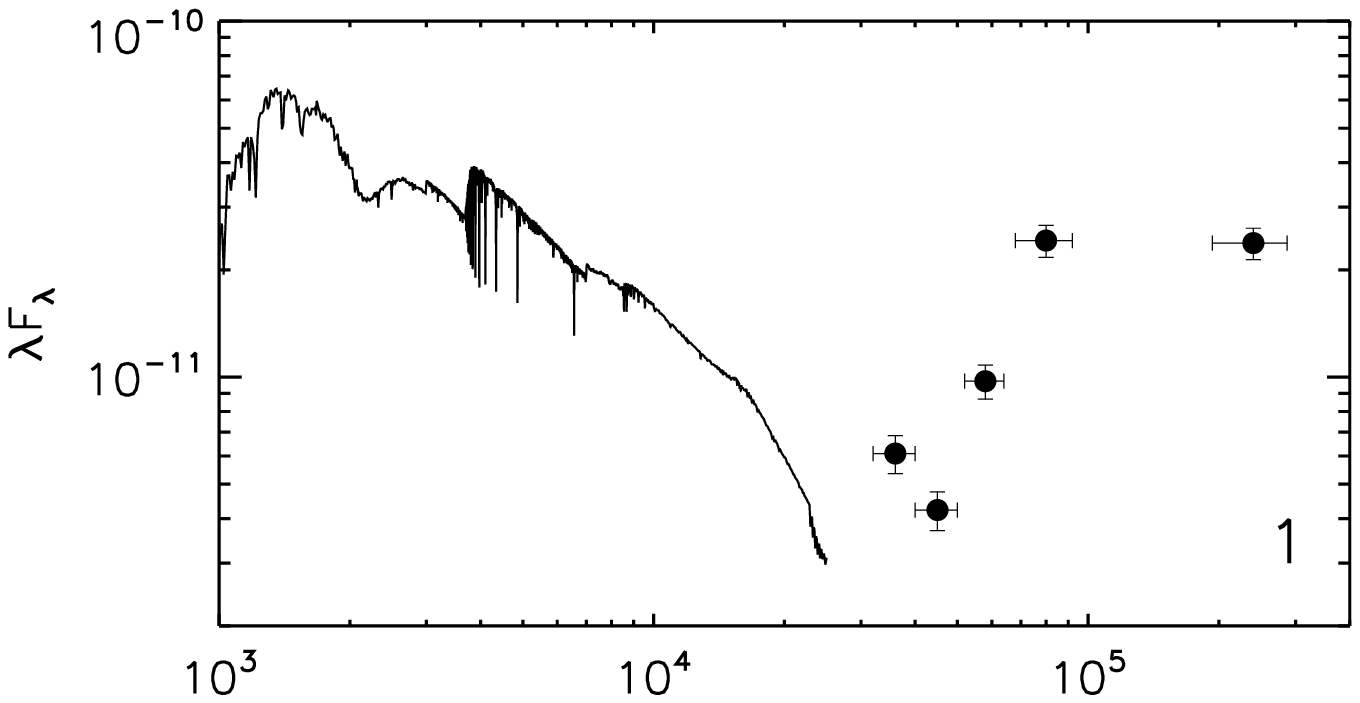}
   \includegraphics[width=4.4cm,bb=45 30 410 230]{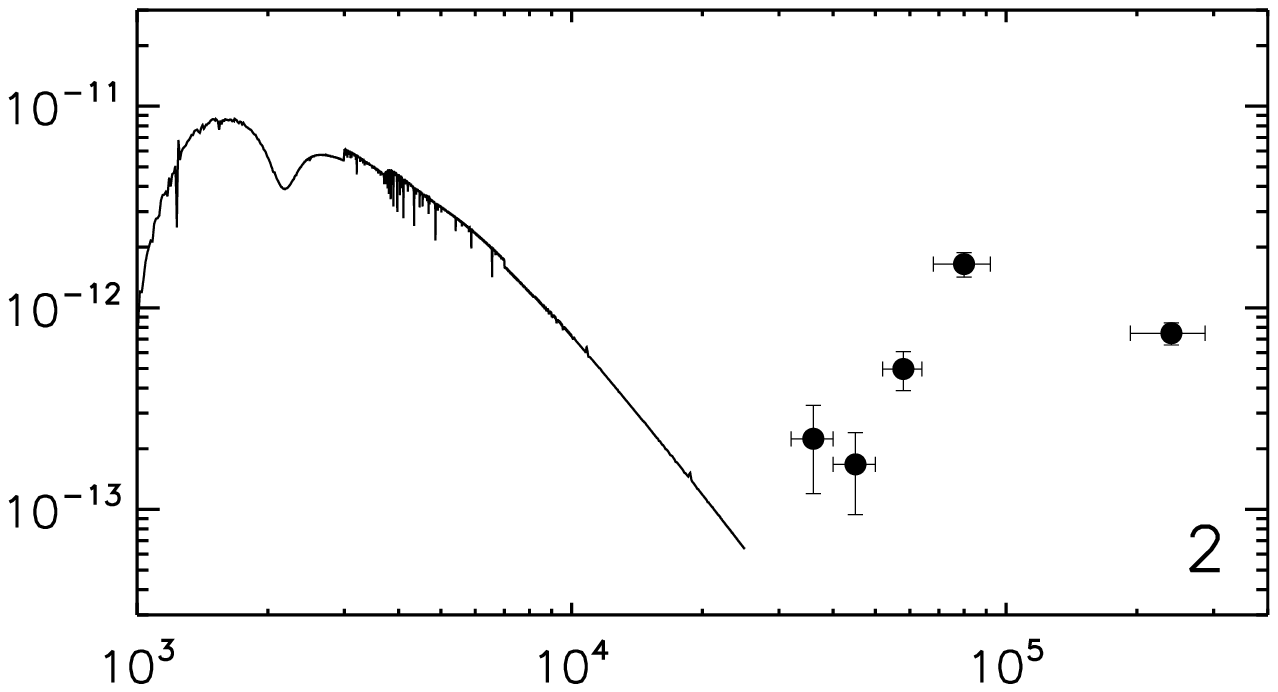}
   \includegraphics[width=4.4cm,bb=45 30 410 230]{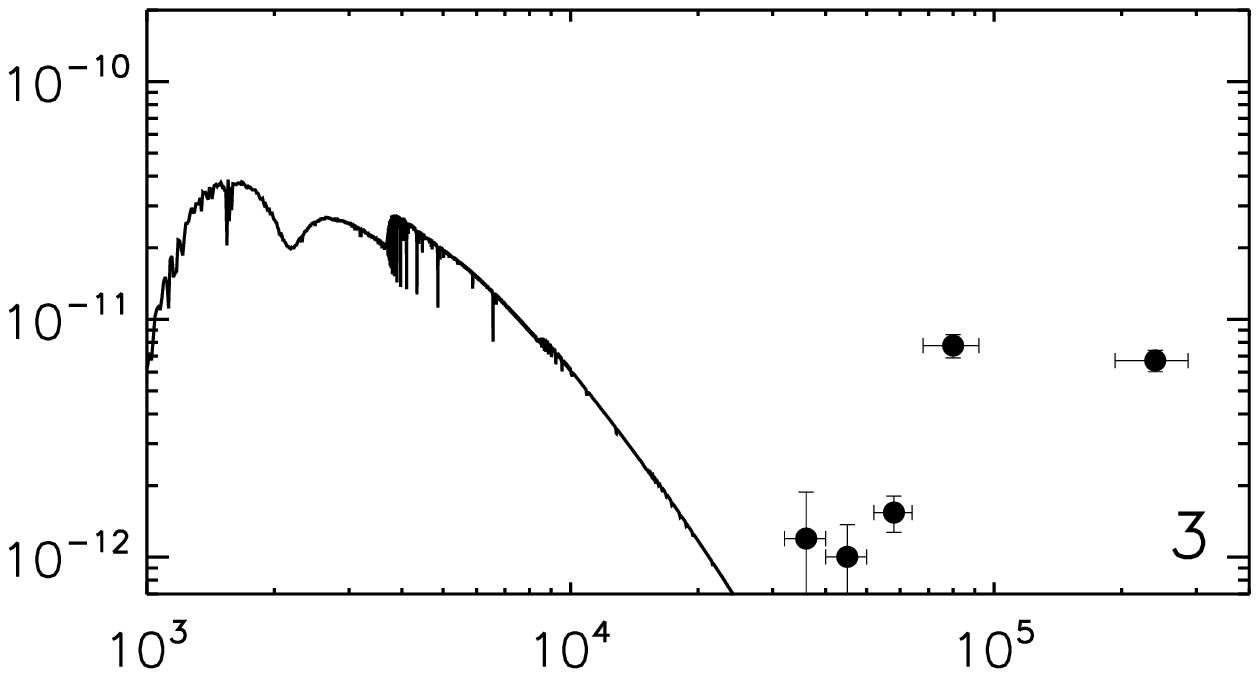}
   \includegraphics[width=4.4cm,bb=45 30 410 230]{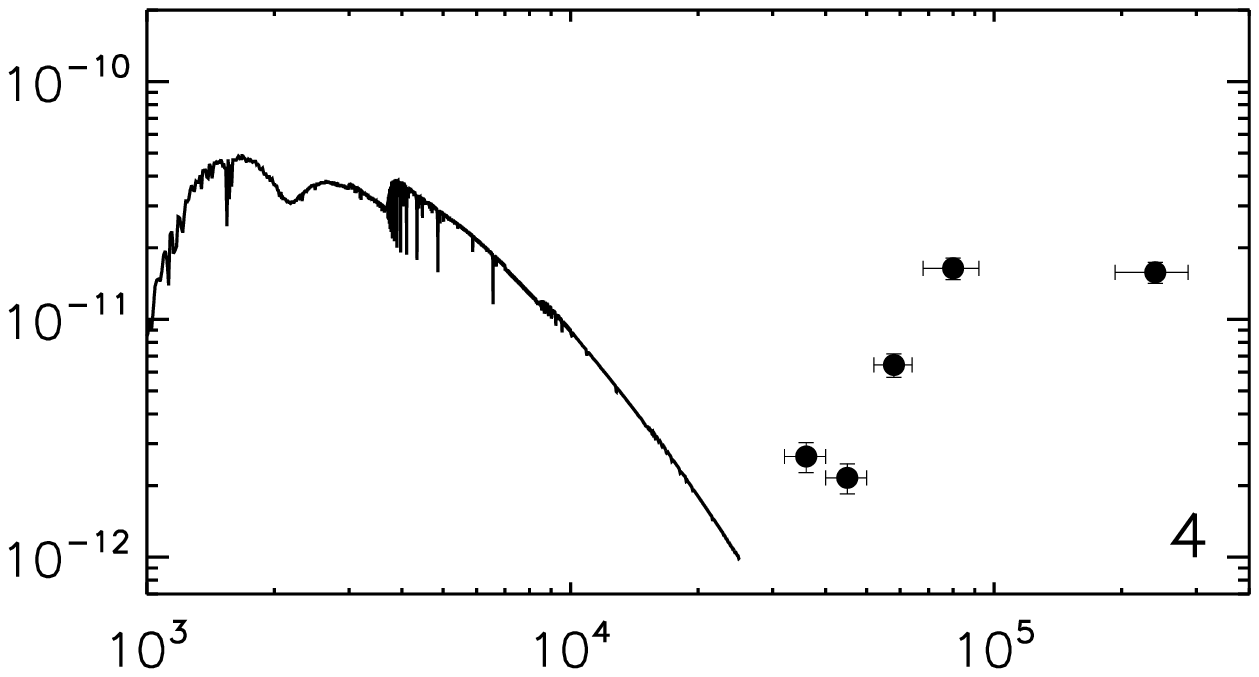}
   \includegraphics[width=4.7cm,bb=20 30 410 230]{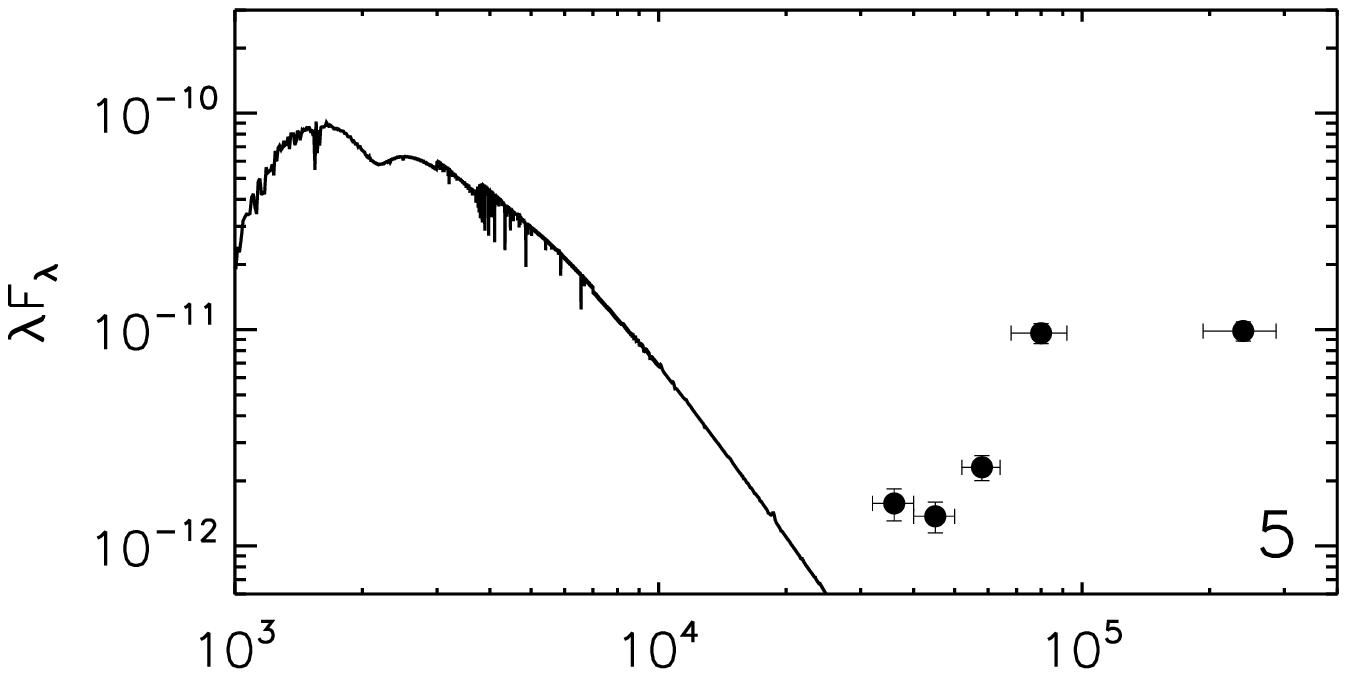}
   \includegraphics[width=4.4cm,bb=45 30 410 230]{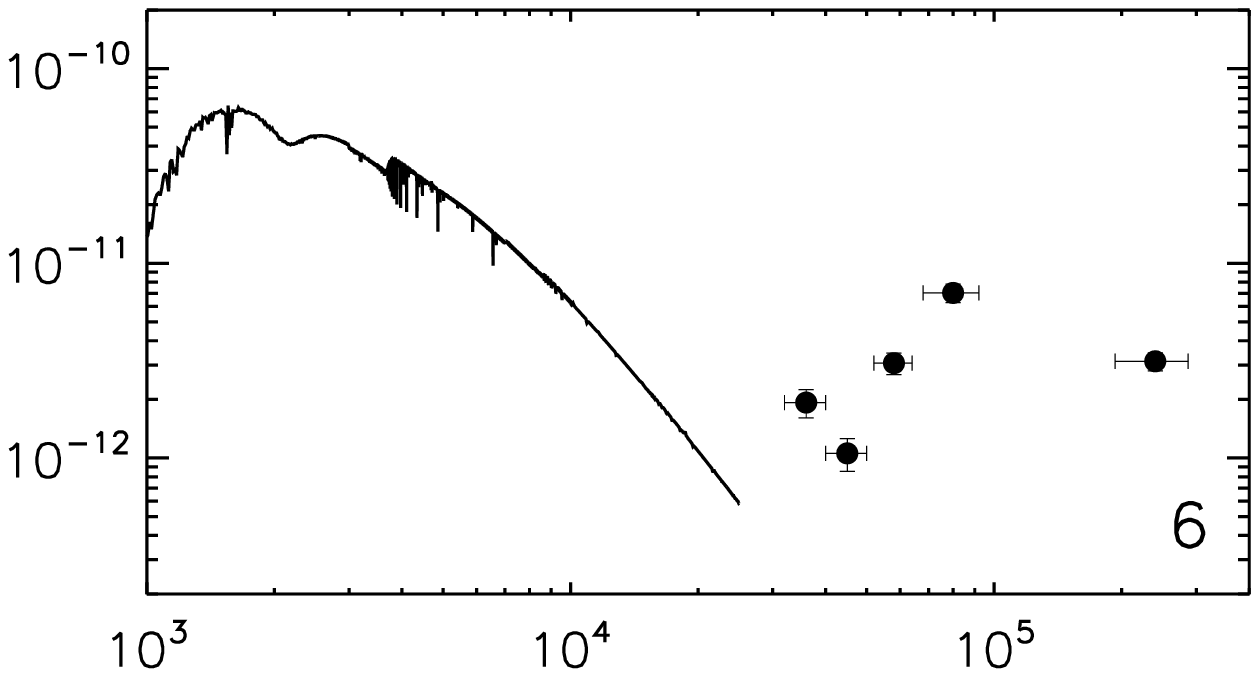}
   \includegraphics[width=4.4cm,bb=45 30 410 230]{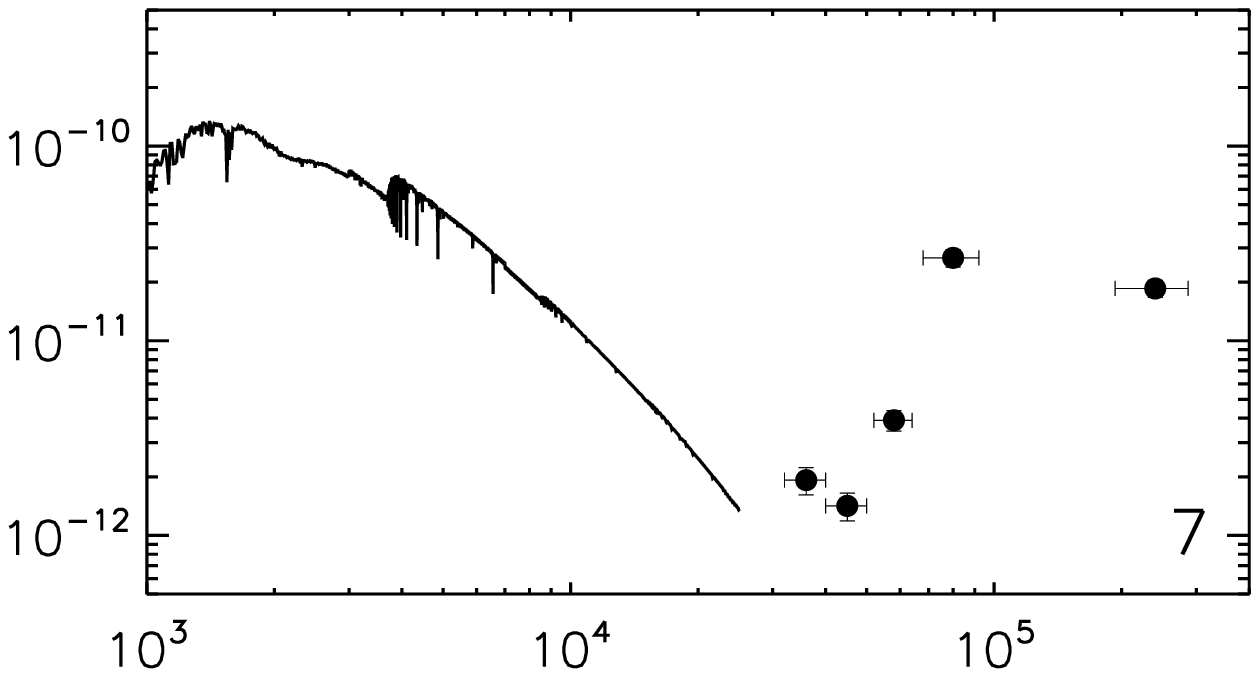}
   \includegraphics[width=4.4cm,bb=45 30 410 230]{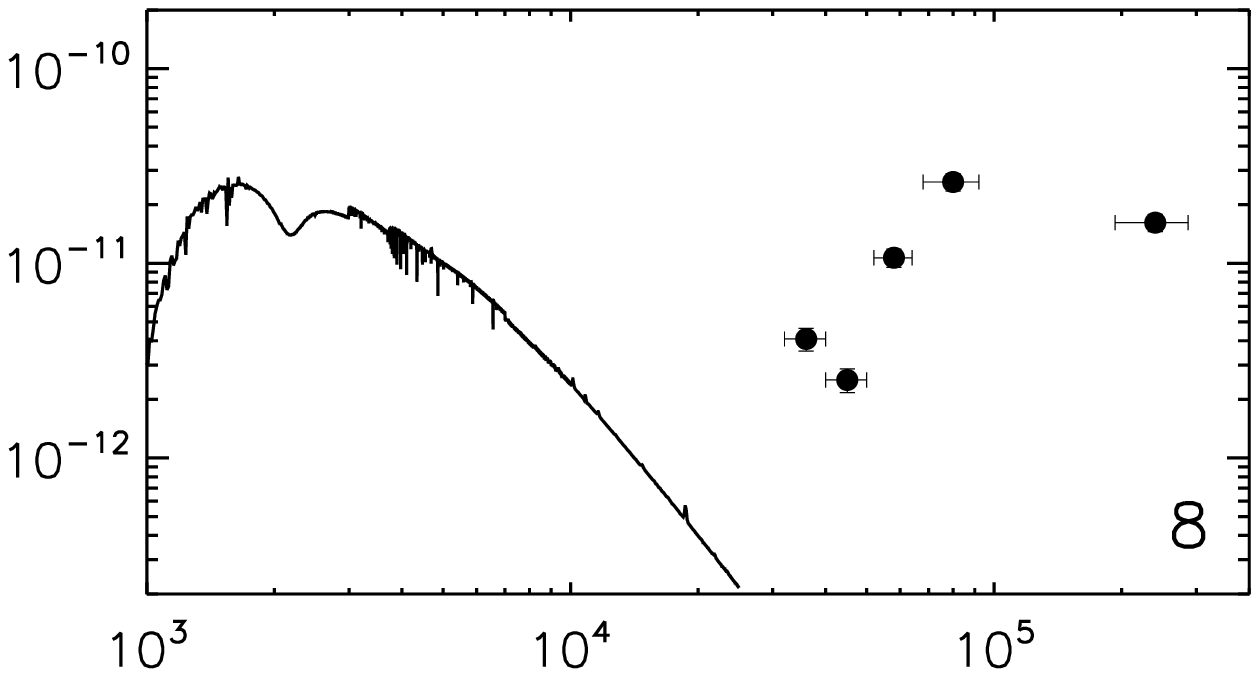}
   \includegraphics[width=4.7cm,bb=20 30 410 230]{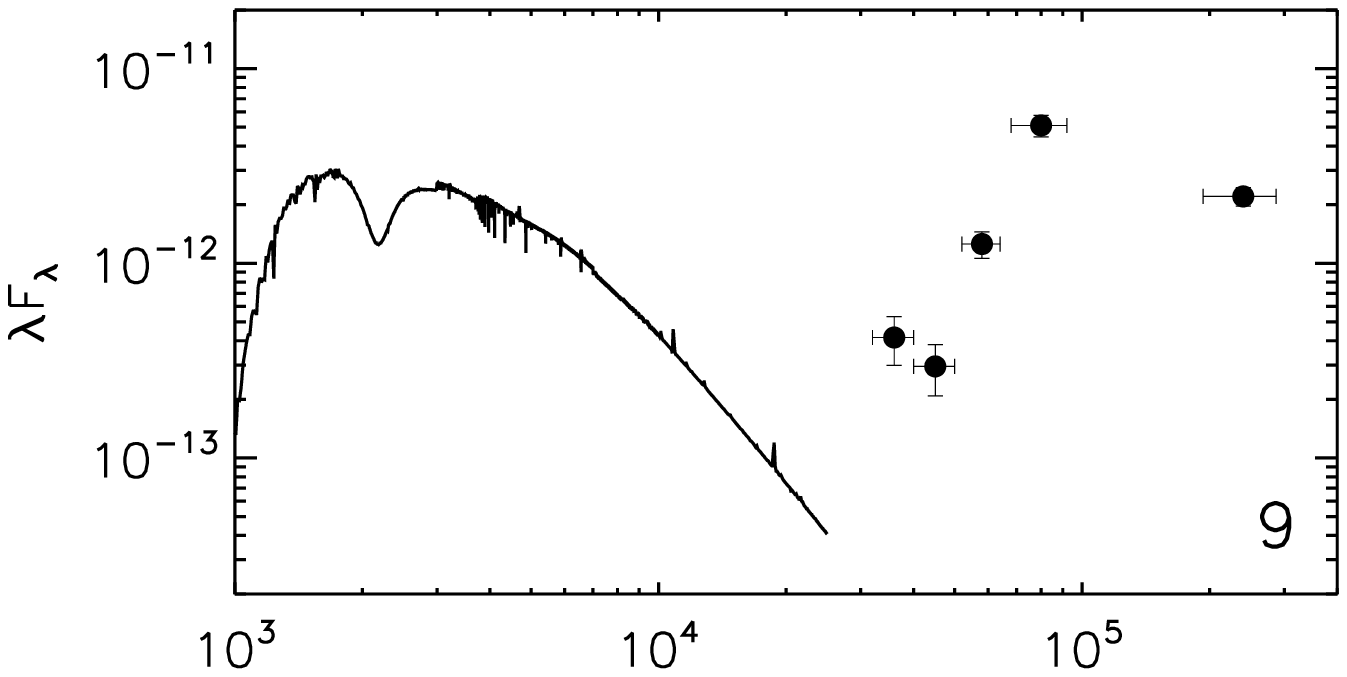}
   \includegraphics[width=4.4cm,bb=45 30 410 230]{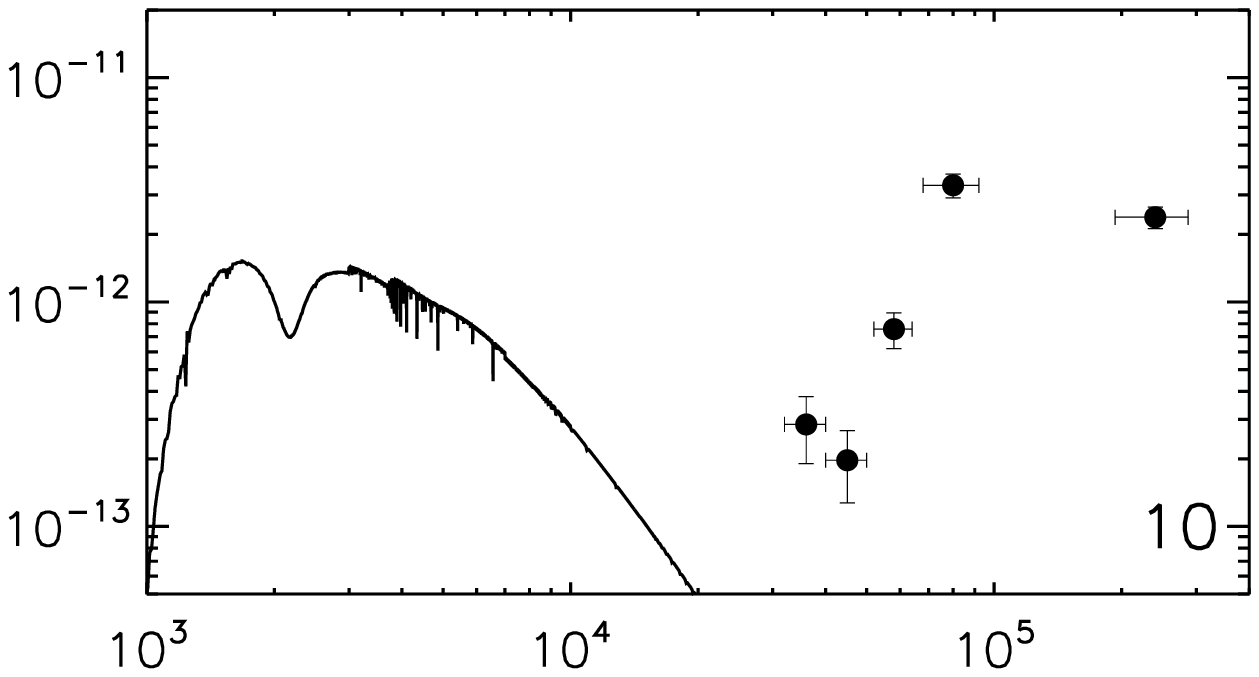}
   \includegraphics[width=4.4cm,bb=45 30 410 230]{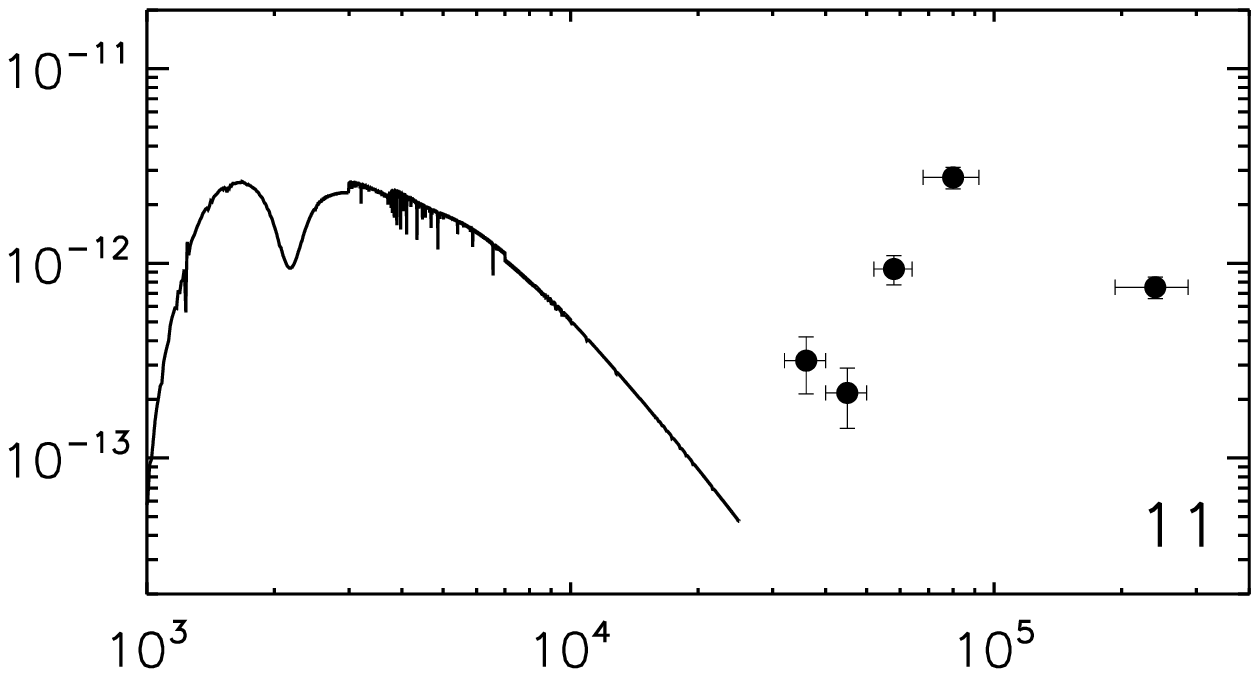}
   \includegraphics[width=4.4cm,bb=45 30 410 230]{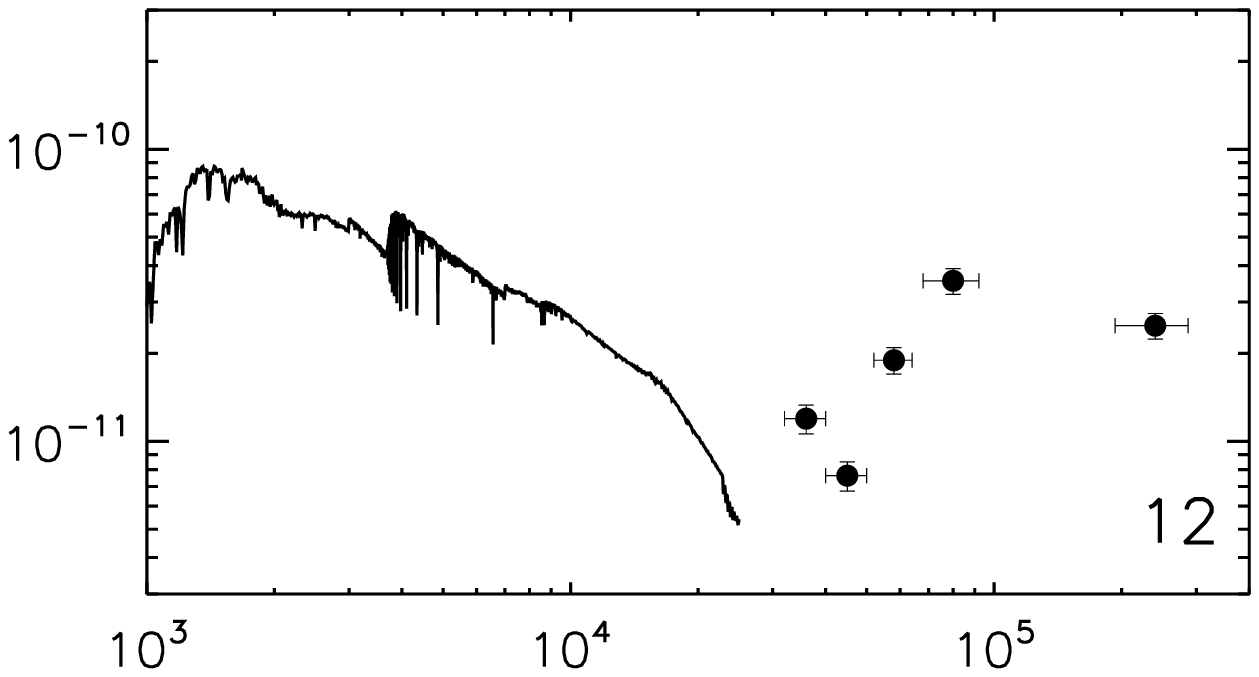}
   \includegraphics[width=4.7cm,bb=20 30 410 230]{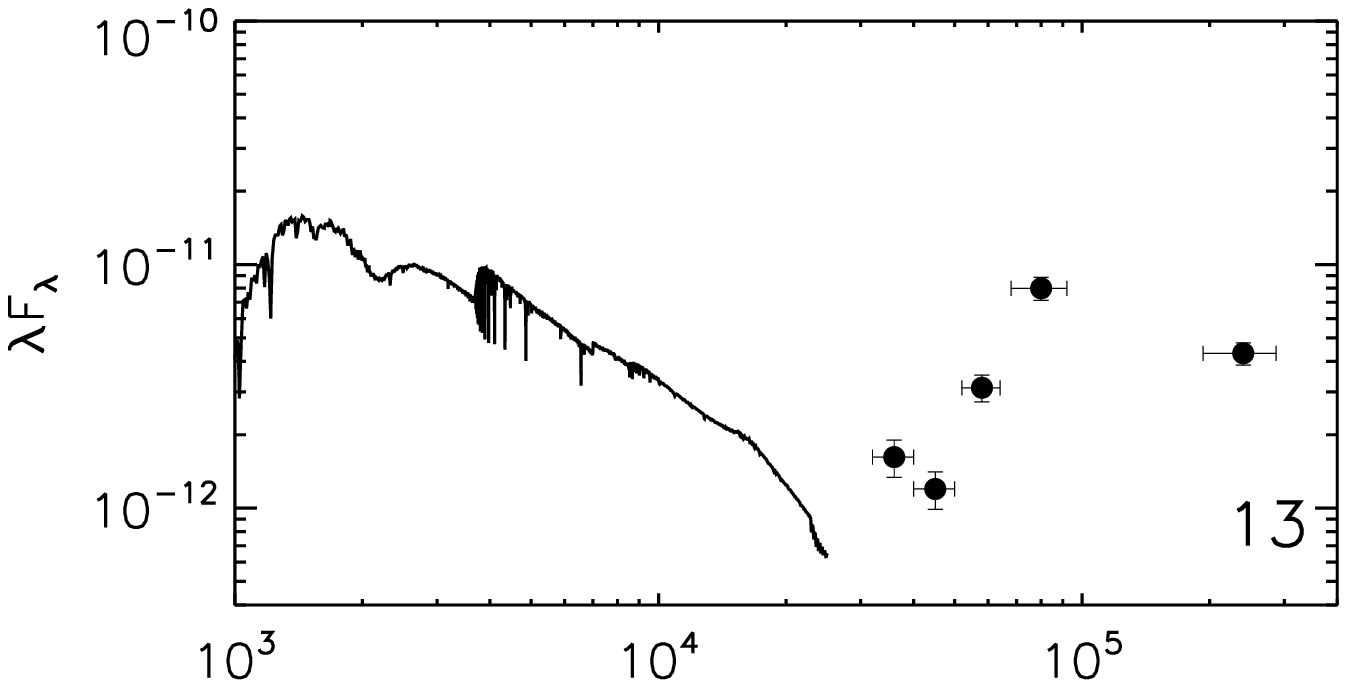}
   \includegraphics[width=4.4cm,bb=45 30 410 230]{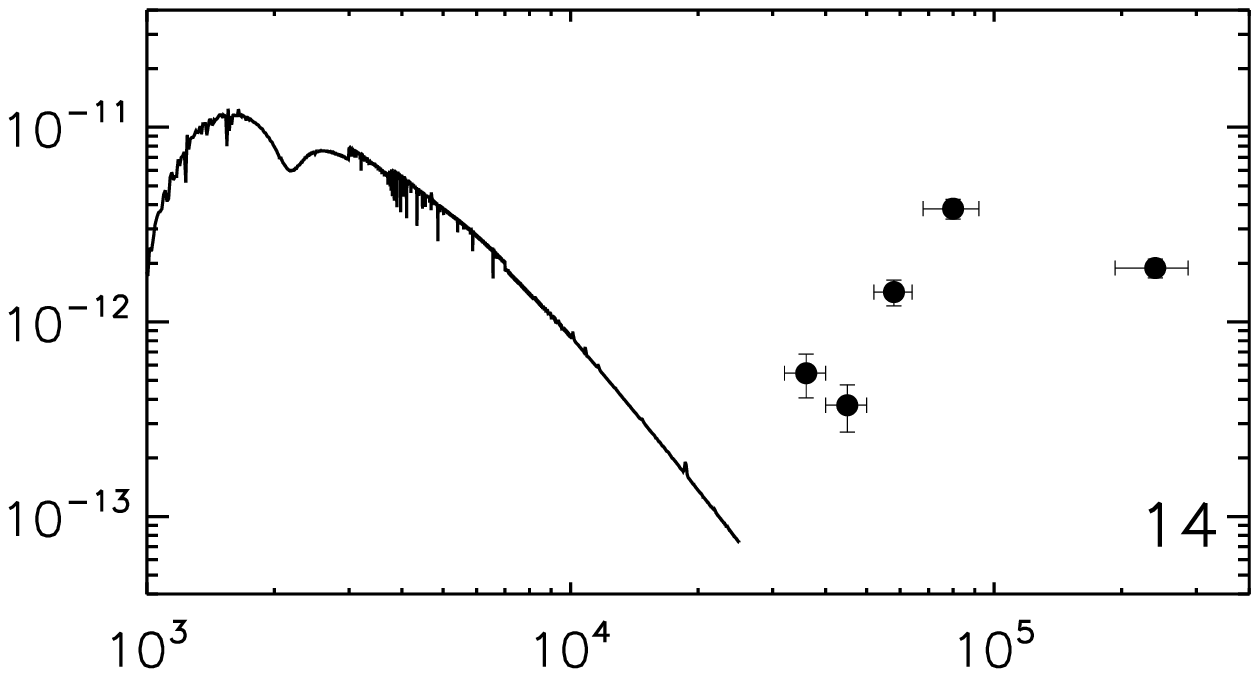}
   \includegraphics[width=4.4cm,bb=45 30 410 230]{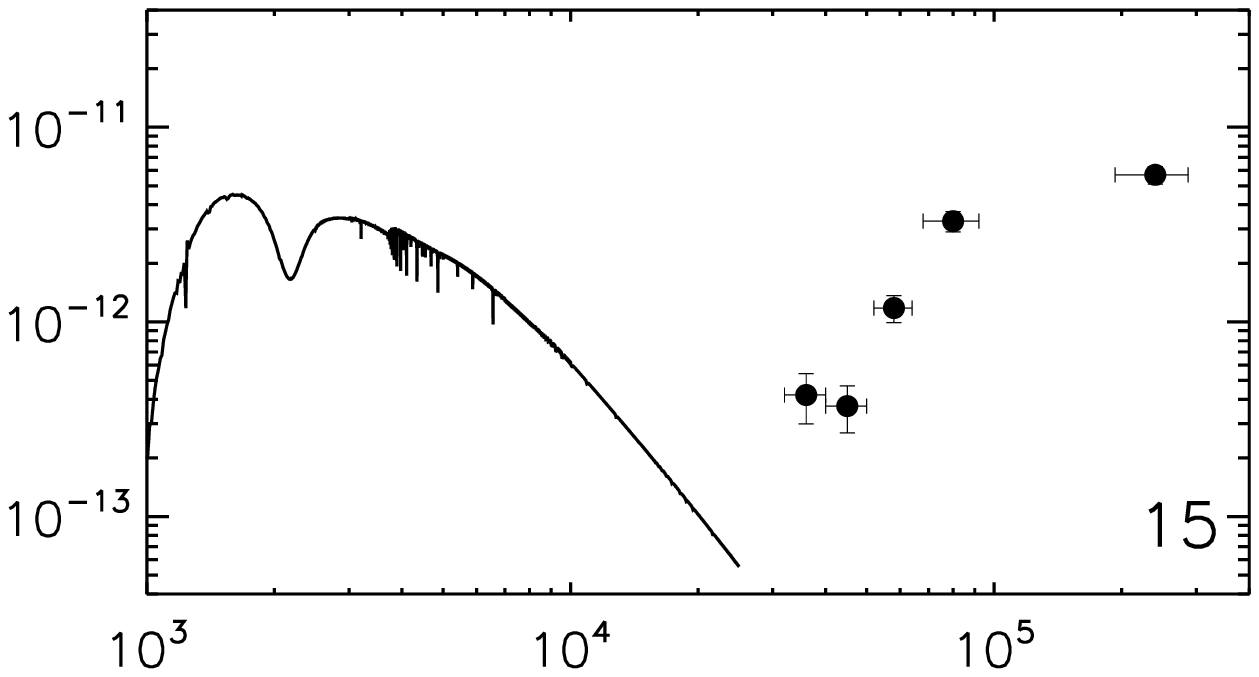}
   \includegraphics[width=4.4cm,bb=45 30 410 230]{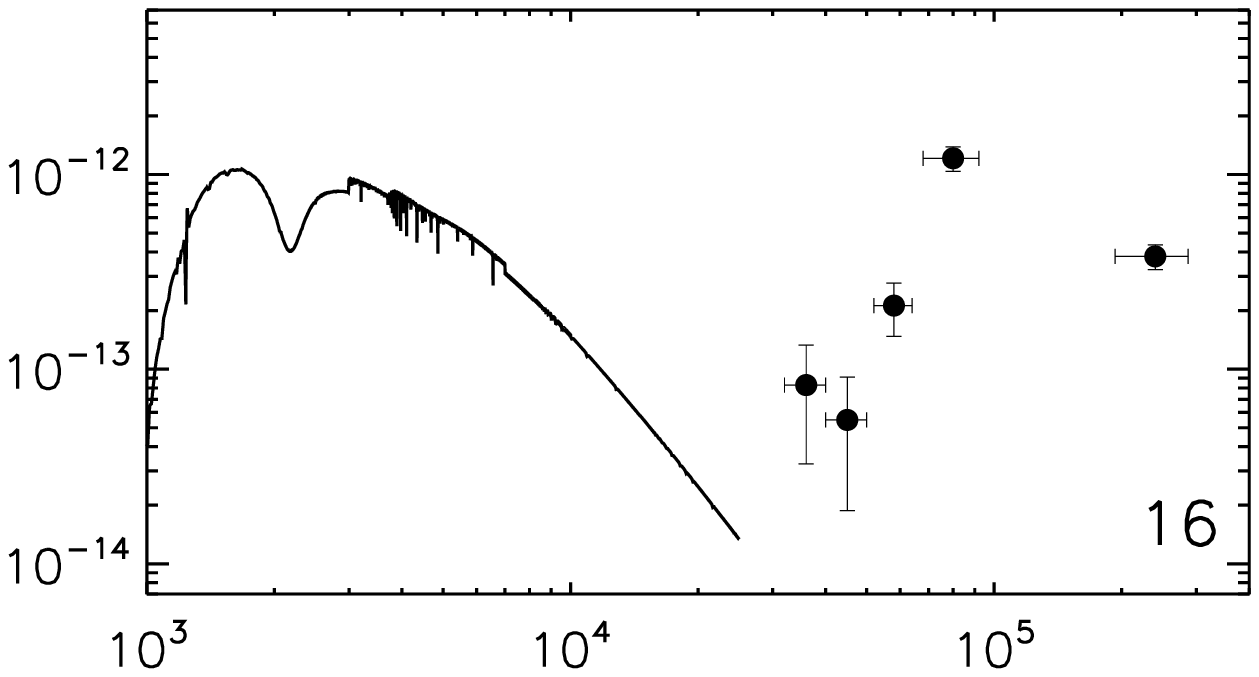}
   \includegraphics[width=4.7cm,bb=20 30 410 230]{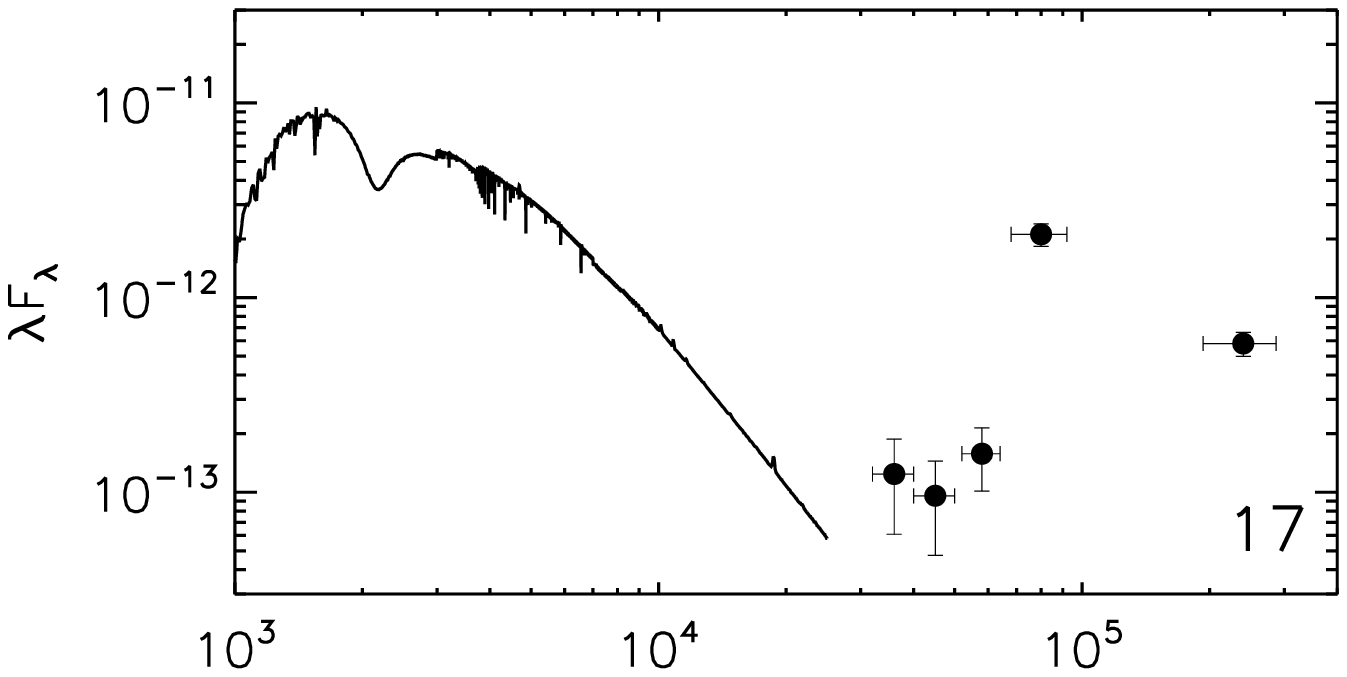}
   \includegraphics[width=4.4cm,bb=45 30 410 230]{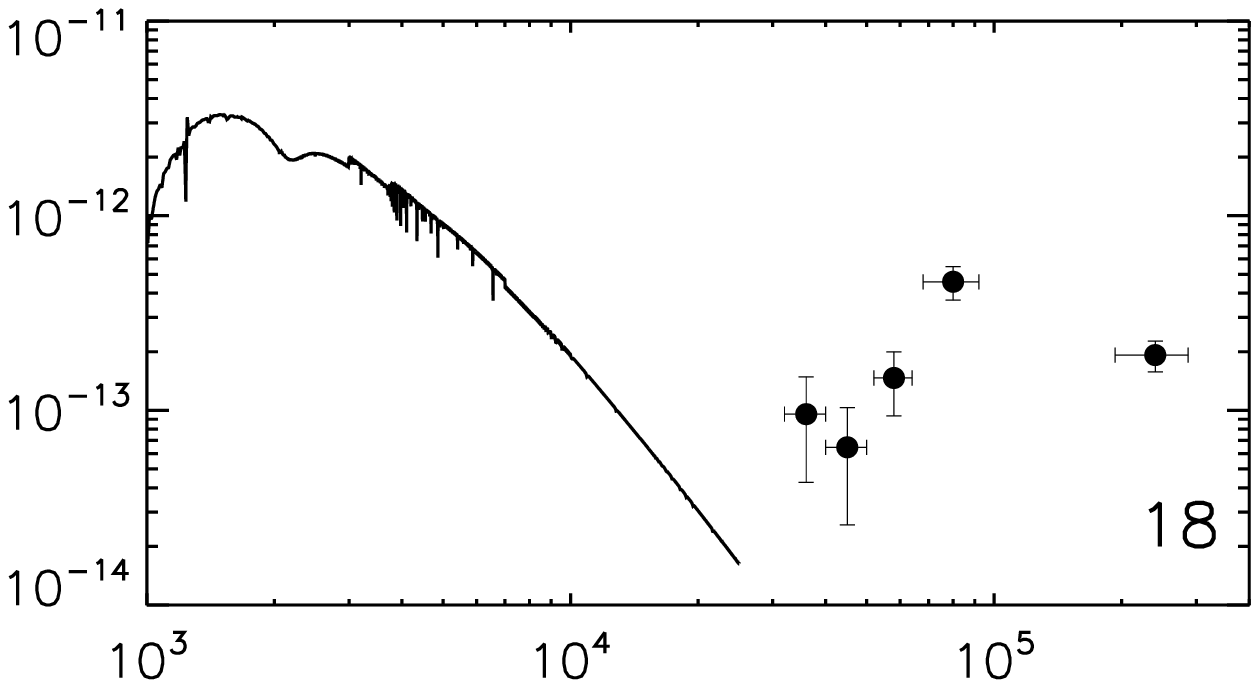}
   \includegraphics[width=4.4cm,bb=45 30 410 230]{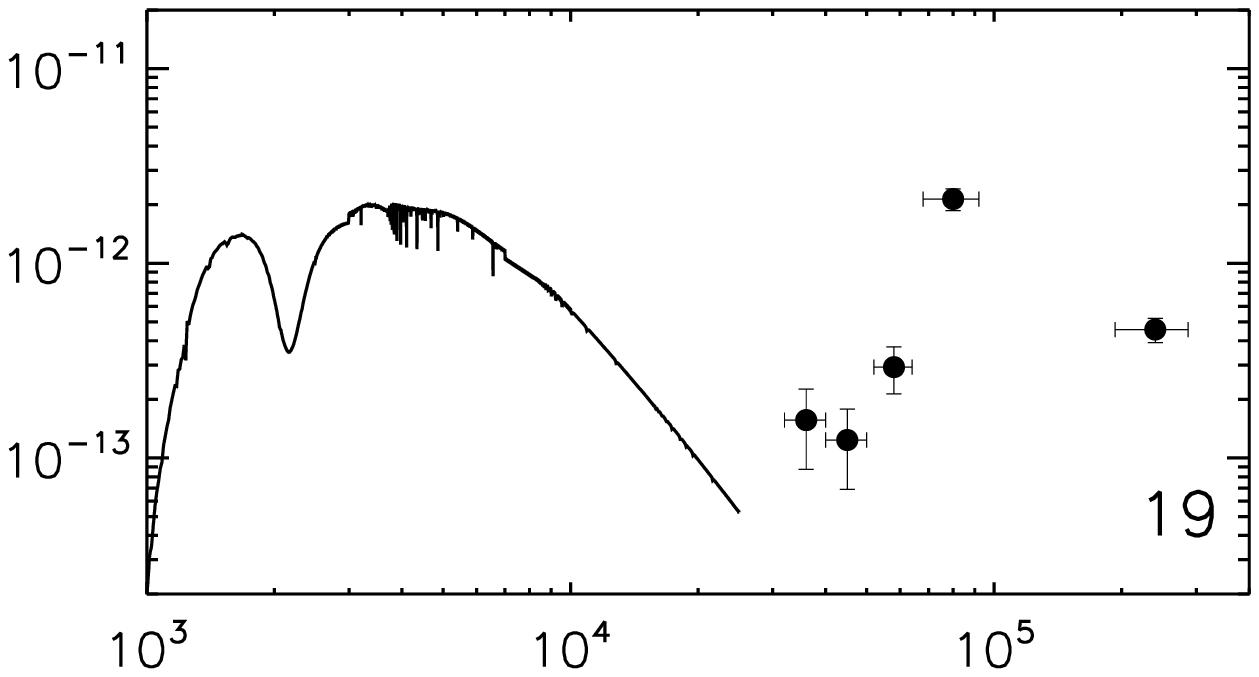}
   \includegraphics[width=4.4cm,bb=45 30 410 230]{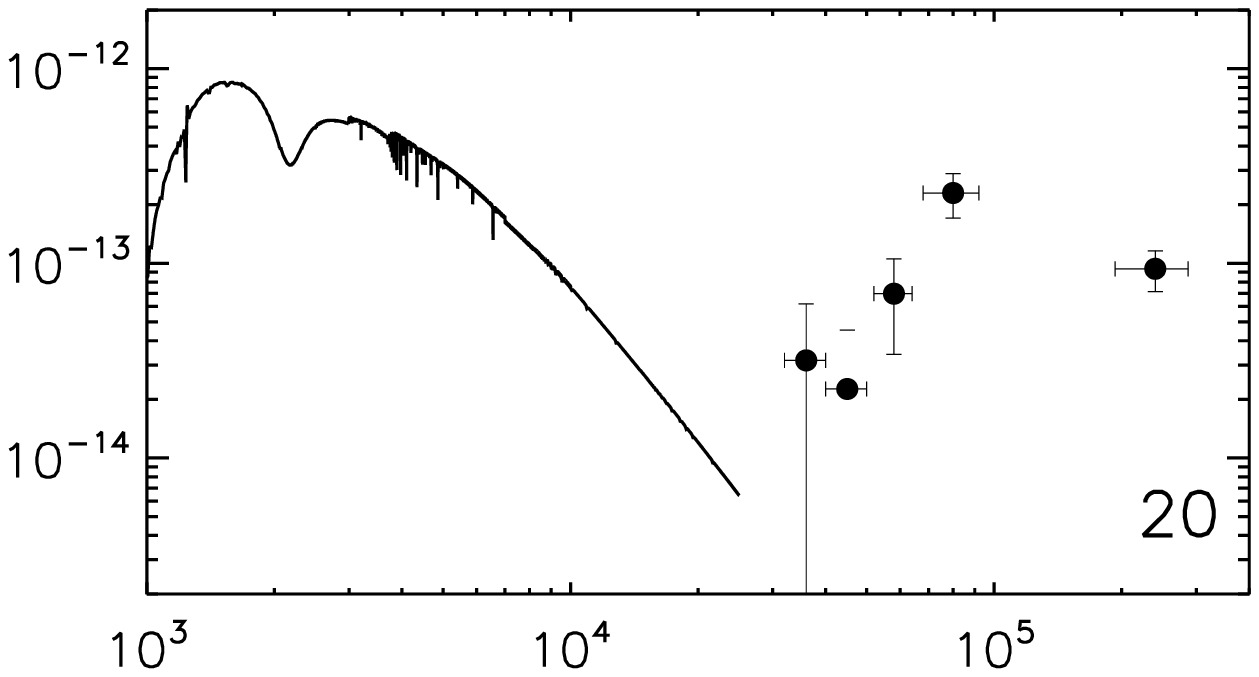}
   \includegraphics[width=4.7cm,bb=20 30 410 230]{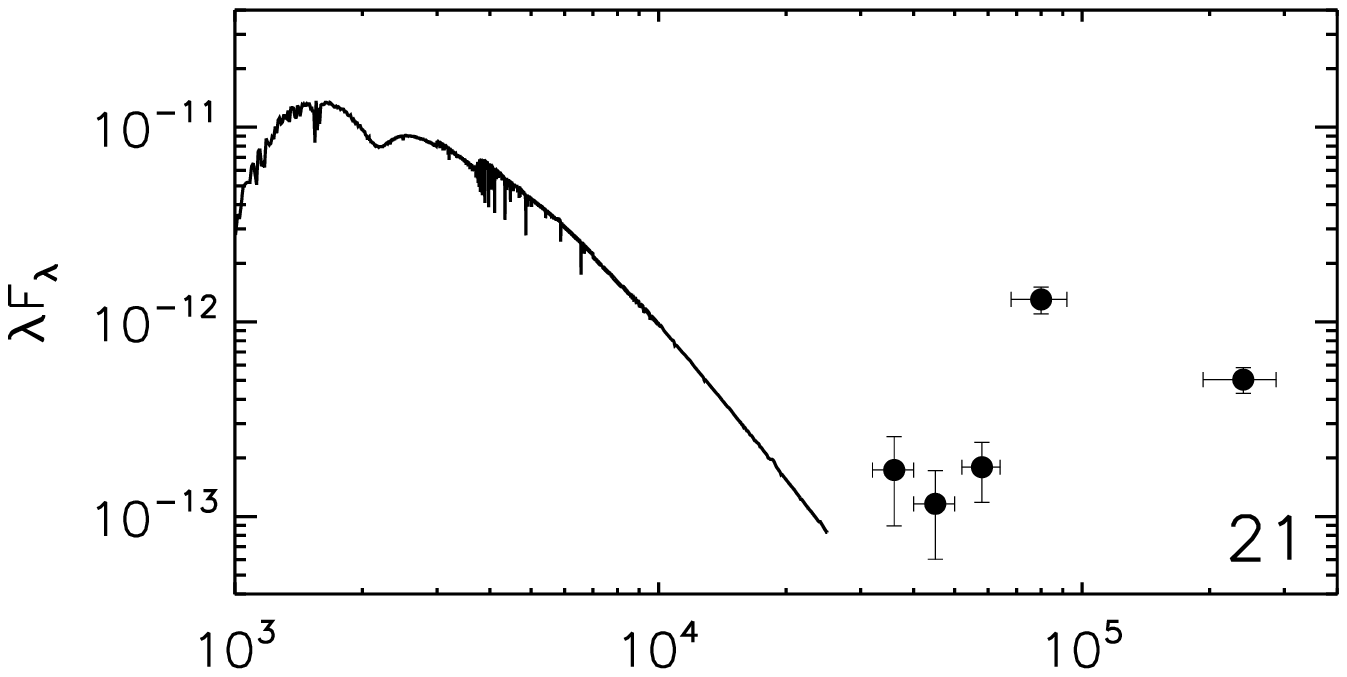}
   \includegraphics[width=4.4cm,bb=45 30 410 230]{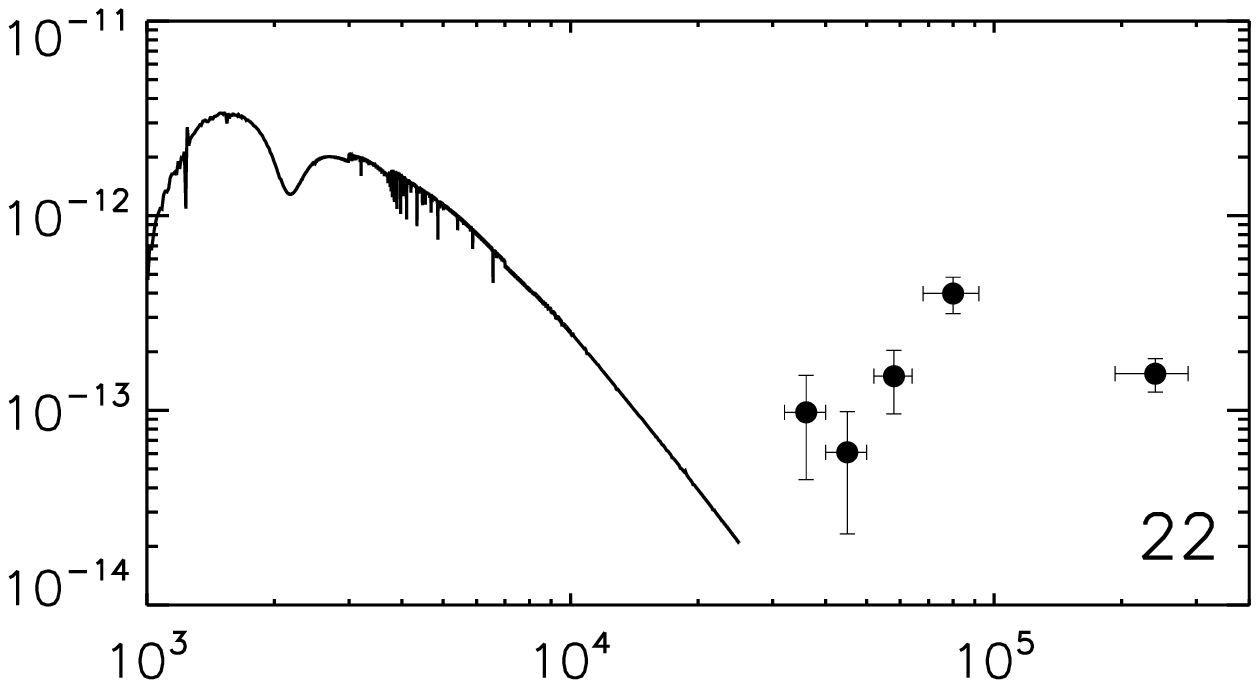}
   \includegraphics[width=4.4cm,bb=45 30 410 230]{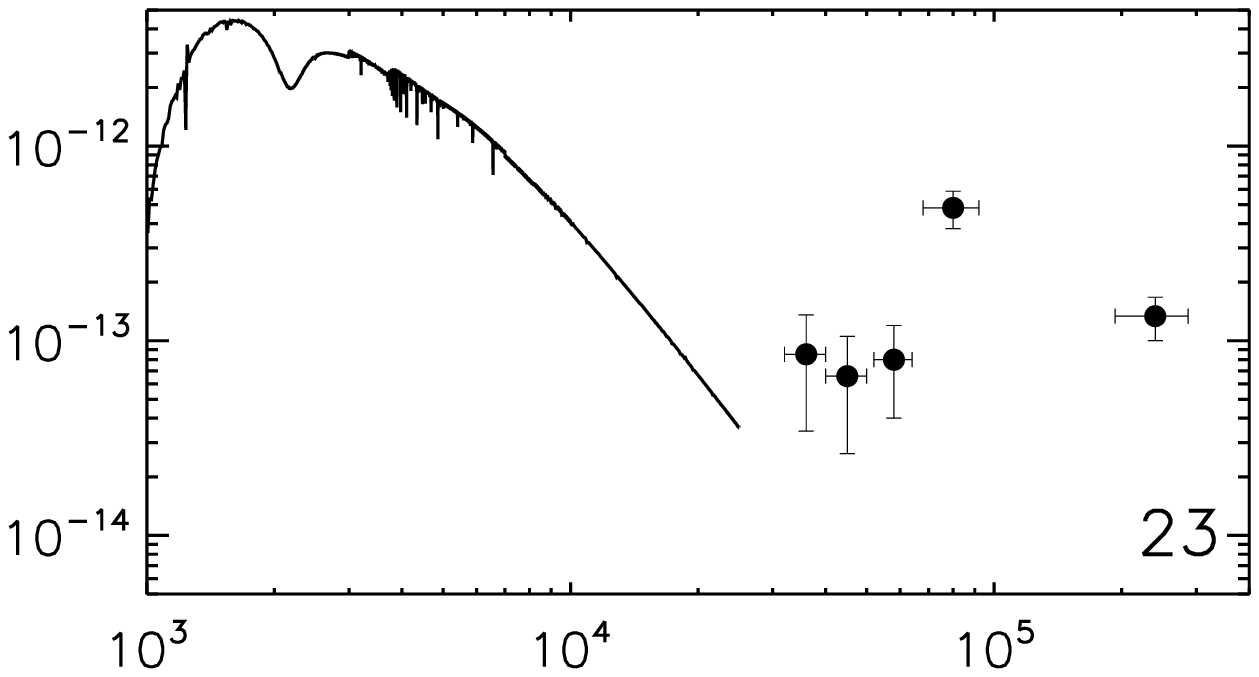}
   \includegraphics[width=4.4cm,bb=45 30 410 230]{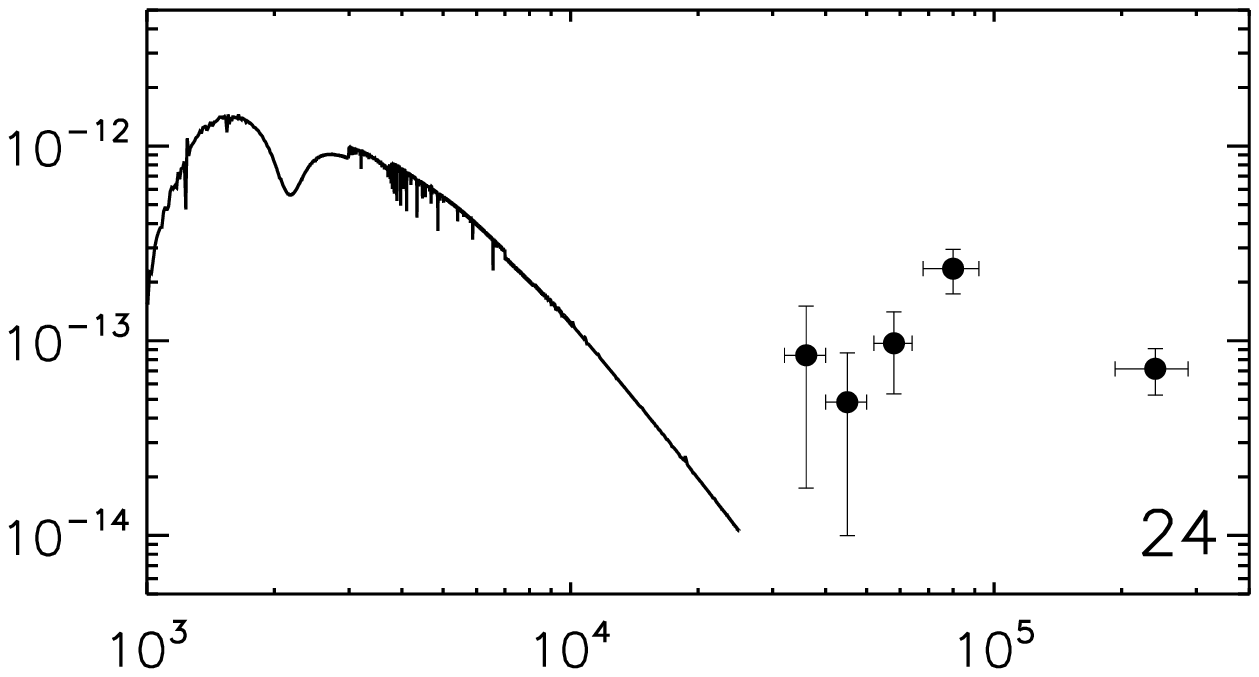}
   \includegraphics[width=4.7cm,bb=20 30 410 230]{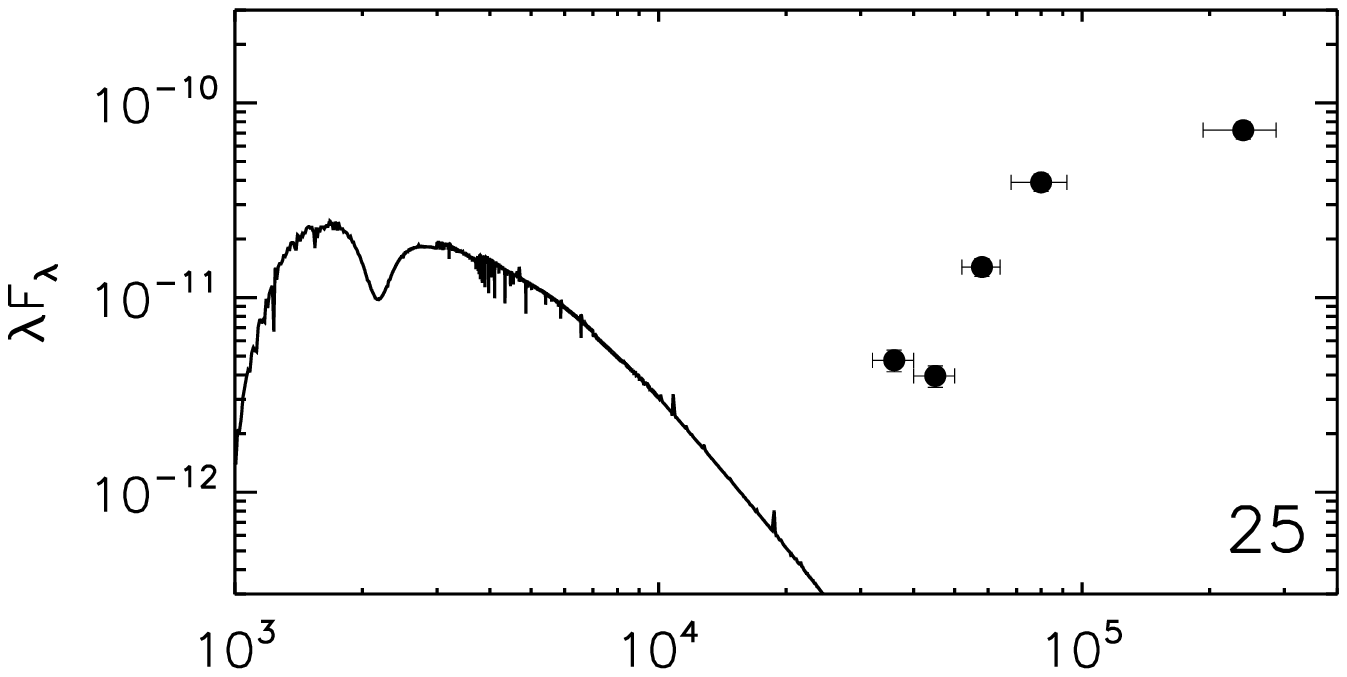}
   \includegraphics[width=4.4cm,bb=45 30 410 230]{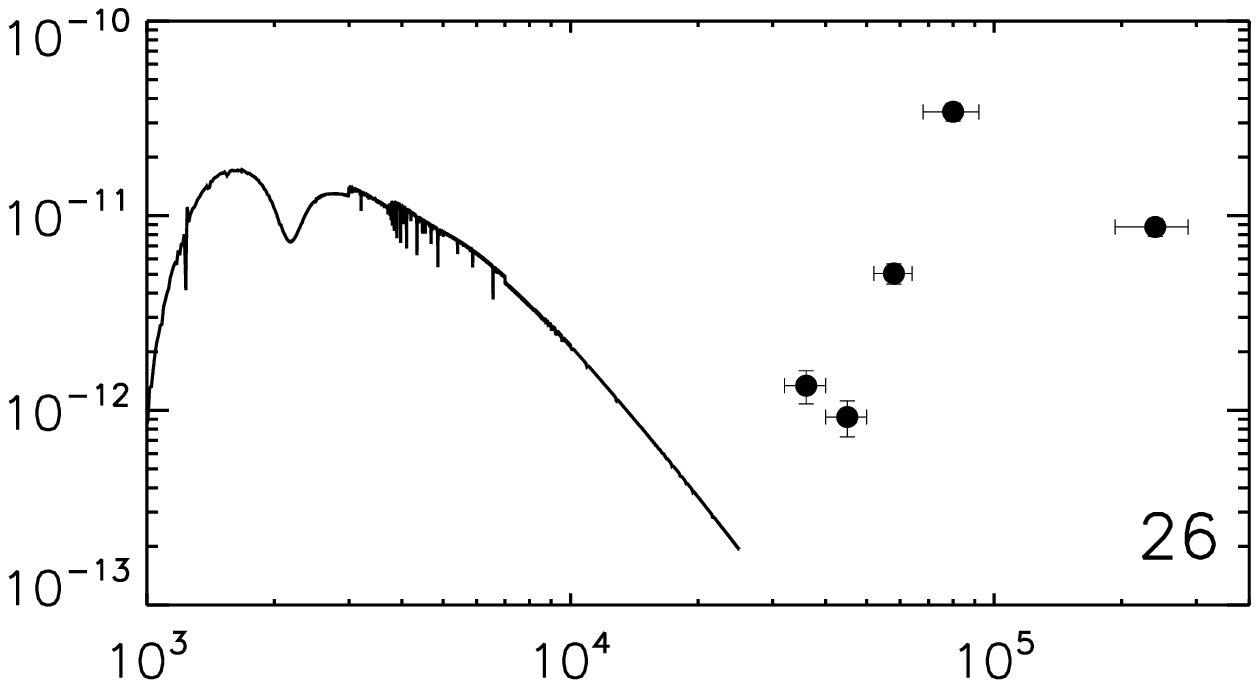}
   \includegraphics[width=4.4cm,bb=45 30 410 230]{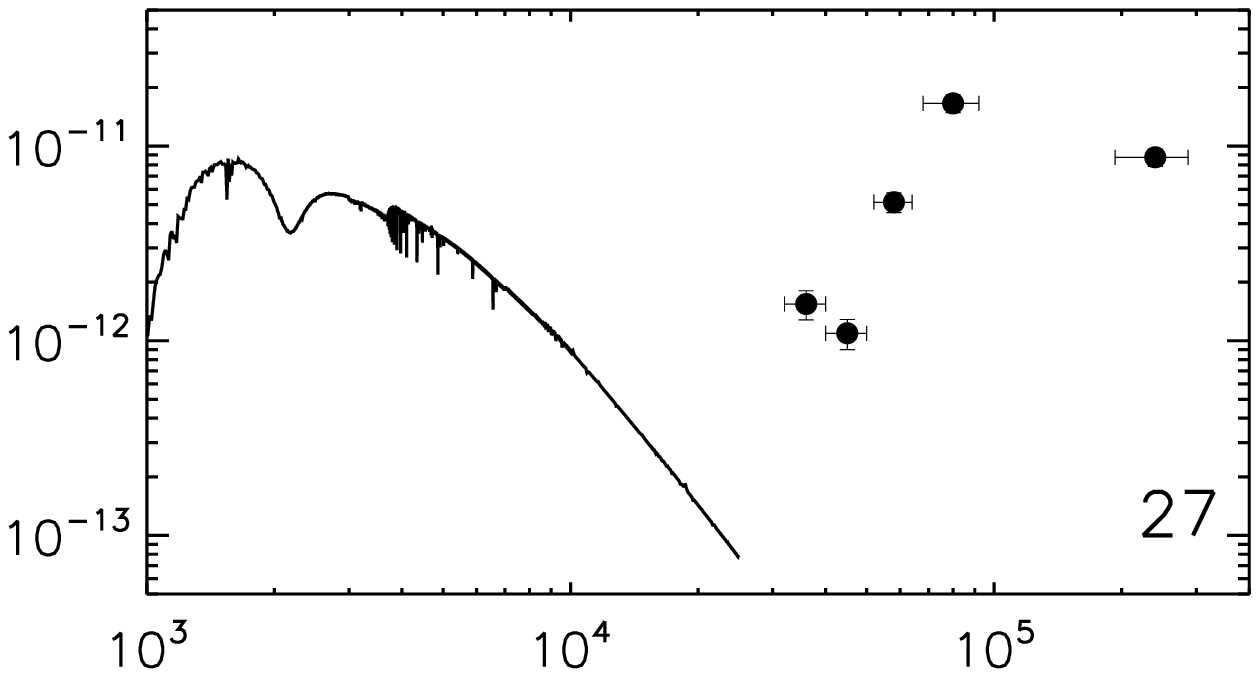}
   \includegraphics[width=4.4cm,bb=45 30 410 230]{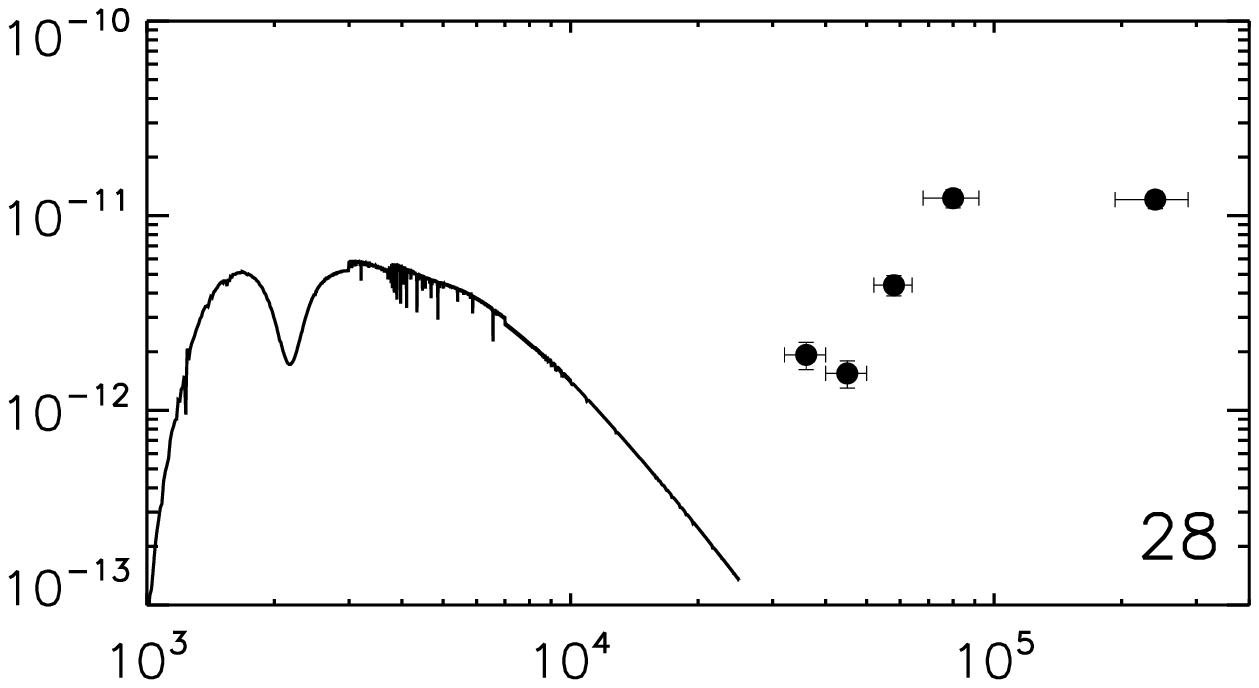}
   \includegraphics[width=4.7cm,bb=20 30 410 230]{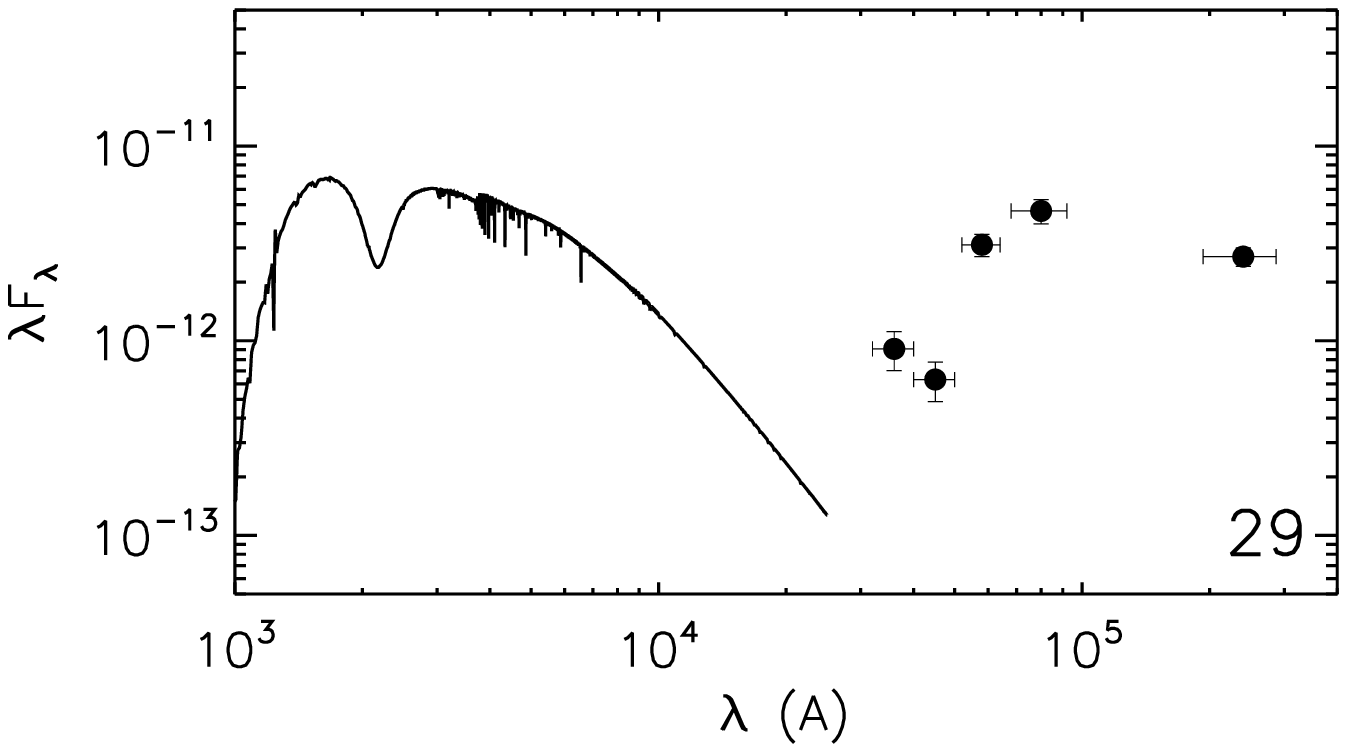}
   \includegraphics[width=4.4cm,bb=45 30 410 230]{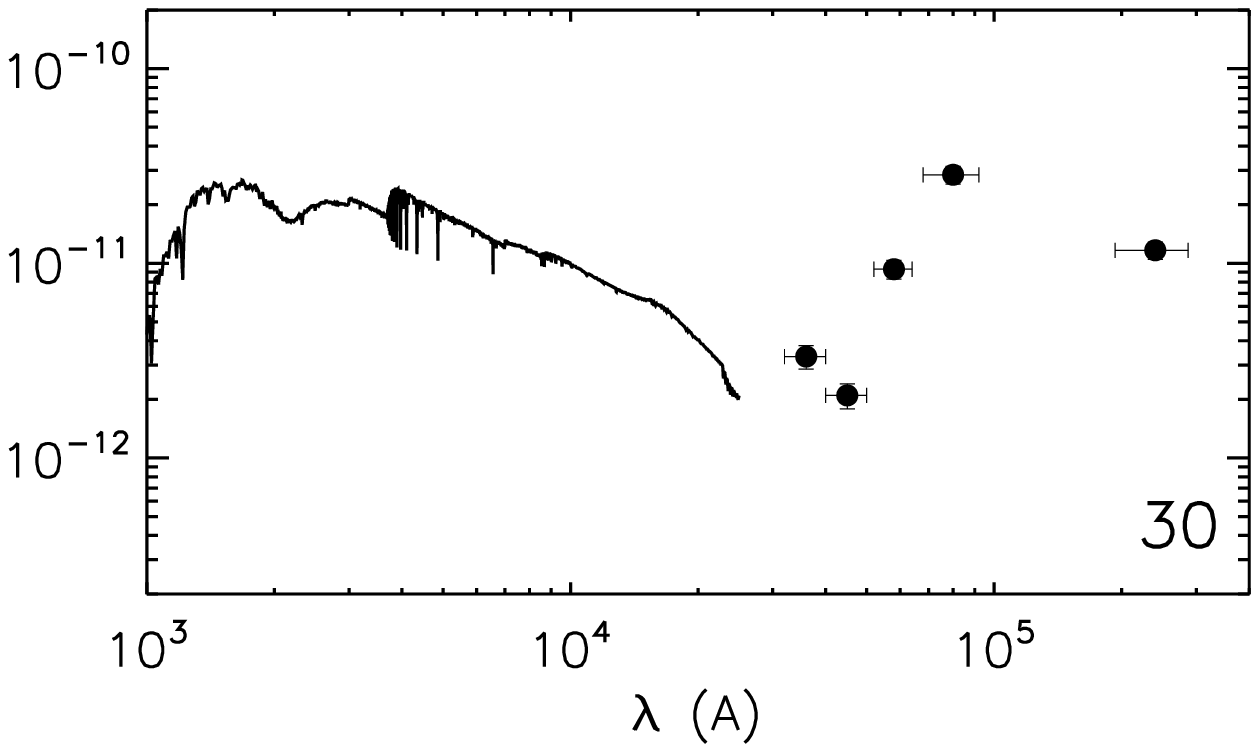}
   \includegraphics[width=4.4cm,bb=45 30 410 230]{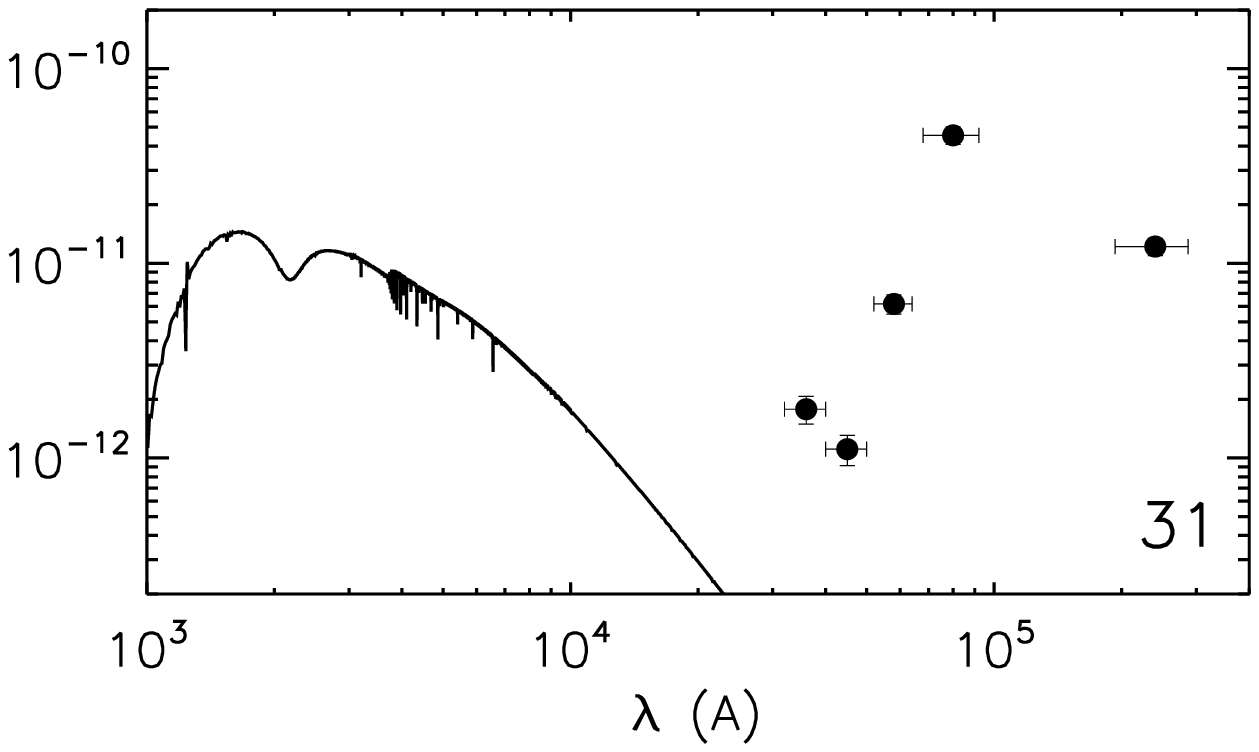}
   \includegraphics[width=4.4cm,bb=45 30 410 230]{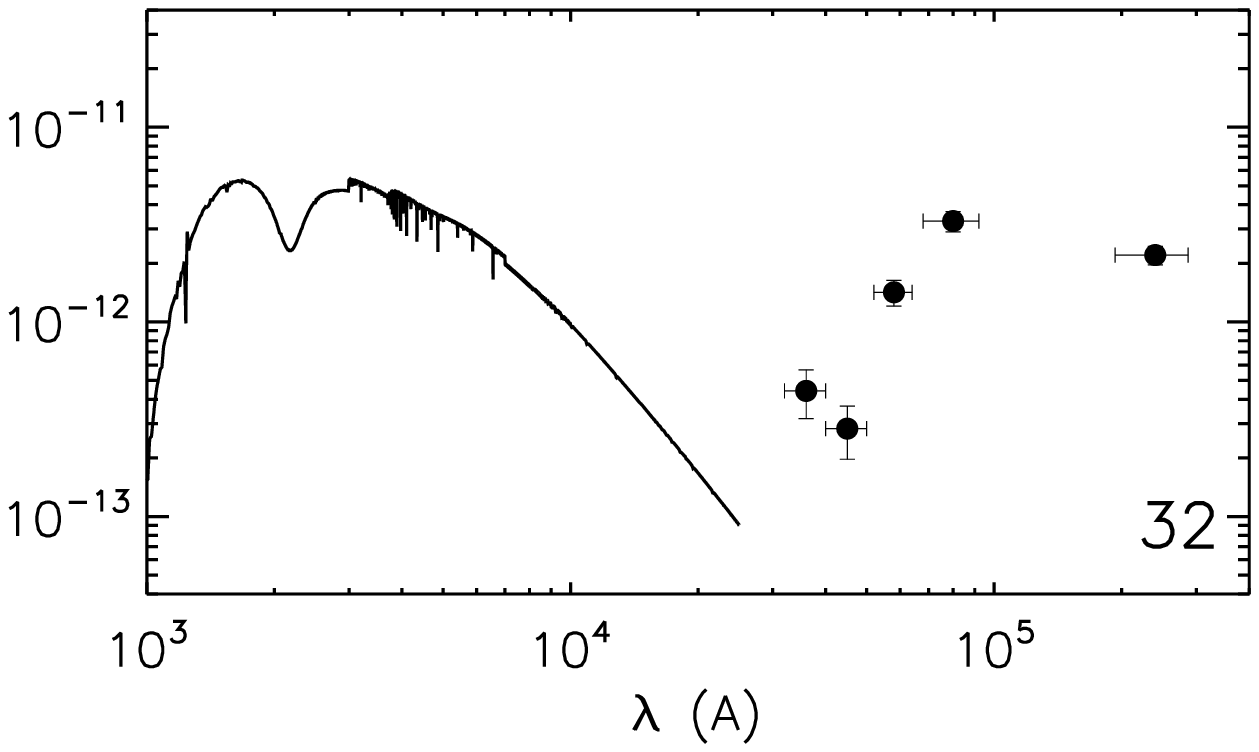}
    \caption{
Model SED for each cluster in flux units ($\lambda$F$_{\lambda}$, erg s$^{-1}$ cm$^{-2}$)
from 1000 to 3 $\times 10^5 \AA$ compared to the {\em Spitzer} IRAC and MIPS photometry.
The numbers at the bottom-right corners of the plots correspond to the cluster
ID given in Table \ref{list_clust}.}  \label{spec2}%
    \end{figure*}

\section{Multi-wavelength photometry}

We performed aperture photometry
within circular regions choosing different apertures for each object.
The aperture radius was defined on the basis of the extent of the $V$ radial
profiles, given the better resolution of the LGS images.
The same position and aperture radius
was applied to extract the photometry in the different bandpasses. The measurements were obtained using the IDL routine APER.
Background subtraction was not straightforward because the clusters are found in different local environments. The contamination of nearby bright stars depends mainly on the object location in the galaxy and on the age of the background stellar population, and this is particulary critical in the $R$, $I$  bands where the contamination by older stars is higher.
The annuli for the background determination were selected individually for each cluster in regions where the contamination of bright nearby objects  was minimal.

Table \ref{phottab} displays the UV and optical photometry of the cluster sample and the corresponding uncertainties, which include the errors from the object count rate, the sky variance, and the zero point calibration.

\subsection{UV photometry}

The flux density in the FUV and NUV filters was measured within circular apertures given in Table \ref{phottab}, and the median background flux was estimated in annuli of 5 pixel width, chosen in a way to reduce the contamination of nearby sources.
Conversion from count rates to the  AB photometric system \citep{1990AJ.....99.1621O} was obtained using the zero-points ZP$_{FUV}^{AB}=18.82$ and ZP$_{NUV}^{AB}=20.08$ \citep{2005ApJ...619L...7M}. UV magnitudes were corrected for foreground Galactic extinction following \citet{1998ApJ...500..525S} before starting the analysis of the cluster SEDs.

\subsection{UBVRI and H$\alpha$ photometry}

$UBVRI$ instrumental magnitudes
were converted to the standard Johnson and Cousin system using stars from the \citet{2006AJ....131.2478M} catalogue.
The colour term in the $UVBRI$ transformation (described in Massey et al. 2006) was ignored, as it is smaller than the errors we report.  The contribution of the Galactic foreground extinction was also subtracted from the optical magnitudes according to \citet{1998ApJ...500..525S}.

The continuum emission in the H$\alpha$ image was removed by subtracting the $R$-band image scaled by a
factor 0.40, which was determined from the aperture photometry of 20 field stars in both filters. The calibration factor of 1.79 $\times 10^{-19}$ erg s$^{-1}$ cm$^{-2}$ given in Massey et al. (2007) was applied to the count rates measured within apertures that included the total H$\alpha$ emission around the cluster.
The fluxes are displayed in Table \ref{MIRphot}. As a further check, H$\alpha$ fluxes were determined for the same apertures on the map of Hoopes \& Walterbos (2000), adopting a calibration factor of 8.2 $\times 10^{-18}$ erg s$^{-1}$ cm$^{-2}$. The resulting measurements agreed with the LGS map within 5\% (1$\sigma$).

H$\alpha$ and oxygen emission lines may significantly contribute to the total flux in the $R$, $V$ and $U$ bands.
We estimated the contamination from \oiii$\lambda$5007 and H$\alpha$ lines to the
photometry in the $V$ and $R$ filters, using the narrowband images provided by
LGS. The correction was obtained by subtracting the flux measured in the narrowband image within the same aperture used for the optical photometry (last column in Table 2). We used the calibration factor of 2.11 $\times 10^{-18}$ and 6.41 $\times 10^{-18}$ erg s$^{-1}$ cm$^{-2}$ \AA$^{-1}$ for the H$\alpha$ and \oiii images respectively, as given in Massey et al. (2007). The $U$ photometry is affected by the contamination from the \oii $\lambda$3727 line, but the LGS does not provide a narrowband image centred on this wavelength.
Thus we multiplied the \oiii narrowband fluxes for the \oii/\oiii line ratio derived for the clusters in the \citet{2008ApJ...675.1213R} catalogue. For those without a direct measurement of this ratio, we applied an average value of \oii/\oiii = 1. In some cases this correction was not sufficient to reduce the excess in the $U$-band photometry.


\subsection{IRAC and MIPS (24 $\mu$m) photometry}

Photometry at 3.6, 4.5, and 5.8 $\mu$m was derived through the same apertures used for the UV-optical bands, while the larger apertures chosen for the H$\alpha$ photometry were adopted to measure 8 and 24 $\mu$m fluxes.
Uncertainties include Poisson noise photon flux, background fluctuations and a 10\% flux calibration uncertainty for both the IRAC \citep{2004ApJS..154...10F, 2005PASP..117..978R} and MIPS bands \citep{2007PASP..119..994E}. Table \ref{MIRphot} displays the fluxes at 3.6, 4.5, 5.8, 8.0, and 24 $\mu$m for the sample.

\begin{table*}
\caption{Results from the SED-fitting technique.}
\label{params}
\begin{minipage}{\textwidth}
\begin{center}
\begin{tabular}{lcccccc}
\hline \hline
ID&     log(age)  &     A$_V$     &   $E(B-V)$    &    log(L$_{Bol}$)                   & log(M) \\
  &     [yr]          &     mag       &    mag        &  [L$_{\odot}$]                        & [M$_{\odot}$]                            \\
\hline \hline
1 & 6.97$\pm$0.01 & 0.28$\pm$0.01 & 0.13$\pm$0.01 & 6.71$\pm$0.01 & 4.31$\pm$0.01  \\
 2 & 6.51$\pm$0.14 & 0.58$\pm$0.11 & 0.26$\pm$0.03 & 6.39$\pm$0.08 & 3.24$\pm$0.08  \\
 3 & 6.68$\pm$0.09 & 0.68$\pm$0.11 & 0.30$\pm$0.05 & 7.23$\pm$0.01 & 4.37$\pm$0.01  \\
 4 & 6.72$\pm$0.03 & 0.59$\pm$0.07 & 0.24$\pm$0.03 & 6.98$\pm$0.05 & 4.20$\pm$0.05  \\
 5 & 6.54$\pm$0.03 & 0.61$\pm$0.02 & 0.23$\pm$0.01 & 7.56$\pm$0.01 & 4.49$\pm$0.01  \\
 6 & 6.65$\pm$0.15 & 0.52$\pm$0.17 & 0.23$\pm$0.07 & 6.76$\pm$0.12 & 3.86$\pm$0.12  \\
 7 & 6.71$\pm$0.02 & 0.37$\pm$0.04 & 0.14$\pm$0.02 & 7.11$\pm$0.03 & 4.31$\pm$0.03  \\
 8 & 6.48$\pm$0.16 & 0.70$\pm$0.09 & 0.29$\pm$0.03 & 6.94$\pm$0.07 & 3.78$\pm$0.07  \\
 9 & 6.40$\pm$0.10 & 0.82$\pm$0.07 & 0.35$\pm$0.02 & 6.21$\pm$0.03 & 3.07$\pm$0.03  \\
10 & 6.35$\pm$0.21 & 0.93$\pm$0.09 & 0.39$\pm$0.02 & 6.00$\pm$0.10 & 2.88$\pm$0.10  \\
11 & 6.36$\pm$0.13 & 0.93$\pm$0.06 & 0.41$\pm$0.01 & 6.37$\pm$0.04 & 3.24$\pm$0.04  \\
12 & 6.99$\pm$0.04 & 0.40$\pm$0.02 & 0.16$\pm$0.01 & 6.95$\pm$0.01 & 4.57$\pm$0.01  \\
13 & 7.18$\pm$0.04 & 0.39$\pm$0.02 & 0.16$\pm$0.01 & 6.19$\pm$0.01 & 4.04$\pm$0.01  \\
14 & 6.44$\pm$0.16 & 0.61$\pm$0.08 & 0.25$\pm$0.02 & 6.52$\pm$0.07 & 3.37$\pm$0.07  \\
15 & 6.38$\pm$0.16 & 0.82$\pm$0.07 & 0.36$\pm$0.02 & 6.51$\pm$0.05 & 3.38$\pm$0.05  \\
16 & 6.36$\pm$0.15 & 0.83$\pm$0.07 & 0.37$\pm$0.02 & 5.85$\pm$0.04 & 2.72$\pm$0.04  \\
17 & 6.35$\pm$0.13 & 0.56$\pm$0.05 & 0.26$\pm$0.01 & 6.44$\pm$0.03 & 3.32$\pm$0.03  \\
18 & 6.43$\pm$0.17 & 0.54$\pm$0.08 & 0.22$\pm$0.02 & 5.89$\pm$0.07 & 2.75$\pm$0.07  \\
19 & 6.22$\pm$0.12 & 1.05$\pm$0.03 & 0.51$\pm$0.01 & 6.49$\pm$0.01 & 3.39$\pm$0.01  \\
20 & 6.35$\pm$0.14 & 0.66$\pm$0.06 & 0.30$\pm$0.01 & 5.57$\pm$0.04 & 2.45$\pm$0.04  \\
21 & 6.53$\pm$0.16 & 0.56$\pm$0.10 & 0.24$\pm$0.04 & 6.46$\pm$0.09 & 3.36$\pm$0.09  \\
22 & 6.37$\pm$0.09 & 0.61$\pm$0.03 & 0.28$\pm$0.01 & 6.11$\pm$0.02 & 2.98$\pm$0.02  \\
23 & 6.43$\pm$0.13 & 0.71$\pm$0.08 & 0.30$\pm$0.02 & 6.28$\pm$0.06 & 3.14$\pm$0.06  \\
24 & 6.39$\pm$0.10 & 0.66$\pm$0.05 & 0.30$\pm$0.01 & 5.75$\pm$0.03 & 2.62$\pm$0.03  \\
25 & 6.48$\pm$0.14 & 0.70$\pm$0.12 & 0.32$\pm$0.03 & 6.99$\pm$0.07 & 3.83$\pm$0.07  \\
26 & 6.42$\pm$0.14 & 0.80$\pm$0.07 & 0.34$\pm$0.02 & 7.01$\pm$0.06 & 3.87$\pm$0.06  \\
27 & 6.50$\pm$0.17 & 0.66$\pm$0.11 & 0.30$\pm$0.03 & 6.50$\pm$0.09 & 3.33$\pm$0.09  \\
28 & 6.33$\pm$0.14 & 1.05$\pm$0.07 & 0.46$\pm$0.01 & 6.84$\pm$0.03 & 3.72$\pm$0.03  \\
29 & 6.20$\pm$0.11 & 0.88$\pm$0.06 & 0.40$\pm$0.01 & 6.71$\pm$0.03 & 3.58$\pm$0.03  \\
30 & 7.16$\pm$0.11 & 0.56$\pm$0.11 & 0.27$\pm$0.09 & 6.60$\pm$0.02 & 4.44$\pm$0.02  \\
31 & 6.46$\pm$0.17 & 0.81$\pm$0.08 & 0.33$\pm$0.03 & 6.90$\pm$0.07 & 3.75$\pm$0.07  \\
32 & 6.49$\pm$0.14 & 0.88$\pm$0.11 & 0.38$\pm$0.03 & 6.57$\pm$0.07 & 3.40$\pm$0.07  \\
\hline \hline
\end{tabular}
\end{center}
\end{minipage}
\end{table*}

   \begin{figure*}
   \centering
   \includegraphics[width=7.6cm,bb=45 15 435 405]{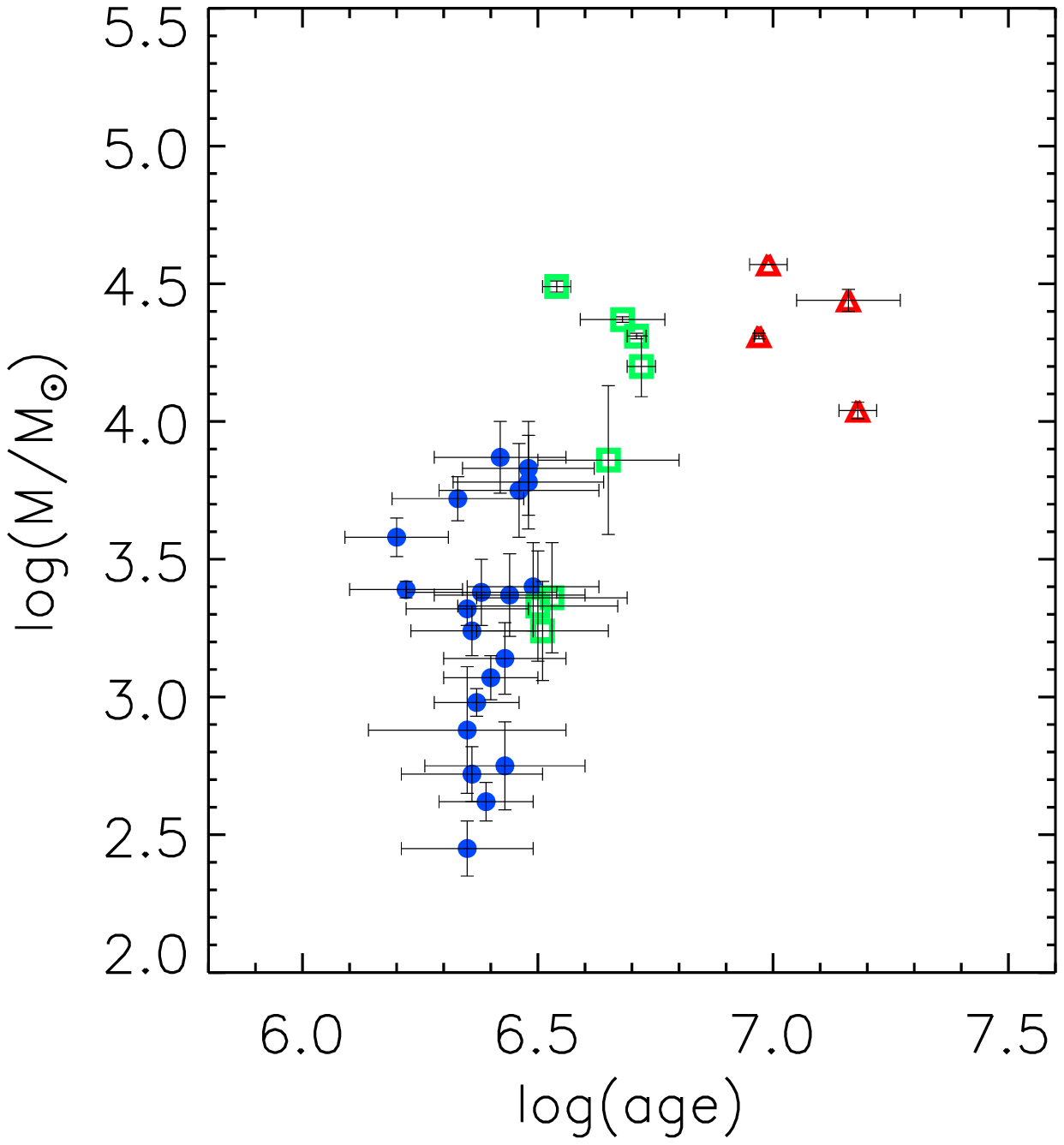}
   \includegraphics[width=7.6cm,bb=45 15 435 405]{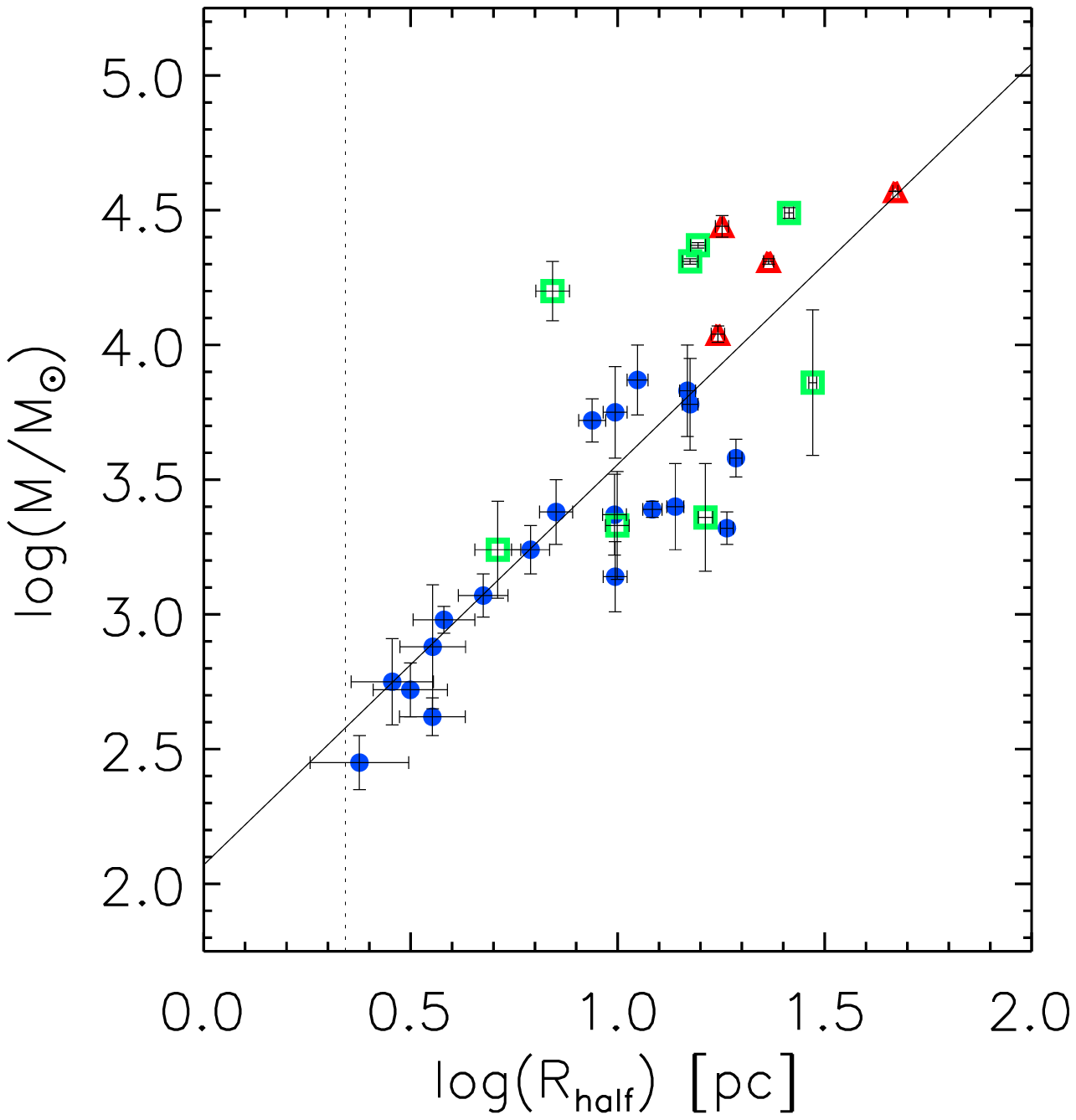}
   \includegraphics[width=7.6cm,bb=45 15 435 405]{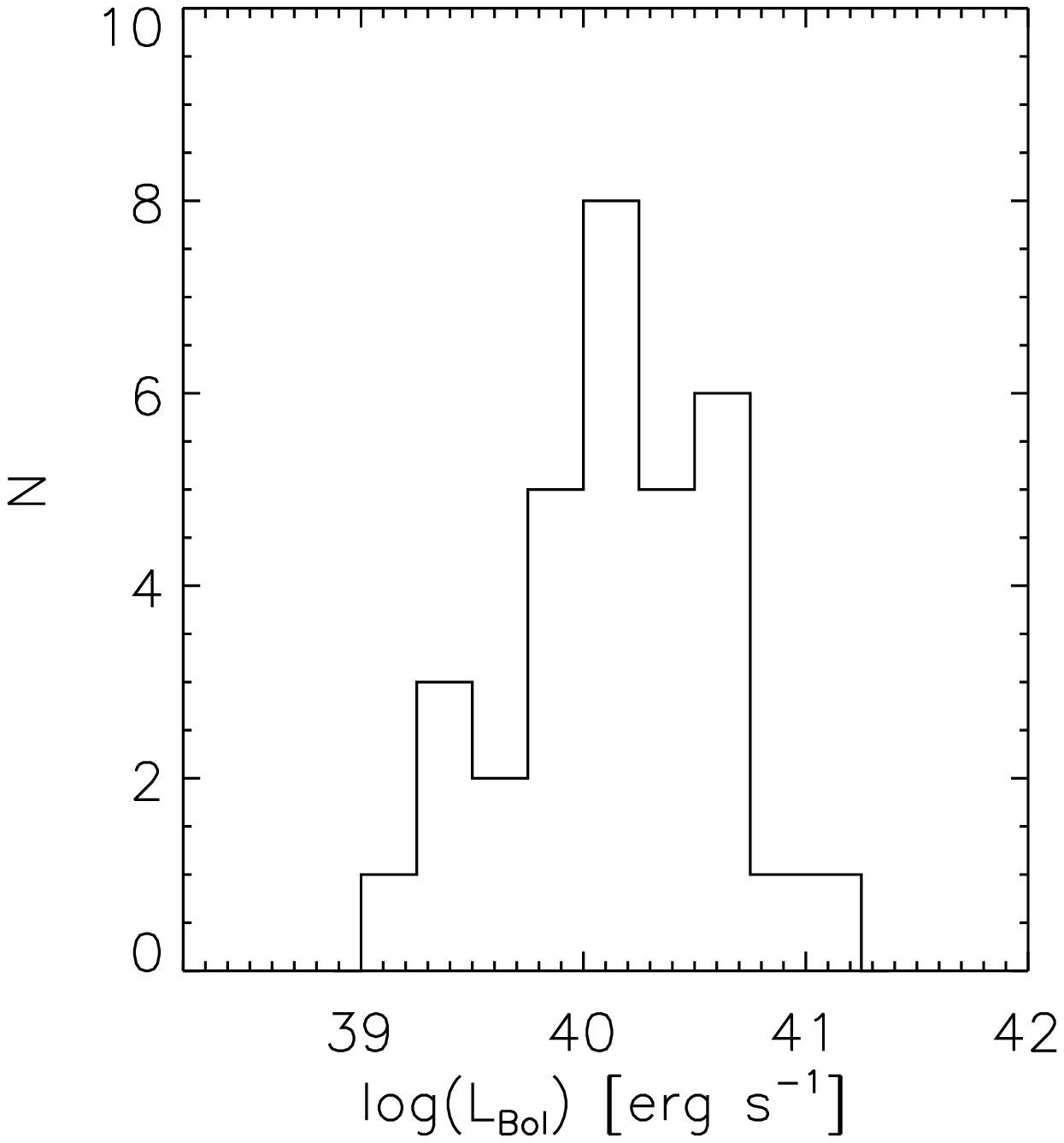}
   \includegraphics[width=7.6cm,bb=45 15 435 405]{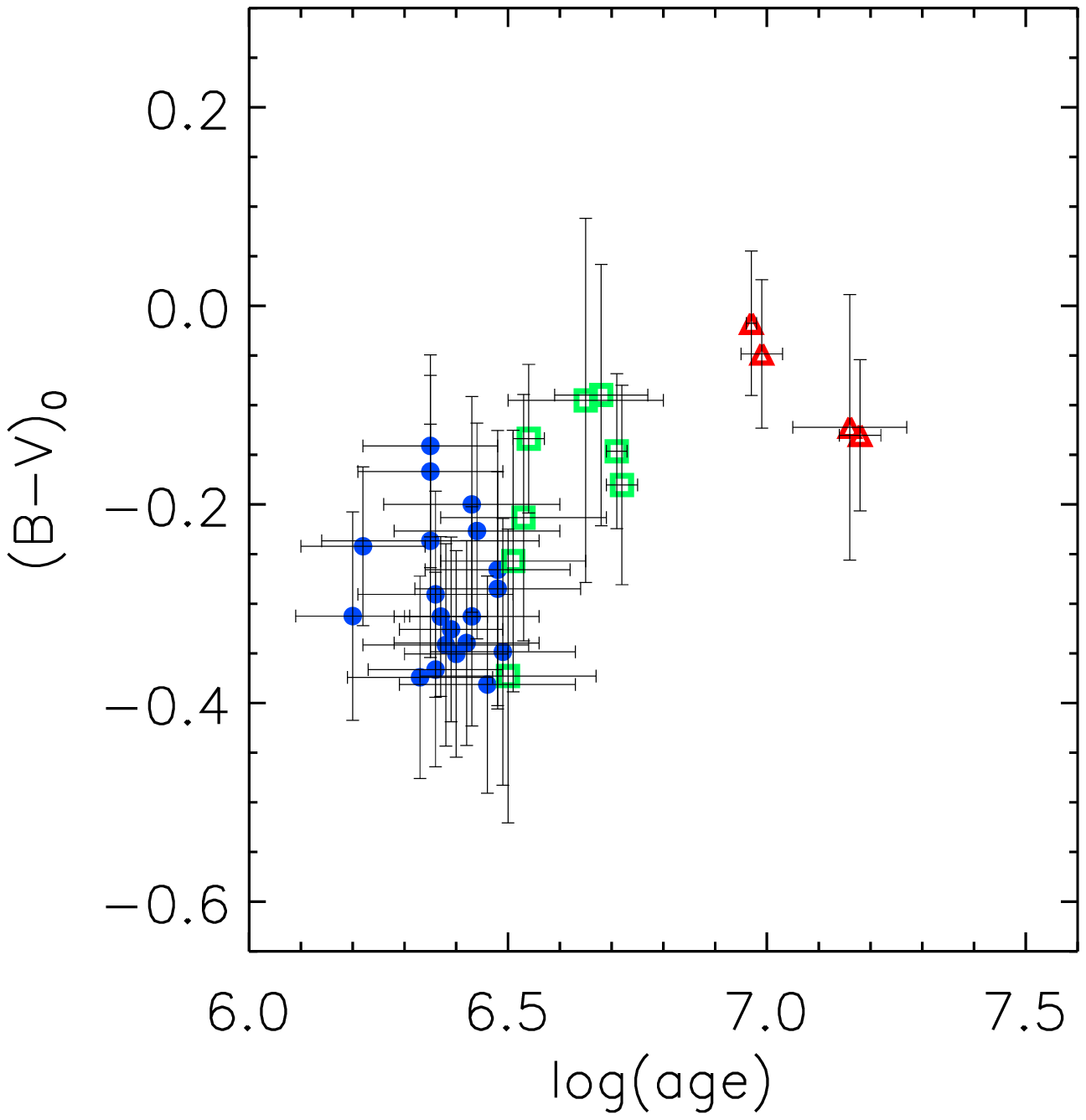}
   \caption{{\em Top-left:} cluster ages and masses determined from
the SED  fitting. {\em Top-right:} mass versus radius relation: the solid line with a slope of 1.5 provides the best linear fits to the data points. The dotted-line shows the minimum size corresponding to the resolution of the $V$-band image.
   {\em Bottom-left:} bolometric luminosity distribution of the sample. {\em Bottom-right:} $(B-V)_0$ colour distribution as a function of age. The symbols correspond to three different age bins:
log(age) $< 6.5$ ({\em filled dots}), $6.5 <$ log(age) $< 6.9$ ({\em squares}), log(age) $> 6.9$ ({\em triangles}).}  \label{age_mass}%
    \end{figure*}

\section{SED fitting with CHORIZOS}

CHi-square cOde for parameteRized modelling and characterIZation of phOtometry and
Spectrophotometry (CHORIZOS) is a $\chi^2$ minimization code that enables one to fit an
arbitrary family of spectral energy distributions to a set of photometric and
spectrophotometric data \citep{2004PASP..116..859M}.
It compares an observed set of magnitudes to a grid of model SEDs and calculates the
probability for a given model to be consistent with the observed photometry.
It  allows one to constrain both intrinsic parameters (such as age and metallicity) and extrinsic ones such as
extinction law and colour excess.
The user selects the SED family and chooses the behaviour of each parameter  to be  fixed,
constrained within a certain range, or to be kept free.
The code uses the monochromatic equivalents of $E(B-V)$ and $R_V$, $E$(4405-5495) and
$R_{5495}$, where 4405 and 5495 are the assumed central wavelengths in Angstrom of the $B$
and $V$ filters. It allows one to choose between the galactic extinction law of
\citet{1989ApJ...345..245C}, the average LMC and LMC2 laws of \citet{1999ApJ...515..128M}
and the Small Magellanic Curve (SMC) extinction law of \citet{1998ApJ...500..816G}.
Then a $\chi^2$ analysis is performed to assign a likelihood to each SED as a function
of the range of parameters.

The code includes a set of STARBURST99 \citep{1999ApJS..123....3L} instantaneous burst
models for a range of  ages and  metallicities, produced using a \citet{1955ApJ...121..161S} stellar initial mass function
(IMF) with masses within the range 0.1-100 \msun. The metallicity of the cluster models was
chosen at Z=0.004, 0.008 and 0.02, while the age varied between 1 and 100 Myr. We
extinguished the theoretical SEDs with all the available extinction laws using a routine
provided with the code.

We ran CHORIZOS for all the objects in our sample using seven magnitudes (from FUV to $I$ bands).
We did not include the 3.6 and 4.5 $\mu$m IRAC bands in the fitting because the contribution of hot dust emission  starts at $\lambda \gtrsim$3 $\mu$m \citep{1988AJ.....95..356C,2003PASP..115..928K} and the STARBURST99 models become inapplicable
above that wavelength \citep{1999ApJS..123....3L}. A more detailed analysis of the MIR region of the cluster SEDs will be presented elsewhere (Giovanardi et al. 2010, in preparation).

We performed the fitting procedure by setting the metallicity at the values displayed in Table 1, leaving the other parameters (i.e. age and intrinsic extinction in M33) unconstrained. Then we reran the code a second time, with a higher resolution grid, to refine the estimate of the parameters. The errors on
the age and the extinction are estimated by the procedure.
To assess the goodness of the fit, we relied on the value of $\chi^2$ provided by CHORIZOS; on
average clusters with $\chi^2 < 10$ were considered as good fits.

Figures \ref{spec1} and \ref{spec2} show our SED-fitting results.
For each cluster we display the output best-fit SED in units of AB magnitudes from 1000 to
10000 $\AA$ overlaid on our measurements to show the quality of the fitting (Fig. \ref{spec1}), and the model
SED in flux units ($\lambda$F$_{\lambda}$, erg s$^{-1}$ cm$^{-2}$) from 1000 to 3 $\times
10^5$ $\AA$ to compare the UV, optical and IR emission of the clusters (Fig. \ref{spec2}). The numbers at the
top-left and bottom-right corners of the plots correspond to the cluster ID given in Table
\ref{list_clust}.
The best-fit SEDs overlaid on the observed magnitudes (Fig. \ref{spec1}) show that in some cases the $U$ photometry disagrees with the model,
despite our attempt to take into account the effects of \oii line contamination,
causing an increase in the value of $\chi^2$.
Comparison between the best-fit model SEDs and the MIR photometry (Fig. \ref{spec2}) shows an excess emission at 3.6 $\mu$m and 4.5 $\mu$m bands, suggesting that this non-stellar emission may be due to a hot
dust component.

\section{Results: comparison of cluster properties}

\subsection{Ages, masses, sizes, and colours}

The SED fittings confirm the young age of the clusters, being younger than $\sim$15 Myr. The results of the fitting procedure are displayed in Table \ref{params}.
Uncertainties in age are
given in Table \ref{params} and are less than 0.2 (in log years). Variations in the IMF slope from the standard Salpeter value, or stochastic effects in the IMF
when sampling small mass
clusters, may increase the quoted  uncertainties \citep{2004A&A...413..145C,2009A&A...495..479C}.
We will discuss the issue of stochastic fluctuations in Sect. 7.1.

CHORIZOS also determines the bolometric correction to the $V$ magnitude for each best-fit model.
Knowing the age, metallicity, and the bolometric luminosity, we inferred the stellar mass using the STARBURST99 evolutionary synthesis models.
Figure \ref{age_mass} ({\em top-left} panel) displays the distribution in age and mass of the clusters. It shows that most of the clusters in our sample are found in a very narrow range of ages, while the mass distribution encompasses two orders of magnitude, between $\sim 3 \times 10^2$ and $\sim 4\times$10$^4$ \msun.  In this and in the following figures we used different symbols (and colours) for clusters within different age bins: filled (blue) dots correspond to ages younger than 3 Myr, (green) squares to clusters with ages in the range 3-8 Myr, while the oldest ones ($>8$ Myr) are indicated with (red) triangles.

Observations of Giant Molecular Clouds (GMCs) in different environments indicate that they follow a mass-radius relation $M \propto R^{2}$ \citep{1981MNRAS.194..809L,
1990ApJ...363..435W, 1994ApJ...429..177H}.
Thus, once we derived the masses, we checked whether the stellar systems under study follow a similar mass-radius relation or not. As an estimate of the size of the cluster we used the half-light radius measured in the $V$ band.
As we show in the {\em upper-right} panel of Fig. \ref{age_mass}, we find a correlation between the half-light radii of the clusters and the mass obtained from the SED fitting,
with a slope of 1.5 $\pm$ 0.2.
The correlation is stronger for clusters with sizes smaller than 10 pc, while the others show a larger scatter around the best-fit line.

Compact star clusters ($R < 10$ pc) with ages between few tens and few hundreds of Myr
show a poor correlation between mass and radius \citep{1999AJ....118..752Z,2004ASPC..322...19L}, implying that the GMC mass-radius relation must be broken at a certain stage during the cluster evolution. \citet{2004ASPC..322...19L} find $M \propto R^{0.1}$  in a sample of relatively young compact clusters (age $<$ 1 Gyr and $R < 10$ pc) in nearby spiral galaxies. On the other hand, cluster complexes with a range of ages more similar to our sample ($\lesssim 10$ Myr) but larger sizes ($R > 85$ pc) follow a similar mass-radius relation to GMCs, $M \propto R^{2.3}$ \citep{2005A&A...443...79B}. Our results seem to indicate that even at scales smaller than those of stellar complexes young stellar systems still display a similar trend to the GMCs,
and that the observed stellar densities are representative of the proto-cluster cloud environment.

The {\em bottom-left} panel of Fig. \ref{age_mass} shows the bolometric luminosity distribution of the clusters, obtained from the SED fitting.
The plot of the dereddened $(B-V)_0$ colour as a function of age ({\em bottom-right} panel) gives an overview of the colour range of the sample. Given their young age, the clusters are characterised by very blue colours and, as expected, ``older'' clusters are redder than the younger ones.

\begin{figure}[h]
   \centering
   \includegraphics[width=7.5cm,bb=65 365 490 780]{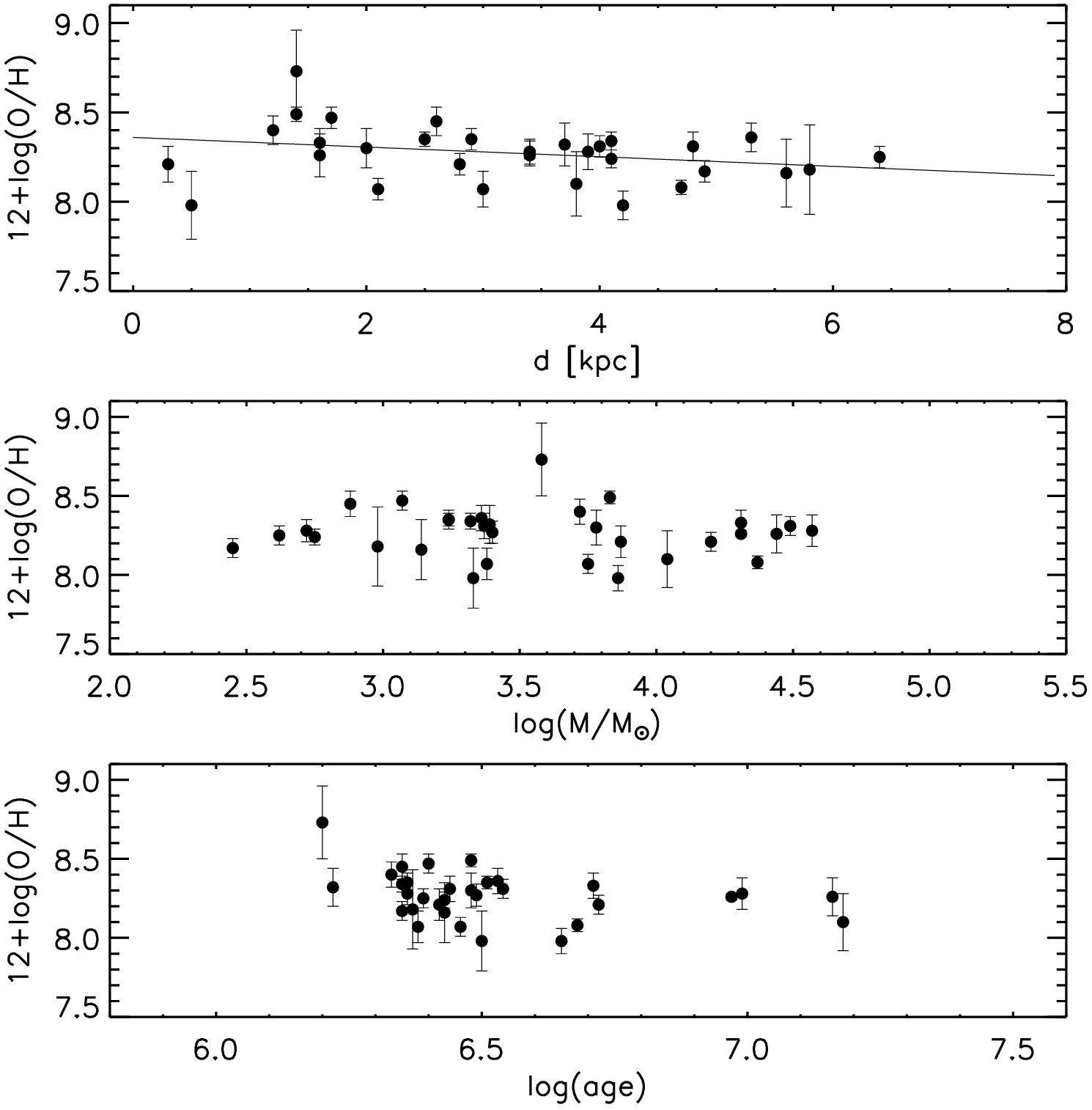}
      \caption{Cluster oxygen abundance versus galactocentric distance, mass, and age. The solid line in the top panel shows the metallicity radial gradient relation found by \citet{2008ApJ...675.1213R}.}
\label{met}%
    \end{figure}

\subsection{Metallicity}

The metallicity
gradient in M33, from 1 to $\sim$7 kpc is rather shallow \citep{1988MNRAS.235..633V,2007A&A...470..865M}, with
a central abundance of 12+log(O/H) =8.36$\pm$ 0.04  and a slope of –0.027$\pm$ 0.012  dex kpc$^{-1}$
 \citep{2008ApJ...675.1213R}.
This can be explained by the global chemical evolution of the M33 disc through an inside-out formation process \citep{2007A&A...470..843M}.
In Fig. \ref{met} ({\em top panel}), the metallicity of the clusters is plotted versus the galactocentric distance, and it is shown that the clusters follow the radial trend found by Rosolowsky and Simon (2008) indicated by the solid line.
However, Rosolowsky and Simon (2008) find a scatter of about 0.1 dex in metallicity at a given radius, which is unexplained by the abundance uncertainties only \citep{2008ApJ...675.1213R}.
A possible explanation would be that self-enrichment could be more efficient in more massive clusters, and  produce the scatter around the average value. If that is the case, we should find a relation between mass and metallicity in our sample.

The {\em central panel} of  Fig. \ref{met} shows that this correlation is lacking in our sample, indicating that the metal abundance is related to the local environmental properties of the clusters rather than being influenced by the cluster mass. Neither do we find a clear trend between the metallicity and age ({\em lower panel}); the Spearman correlation coefficient is -0.32, indicating a poor degree of correlation between these two parameters.

\subsection{Extinction}

\citet{1983ApJ...275..578M} first concluded from the analysis of the IUE spectra of M33 \hii regions that its extinction curve is different from the galactic one, and that it is
characterised by a weak 2175 $\AA$ hump similar to that of the LMC.
\citet{1983MNRAS.203..157D} find a similar result after a study of the \hii region IC132. For the brightest \hii region of M33, NGC604, a reddening parameter E(B-V)=0.12 has been obtained assuming a LMC curve \citep{2000MNRAS.317...64G}.

The $E(B-V)$ values we derive for the M33 clusters range between 0.1 and 0.5 mag, corresponding to a visual extinction $A_V$ between 0.3 and 1 mag.
Chandar et al. (1999) obtain $E(B-V)$ values between 0.06 and 0.33 mag, with the majority of the clusters being found around 0.1 mag. \citet{2007AJ....134.2168P} derived reddening values between 0.05 and 0.20 mag. Yet both samples are different from ours because they have a large range of ages as mentioned in Sect. 1.
The LMC average  and LMC2 curves of Misselt et al. (1999) were the ones providing the best fits in the majority of the cases, even though the SMC extinction curve was adopted for one cluster (C400).
Additional evidence supporting the choice of
either the LMC average or the LMC2 extinction curve comes from the UV colours.
The SMC extinction law predicts larger variations in the FUV-NUV colour than what
was observed in our sample.

\begin{figure*}
   \centering
   \includegraphics[width=6cm,bb=45 15 420 405]{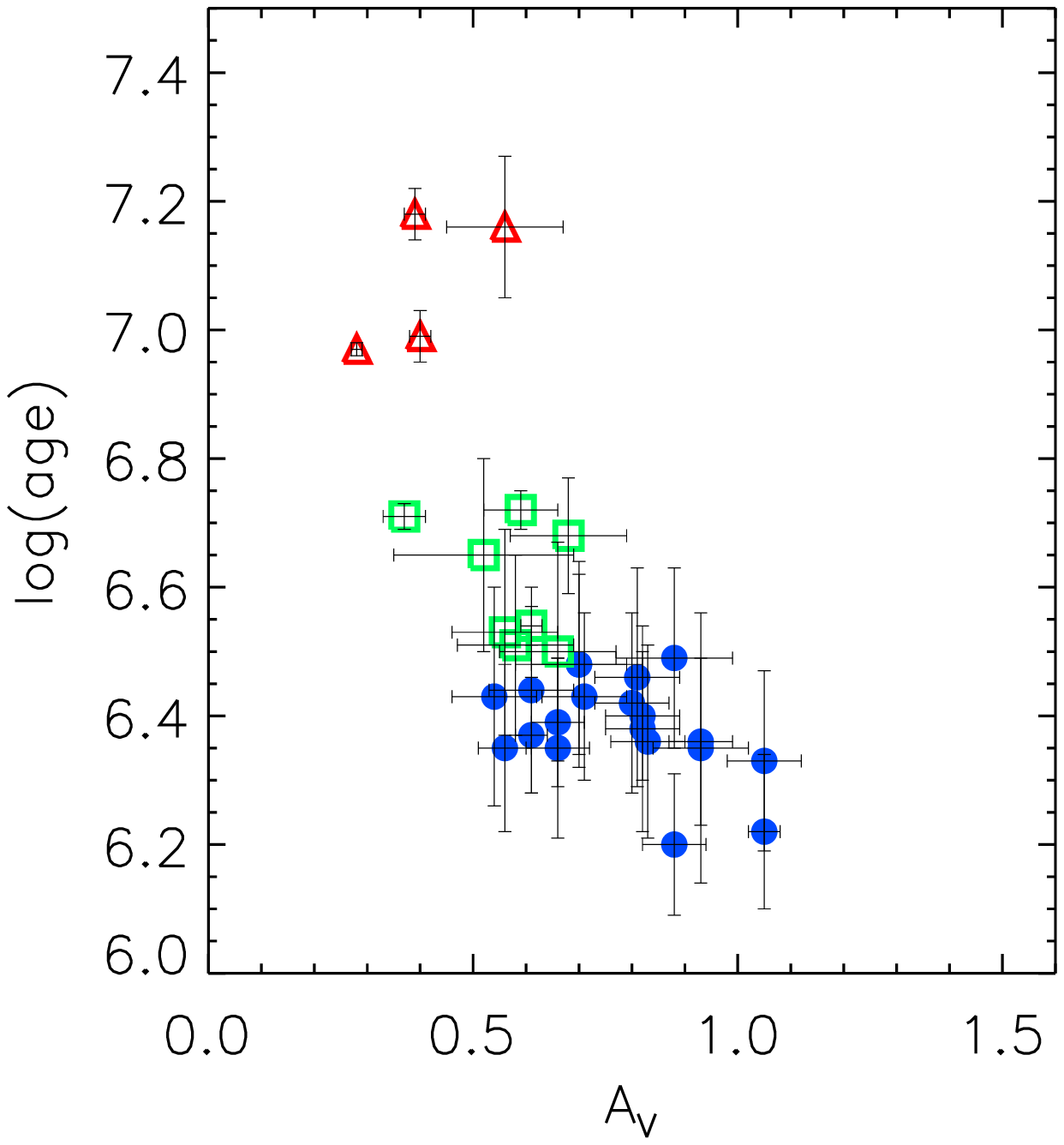}
   \includegraphics[width=6cm,bb=45 15 420 405]{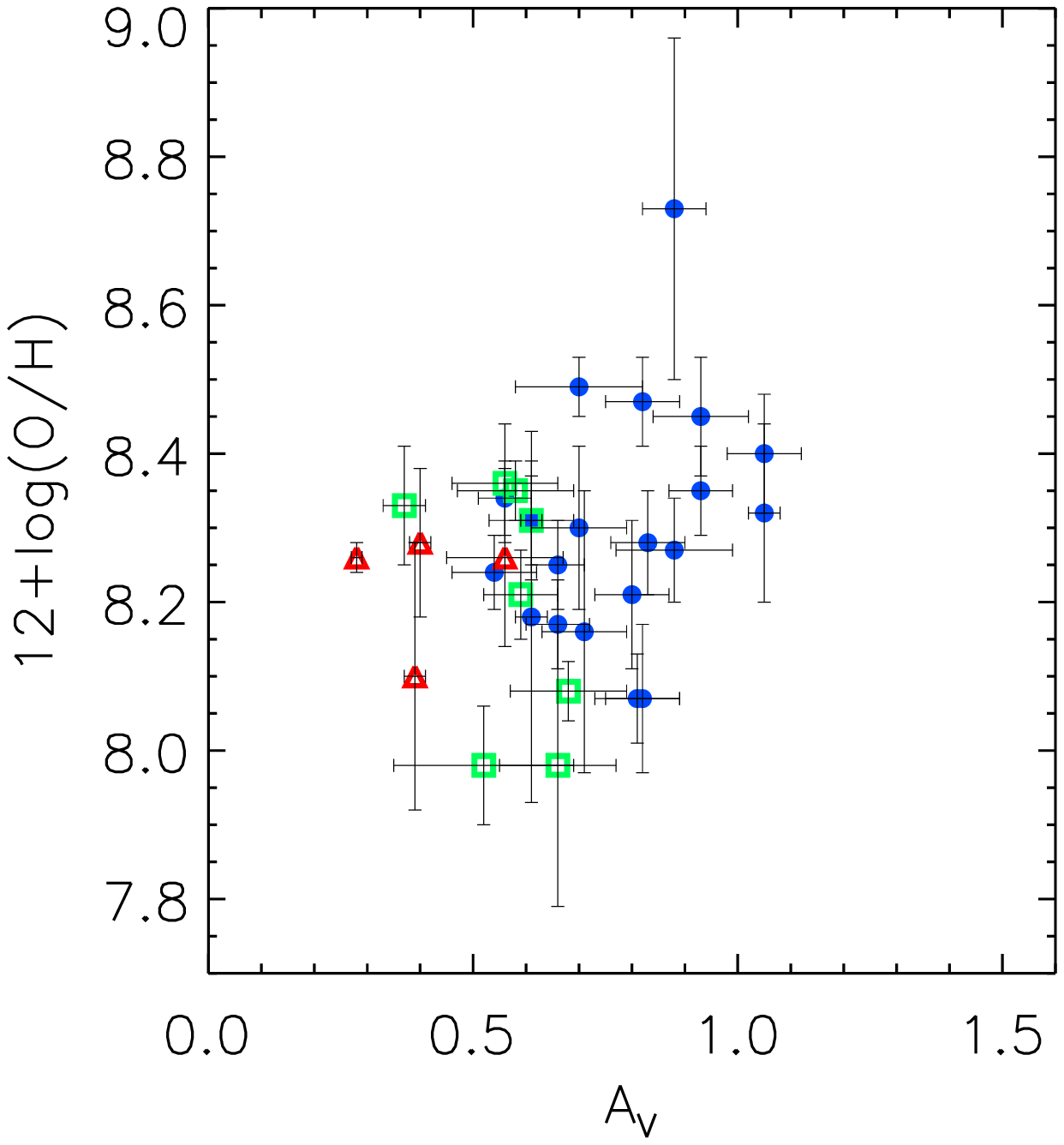}
   \includegraphics[width=6cm,bb=45 15 420 405]{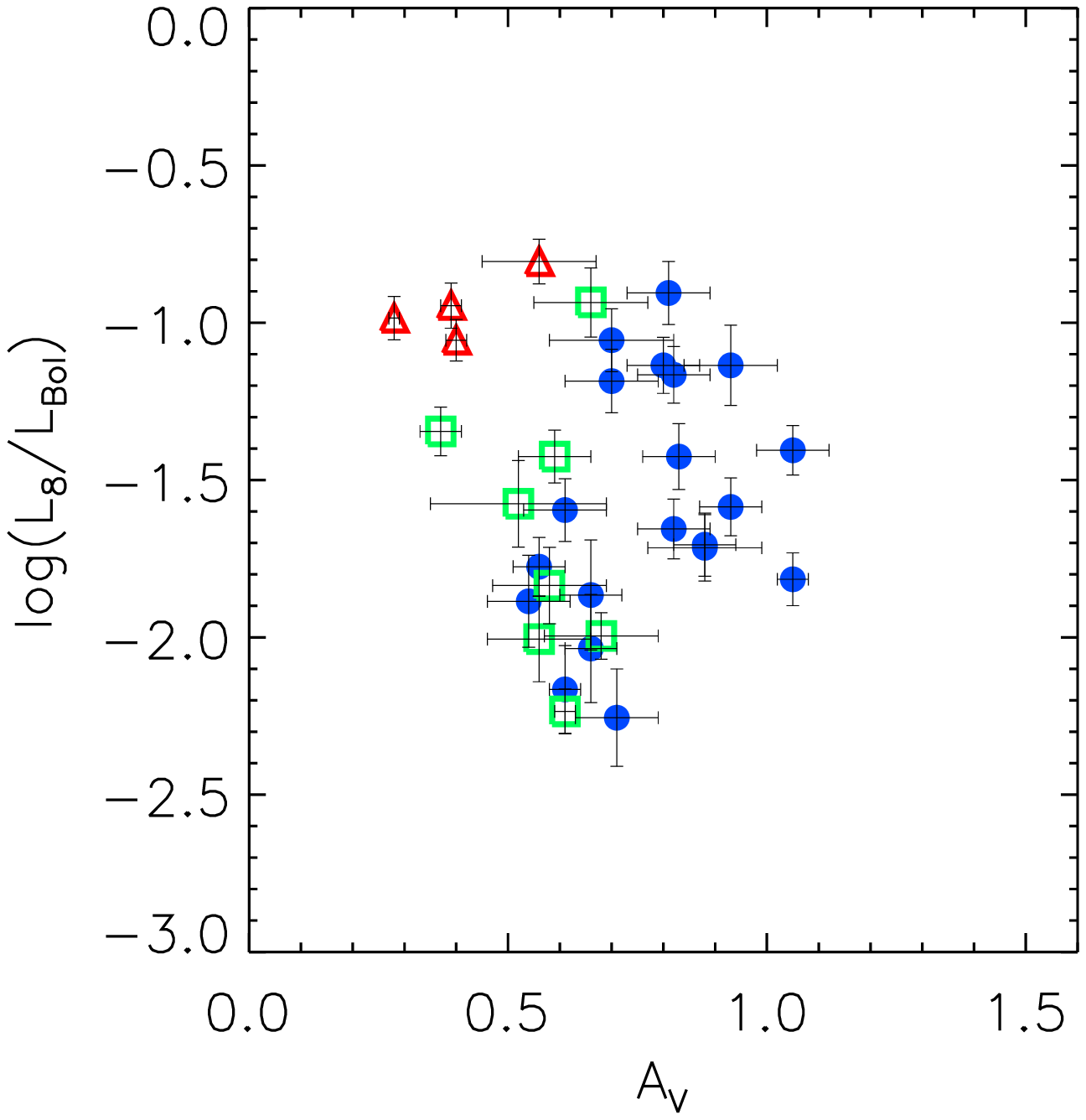}
   \includegraphics[width=6cm,bb=45 15 420 405]{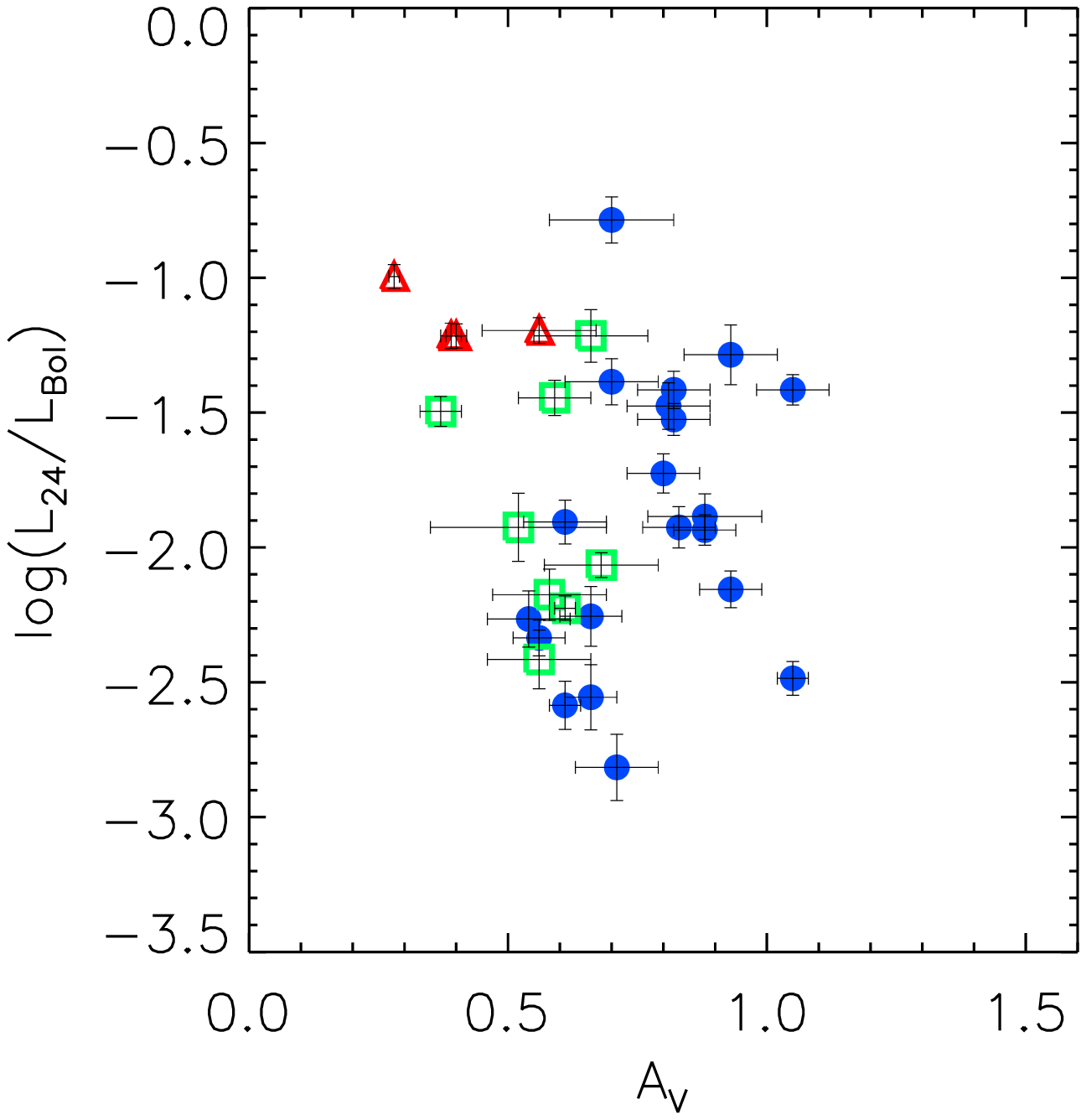}
   \includegraphics[width=6cm,bb=45 15 420 405]{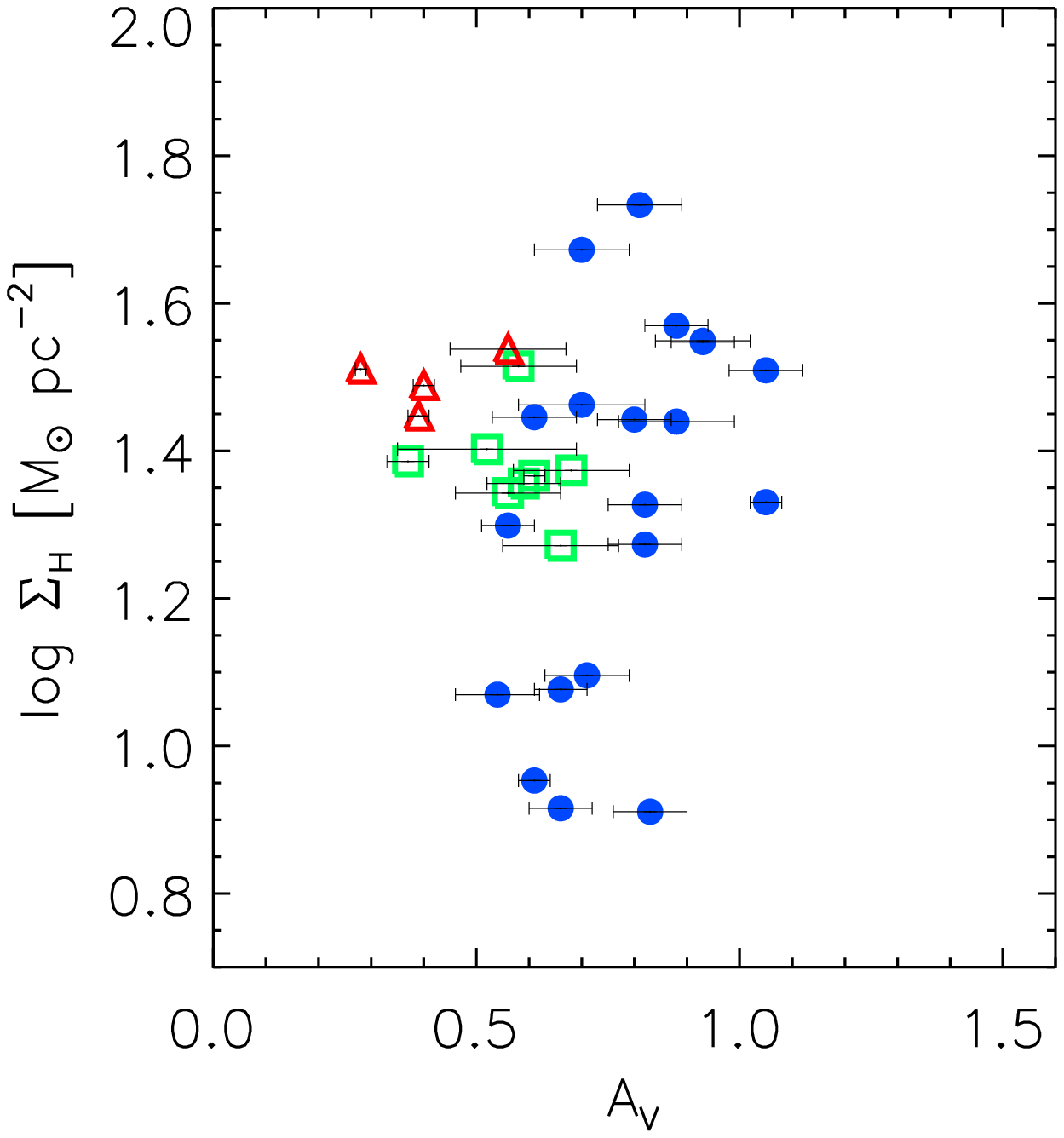}
   \caption{Visual extinction (A$_V$) versus age ({\em top-left}),
the measured oxygen abundance ({\em top-centre}),
the ratio of the the 8 $\mu$m to the bolometric luminosity ({\em top-right}),
the ratio of the the 24 $\mu$m to the bolometric
luminosity ({\em bottom-left}), the total gas density within a region of about 200 pc around the cluster ({\em bottom-right}). Symbols are the same as for Fig. \ref{age_mass}.}
\label{ext}%
    \end{figure*}

Figure \ref{ext} shows that there is a correlation between the cluster age and $A_V$ ({\em
top-left}). Younger clusters are more reddened, because they are still partially embedded in the parent molecular cloud where they formed.
$A_V$ does not appear to clearly correlate with the metallicity ({\em top-centre}), and neither does it correlate with the 8 $\mu$m ({\em top-right}) nor the 24 $\mu$m luminosities ({\em bottom-left}) normalised to the bolometric one.
A trend between the extinction
and the total hydrogen column density is also lacking (H$_2$+\hi+\hii; see also
Sect. 8), as displayed in the {\em bottom-right} panel.

\subsection{Radial gradients}

  \begin{figure*}
   \centering
   \includegraphics[width=6cm,bb=28 10 400 400]{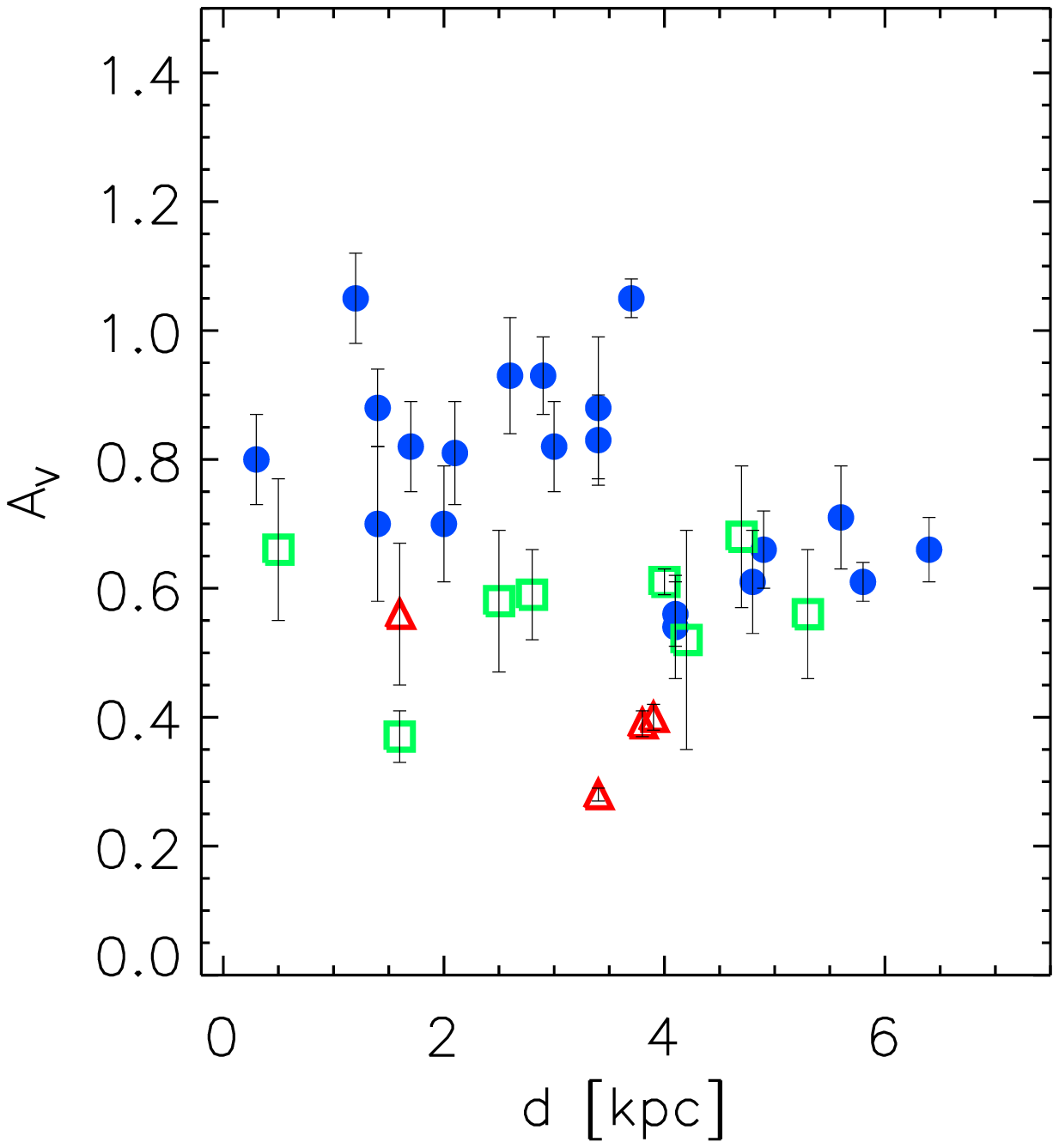}
   \includegraphics[width=6cm,bb=28 10 400 400]{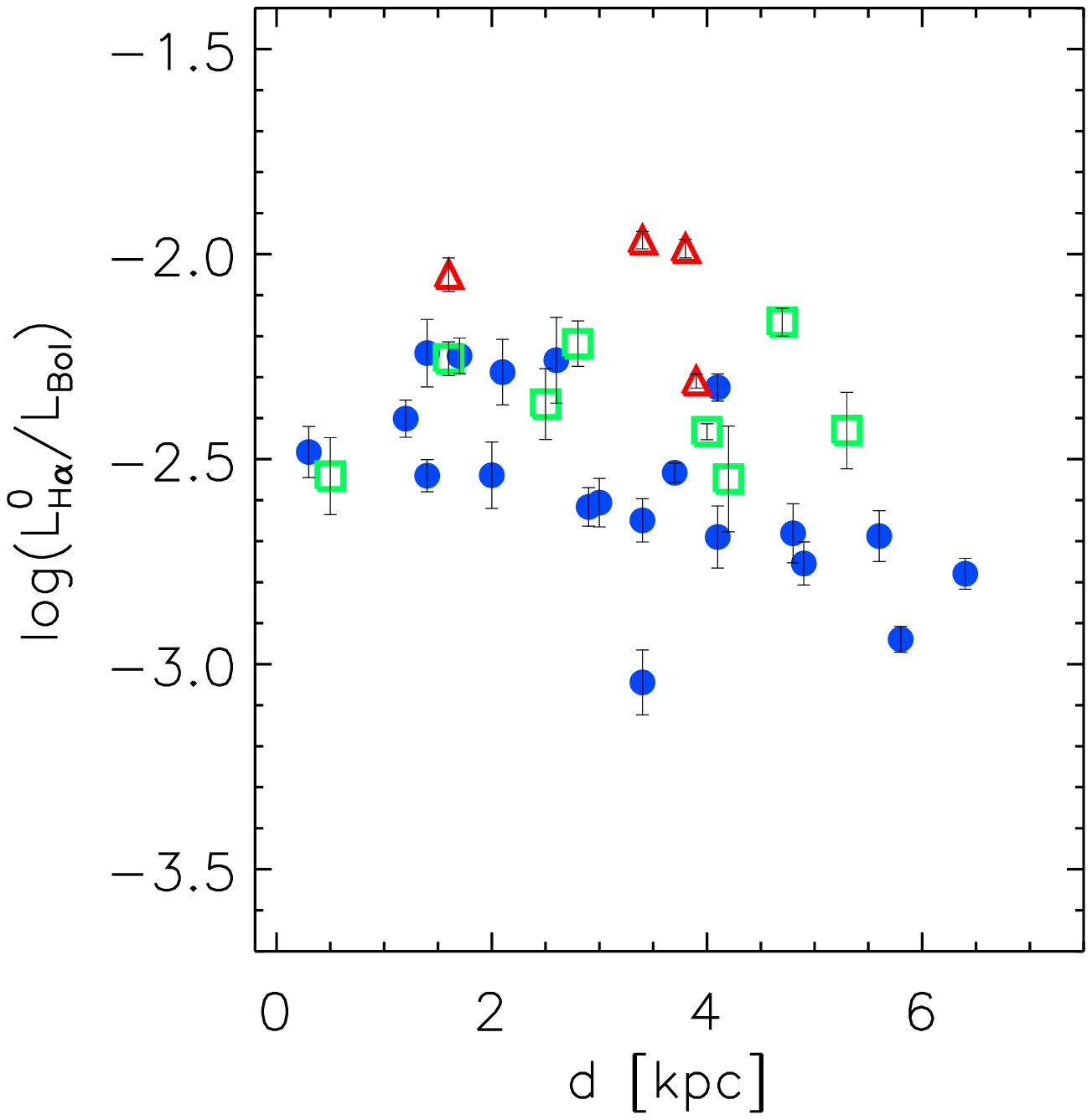}
   \includegraphics[width=6cm,bb=28 10 400 400]{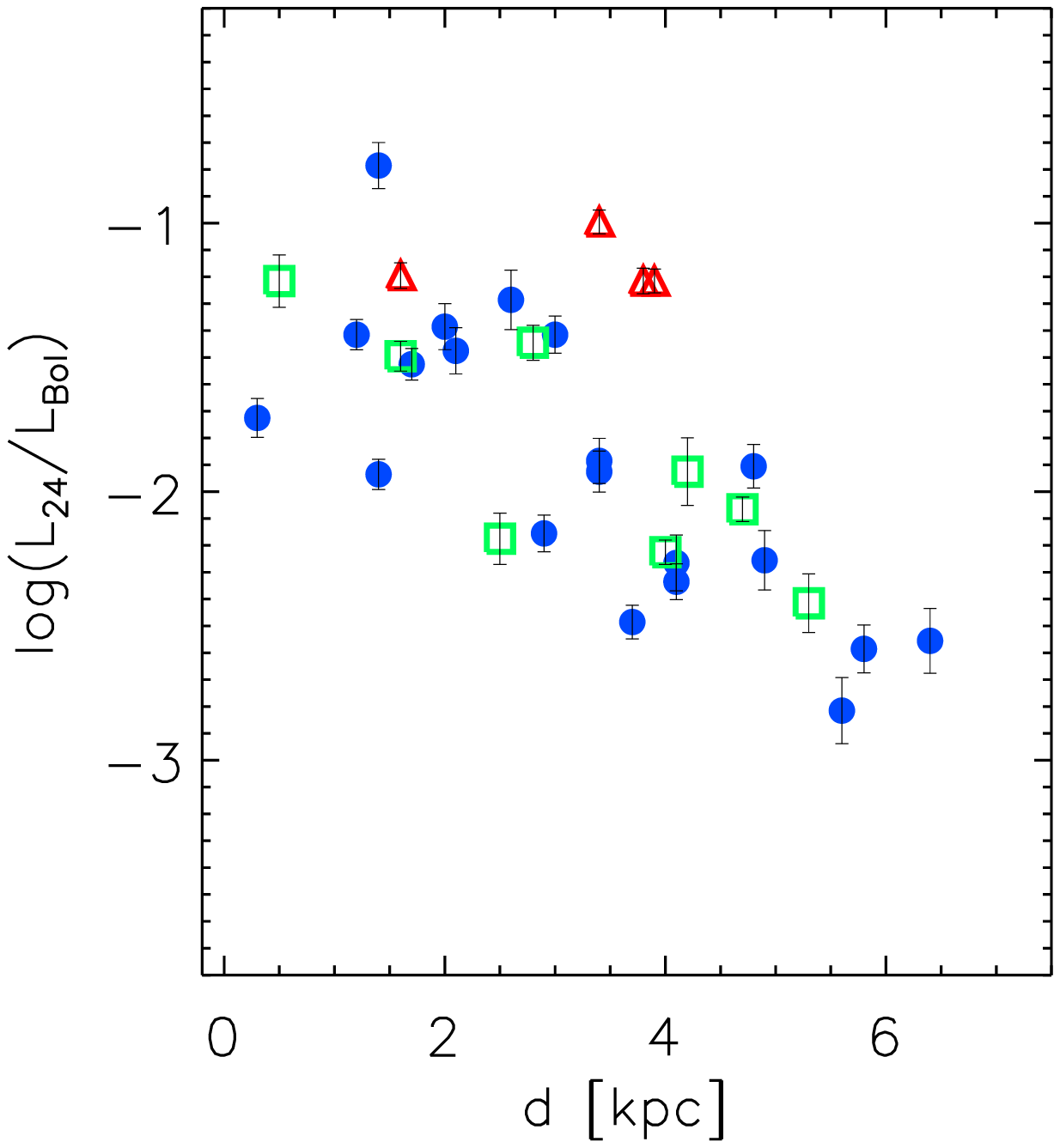}
   \includegraphics[width=6cm,bb=28 10 400 400]{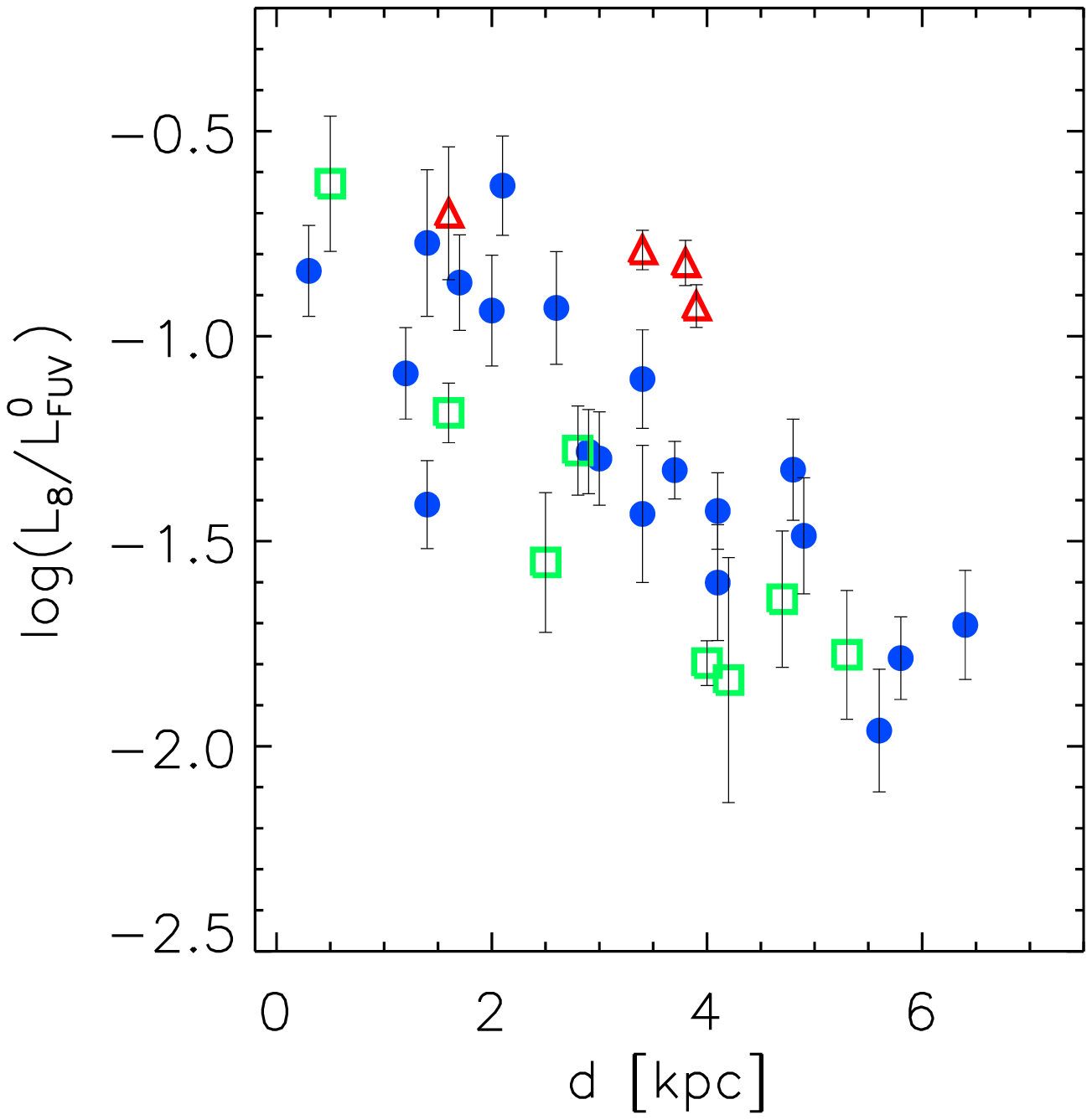}
   \includegraphics[width=6cm,bb=28 10 400 400]{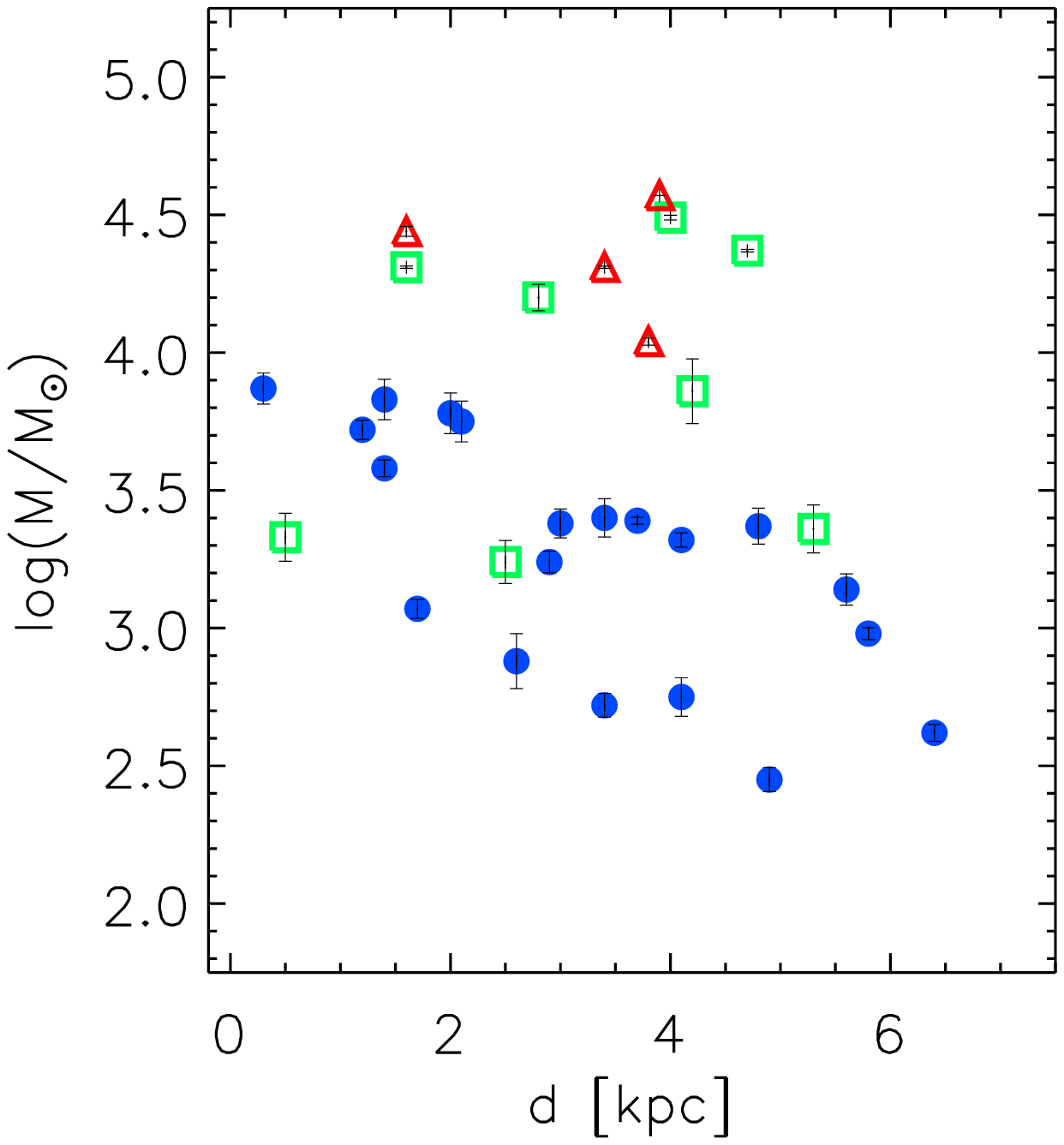}
   \caption{Cluster galactocentric distance versus $A_V$ ({\em top-left}), the L$^0_{H\alpha}$/L$_{Bol}$
   ratio (H$\alpha$ corrected for extinction, {\em top-centre}); the L$_{24}$/L$_{Bol}$ ratio ({\em top-right}); the
L$_{8}$/L$^0_{FUV}$ ratio (FUV corrected for extinction, {\em bottom-left});
the total stellar mass ({\em bottom-right}). Symbols are the same as for Fig. \ref{age_mass}.}
\label{rad_grad}%
    \end{figure*}

We searched for possible variations of the cluster properties with the galactocentric radius, especially with regard to their  UV, H$\alpha$ or MIR emissions (see Fig. \ref{rad_grad}).

Neither $A_V$ ({\em top-left}) nor the L$_{FUV}/$L$_{Bol}$ ratio (not displayed) show a clear trend with the galactocentric distance. A radial change in $A_V$ would imply a decreasing dust attenuation, as it is found for example in M51 \citep{2005ApJ...633..871C}. In this galaxy the UV colour, $A_V$,
and the age of the clusters do vary with the distance, with the clusters becoming younger and less extinct as one moves to the outer regions of the galaxy.  Extinction is related to the dust-to-gas ratio, which depends on the metallicity of a galaxy. Because steep metallicity gradients are observed in many galaxies, one would expect that the attenuation properties would also change with the galactocentric distance.
The shallow metallicity gradient in M33 \citep{2007A&A...470..865M,
2008ApJ...675.1213R} might explain the lack of a clear correlation between $A_V$, or L$_{FUV}/$L$_{Bol}$, and the galactocentric radius.

The L$^0_{H\alpha}$/L$_{Bol}$ ratio ({\em top-centre}, H$\alpha$
corrected for extinction) 
decreases with galactocentric distance for the youngest clusters (filled dots). However, since the youngest age bin includes several low-mass clusters, this might be due to a selection effect: most of the selected low-mass clusters are at large galactocentric radii in regions that are not too crowded in the H$\alpha$ map (see the last panel in the figure).
We will discuss this  in more detail in Sect. 7.
Note that the H$\alpha$ luminosity of the clusters with ages around 10 Myr is more than one order of magnitude higher than what STARBURST99 predicts, implying that
subsequent star-formation episodes must have occurred in order to
explain the observed H$\alpha$ emission.

\begin{figure}
   \centering
   \includegraphics[width=8cm,bb=50 20 414 400]{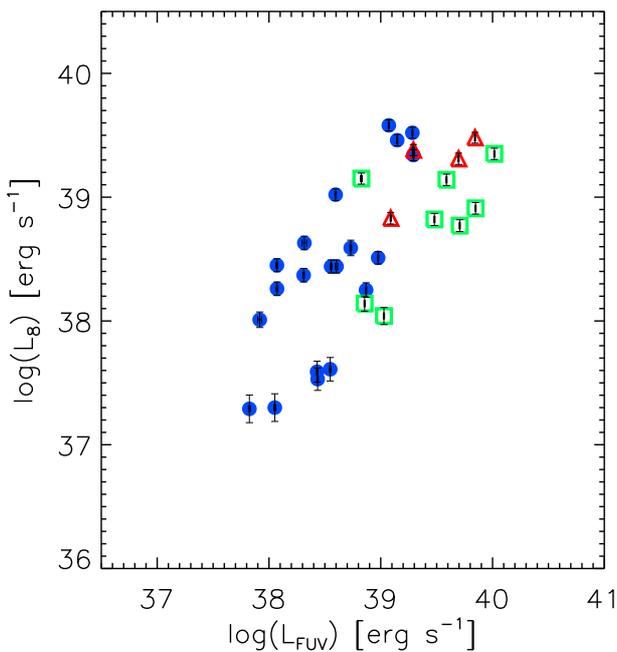}
\caption {FUV luminosity, L$_{FUV}$, versus 8 $\mu$m luminosity, L$_8$, for the cluster sample. Symbols are the same as for Fig. 4.}
\label{fuv8}%
    \end{figure}

We see a well defined radial trend for the MIR fractional
luminosities. Both the L$_{24}$/L$_{Bol}$ ({\em top-right}) and the L$_8$/L$^0_{FUV}$
ratios ({\em bottom-left})  decrease by at least one order of magnitude at large galactocentric radii, at the edge of the star-forming disc.
The MIR emission at 24 $\mu$m is mainly due to very small grains with effective radii
between 15 and 40 $\AA$ \citep{2007ApJ...657..810D}.
Variations in the abundance of small grains with respect to big ones at larger radii,
or, more generally, a radial gradient in the dust-to gas ratio or in the dust temperature, may also contribute to the decrease of the 24 $\mu$m emission \citep{2001A&A...373..702H,2002A&A...389..239A,2005ApJ...633..871C}.
The luminosity at 8~$\mu$m is likely dominated by Polyciclic Aromatic Hydrocarbons (PAH) features around star-forming regions excited by the cluster UV radiation, and we find
a good correlation between 8 $\mu$m and the FUV luminosities of the clusters (Fig. \ref{fuv8}). The PAH emission is reduced in low-metallicity objects, but over the small range of metal abundances in our sample there is no clear dependence of the
L$_8$/L$_{FUV}^0$  ratio on the metallicity. At larger radii the volume density of
the ISM decreases, therefore one would expect that dust and PAHs may be more efficiently removed from star-forming sites in the outer regions of the disc \citep{2005ApJ...632..227R}.

The total stellar mass does not seem to be clearly related to the cluster position within the galaxy as shown in the {\em bottom-right} panel of Fig. \ref{rad_grad}, even though the smallest mass clusters of the sample (M $< 10^3$ M$_{\odot}$) are mainly located at large galactocentric distances.

\subsection{Bolometric luminosity}

The SED-fitting method provides the value of the bolometric luminosity
for each cluster, L$_{Bol}$ (see Table 5).
We compared this parameter with the bolometric luminosities estimated from observations of the total IR and UV
emission, i.e. L$_{Bol}^{obs}=$  L$_{FUV}$ + L$_{NUV}$ + L$_{TIR}/(1-\omega)$, where $\omega \sim 0.5$ is the albedo for a standard
dust composition and grain size distribution \citep{2001ApJ...548..296W}.

The sum of the FUV and NUV luminosities provides a measure
of the cluster emission below 3000~\AA, which is close to the bolometric luminosity
if the cluster is young and unobscured by dust \citep{2005ApJ...625...23B, 2007ApJS..173..572T}.
If extinction corrections are unknown, one can use L$_{TIR}$ to estimate the
fraction of the UV emission absorbed by the dust. Therefore L$_{Bol}^{obs}$
is considered to provide a good approximation of the bolometric luminosities of young
clusters \citep[see][]{2007ApJS..173..572T, 2009A&A...495..479C}.
Verley et al. (2009) have shown that the total IR luminosity in M33 can be estimated from the 24 $\mu$m emission and it is approximately L$_{TIR} \sim $ 12 $\times$ L$_{24}$ (see Eq. 4 in Verley et al. 2009).

\begin{figure}
   \centering
   \includegraphics[width=8cm,bb=110 365 470 765]{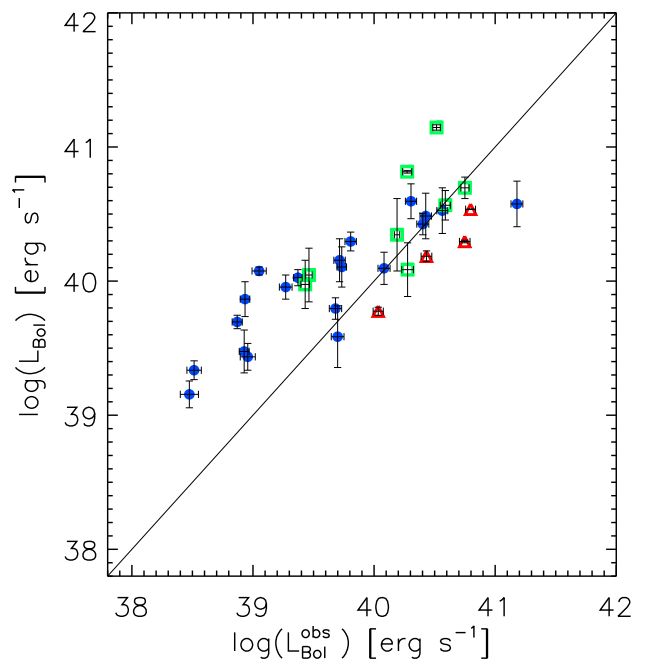}
\caption{
Comparison between the cluster bolometric luminosity determined from
the SED spectral fitting (y-axis) and the observed bolometric luminosity derived by adding the
measured FUV, NUV, and the total IR luminosity estimated following  Verley et al. (2009) as
L$_{TIR} \sim 12 \times$ L$_{24}$. Symbols are the same as for Fig. \ref{age_mass}.
}
\label{lumbol}%
    \end{figure}

Figure \ref{lumbol} shows that the
model bolometric luminosities are higher than the observed ones for L$^{obs}_{Bol}$ $\lesssim 3\times 10^{39}$ erg s$^{-1}$.
This difference could be due to two main reasons: a) the assumption of a complete IMF may be incorrect when we consider clusters with bolometric luminosity (and mass) below a certain threshold, leading to an overestimate of the total cluster luminosity; b)
the estimated total IR emission of the clusters is lower than the amount of energy missing in the non-ionising UV region of the spectrum, implying that there is a deficit of non-ionising radiation in young clusters with bolometric luminosities lower than $3\times 10^{39}$ erg s$^{-1}$. We will discuss these issues in more detail in the next section.

\section{The deficit of ionising radiation}

In the previous section we pointed out
that the total IR emission of the clusters with log($L_{Bol}$) $< 39.5$
is lower than the amount of energy missing in the non-ionising UV region of the
spectrum. Here we want to compare the observed H$\alpha$ luminosity
with that predicted by an instantaneous burst model to check whether there is a deficit of UV-ionising radiation in the sample of clusters under study.

\begin{figure*}
   \centering
   \includegraphics[width=6cm,bb=70 360 450 770]{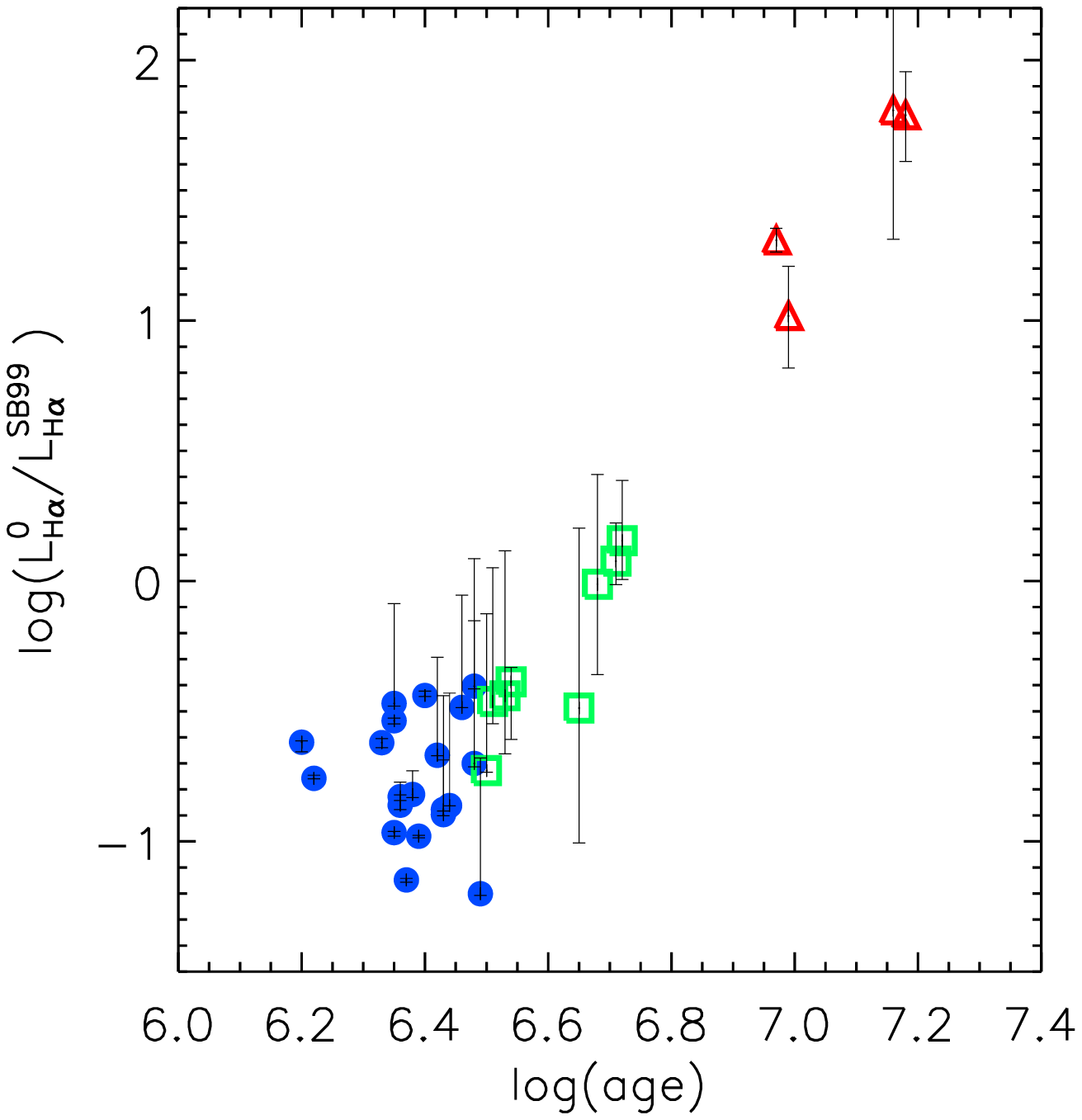}
   \includegraphics[width=6cm,bb=75 360 455 770]{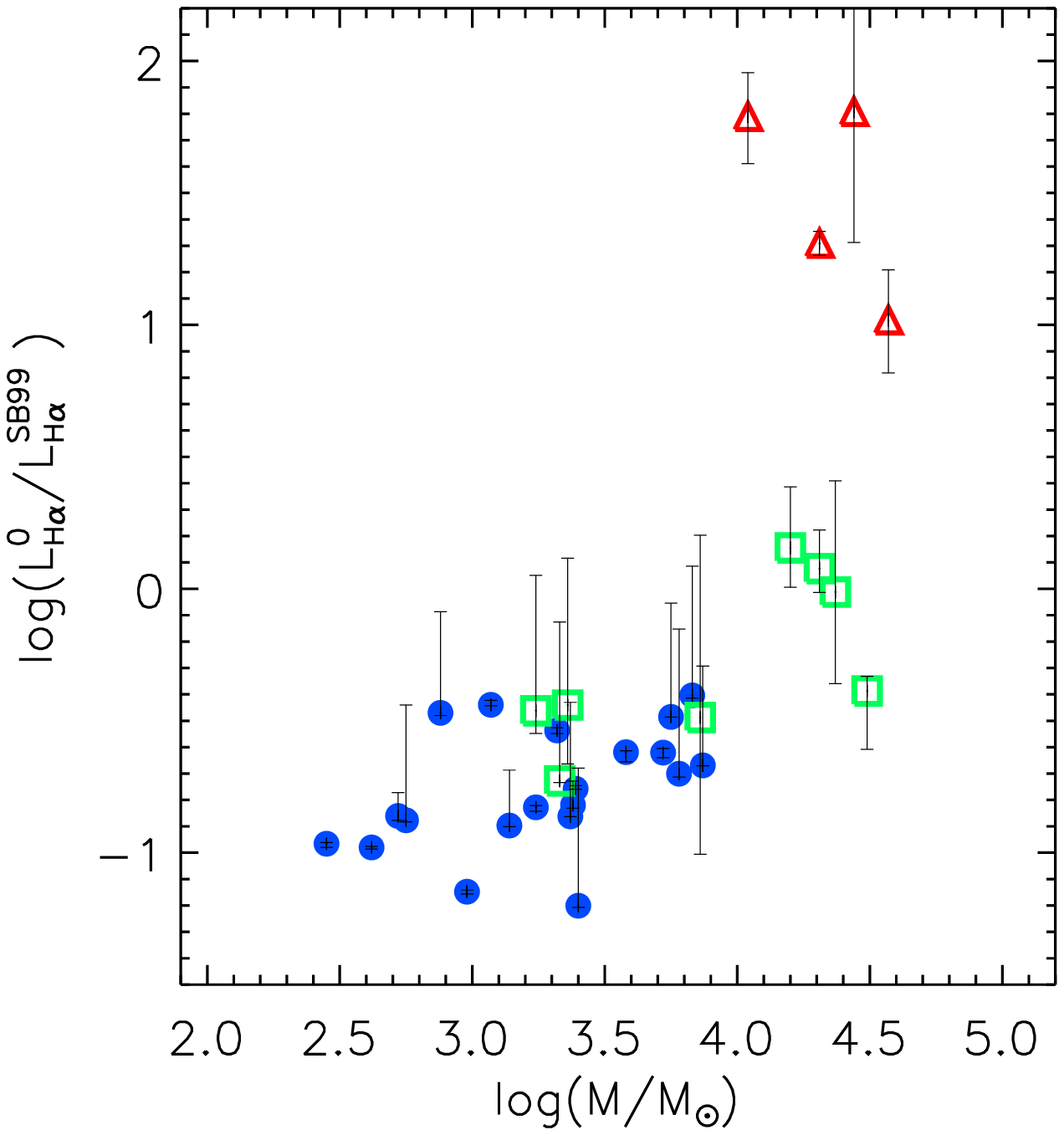}
   \includegraphics[width=6cm,bb=60 360 440 770]{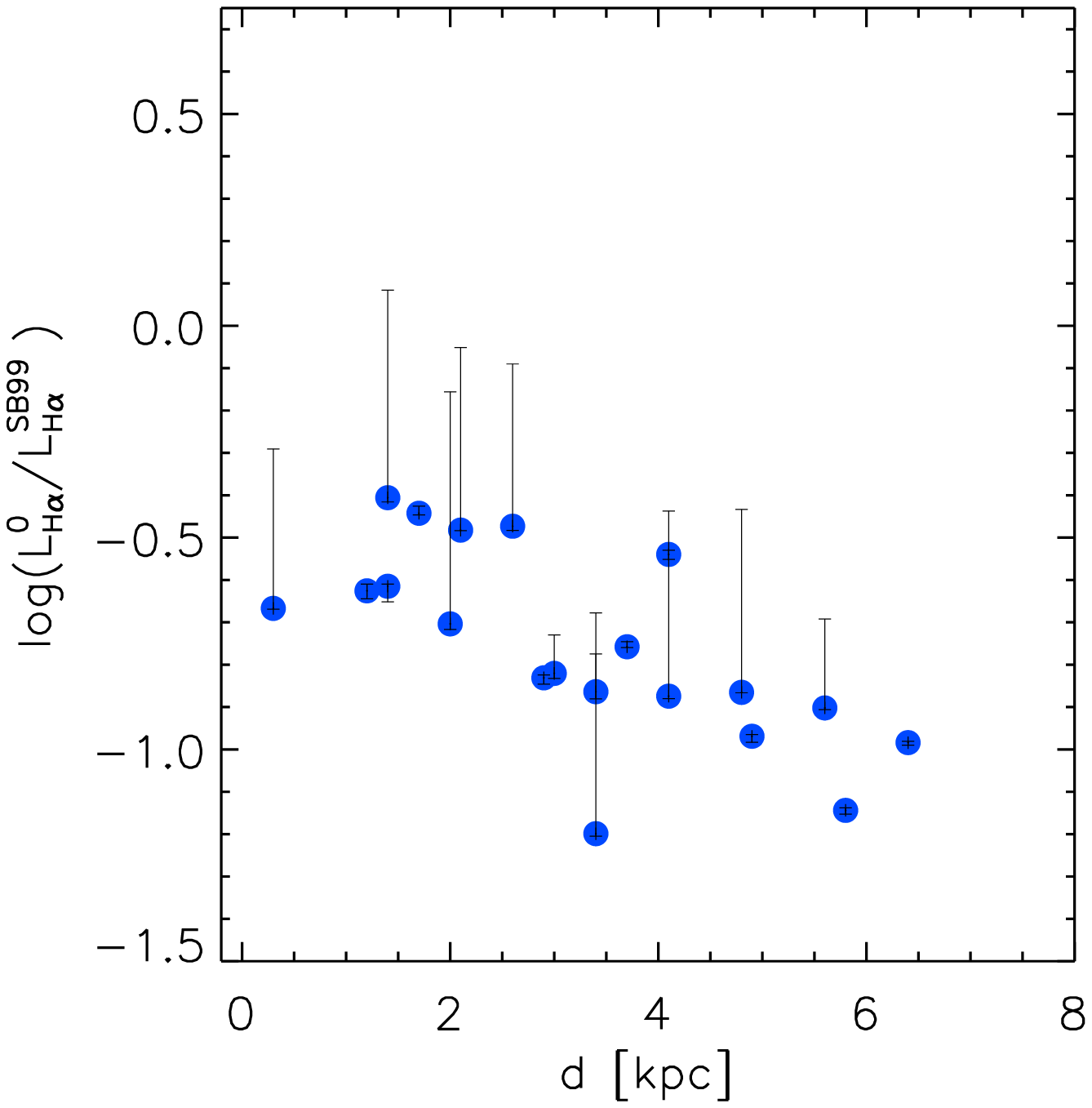}
   \caption{Ratio of the observed (extinction corrected) to the
model H$\alpha$
luminosity determined with  STARBURST99 vs. the age of the clusters ({\em left-panel}), the mass of the clusters ({\em central-panel}), and  the galactocentric distance for clusters with log(age) $<
6.5$ ({\em right-panel}). The error bars give the range of the H$\alpha$ luminosity predicted by STARBURST99 for each cluster due to uncertainty on the age. Symbols are the same as for Fig. \ref{age_mass}.}
\label{leak} %
    \end{figure*}

We ran STARBUST99 to obtain the expected H$\alpha$ luminosity of each
cluster at the age, metallicity, and mass we determined, and we plot
in Fig. \ref{leak} the ratio of the
extinction-corrected observed H$\alpha$
luminosity to the model one -- L$^0_{H\alpha}/$L$^{SB99}_{H\alpha}$ -- versus
the age ({\em left-panel}) and the mass ({\em central-panel}) of the clusters.
The figures show that this ratio
decreases with age and mass, while the {\em right-panel} shows a shallow radial trend with the galactocentric radius for the youngest clusters.
There is a significant discrepancy between the observed and model H$\alpha$ emission especially in young ($t <$ 3 Myr) and low-mass (log(M/M$_{\odot}$) $< 3.5$) systems, for which L$^0_{H\alpha}/$L$^{SB99}_{H\alpha}$ is less than 0.5, or -0.3
in logarithmic units. 
This would imply that
more than 50\% of the ionising flux is missing from these clusters,
a higher percentage than what was found in other studies (see Sect. 7.3).

The error bars in Fig. \ref{leak} show how the ratio of H$\alpha$ luminosities varies if one takes into account the uncertainties of the best-fit ages (Table 4).
The upper errors bars are larger for clusters older than 2.5-3 Myr, because the model H$\alpha$ emission rapidly decreases after this epoch.
For the youngest clusters ($t <$ 3 Myr), the mean L$^0_{H\alpha}/$L$^{SB99}_{H\alpha}$ ratio is 0.19$^{+0.17}_{-0.002}$, while for those with an age between 3 and 8 Myr
the mean is L$^0_{H\alpha}/$L$^{SB99}_{H\alpha} = 0.65^{+0.83}_{-0.22}$.
On the other hand, clusters older than 8 Myr (triangles in Fig. \ref{leak})
are very luminous at H$\alpha$ wavelengths for their age, because L$^0_{H\alpha}$ is more than 10 times higher than the prediction of simple stellar population models.
It is likely that in these cases the assumption of an instantaneous burst is incorrect and an extended star-formation activity is required to explain the observed H$\alpha$ emission.

Hence, even taking into account the uncertainties of the cluster ages, the missing
fraction of ionising radiation is rather large in young clusters, and we examine
four different hypotheses to explain this issue.

First we discuss how the incomplete sampling of the IMF and statistical fluctuations can affect the prediction of cluster luminosities. Then we consider the effects related to the cluster environment such as dust absorption and photon-leakage (i.e. a large fraction of UV radiation escapes the \hii region). Finally we briefly discuss whether
the delayed formation of the most massive stars
or a truncated IMF can offer alternative explanations for the low H$\alpha$ luminosity observed in very young clusters.

\subsection{The consequence of sampling effects in the IMF}

The library of models that we used for the SED fitting assumes a standard
Salpeter IMF, fully populated up to a high-mass end of 100 \msun. 
It is possible that the
difference between the expected and observed H$\alpha$ luminosities in the
youngest clusters is because we are overestimating the number of massive stars 
owing to the incompleteness of the IMF at
the high-mass end.

Incomplete sampling is often neglected, and it is assumed that the mass
spectrum of stellar clusters can be represented by a continuous  function
given by the underlying IMF. On the other hand,  numerical simulations
of synthetic clusters by a stochastic generation of stars show that this assumption is correct for massive systems (M $> 10^4$ \msun), whereas large deviations at the high-mass end occur for small clusters \citep{2009A&A...495..479C}. The critical mass below which
statistical fluctuations become relevant 
depends on the cluster age and metallicity, and on the spectral range under study \citep{2002A&A...381...51C,2003A&A...407..177C}.
For example, according to \citet{2008ApJ...675.1361F}, sampling
effects dominate in the FUV range when extinction corrected FUV fluxes are below 10$^{39}$ erg s$^{-1}$. Most of our selected objects lie above this threshold, but the situation might change if we are interested in the UV ionising radiation, because it has a steeper dependence on the stellar mass than the non-ionising continuum. Hence, to estimate the H$\alpha$ luminosities of the smallest clusters in our sample, sampling effects need to be considered.

\begin{figure}
\centering
\includegraphics[width=8cm,bb=85 365 475 775]{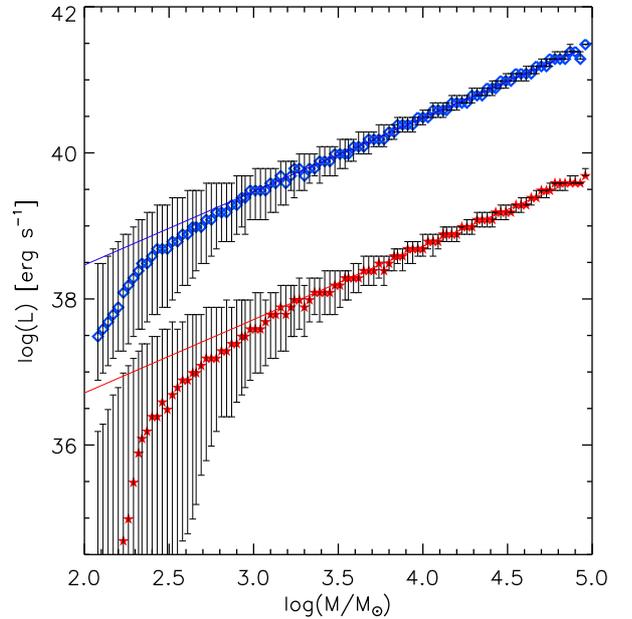}
\caption{Expected median bolometric (diamonds) and H$\alpha$ (stars)
luminosity of stellar clusters as a function of cluster mass and their corresponding uncertainties.}
\label{imf}
\end{figure}

In Fig. \ref{imf} we show how the median bolometric (diamonds) and H$\alpha$
(stars) luminosities of a cluster depend on its mass. 
The simulation was carried out as described in \citet{2009A&A...495..479C},
but here we assumed a maximum stellar mass of 100~\msun to be consistent with the
fitting procedure used in this paper. The luminosities are computed for zero-age main-sequence stars.

Low-mass clusters have a larger dispersion around the median H$\alpha$ luminosity
than around the bolometric luminosity. Moreover, L$_{H\alpha}$ deviates considerably from a simple linear relation (solid line) at masses lower than 1450~\msun. Below this limit the IMF is not fully populated at its high-end, and the stronger dependence of L$_{H\alpha}$ on the stellar mass implies that the mean value of L$_{Bol}$/L$_{H\alpha}$ increases, despite  the occasional presence of outliers.

The IMF statistical fluctuations introduce a large dispersion in the observable properties of the low-mass clusters (M $<$ 1450~\msun), when the linear relation between cluster mass and L$_{H\alpha}$ breaks down.
Thus, to account for the incomplete sampling of the IMF in these clusters, we will use the H$\alpha$ luminosity and the dispersions derived with our simulation (see Fig. \ref{imf}). Note that all clusters with $M < 1450$ \msun in our sample are quite young, hence their expected H$\alpha$ emission is well represented by the simulated data in Fig. \ref{imf}. For older and massive clusters we shall use the luminosities obtained for a complete IMF (STARBURST99) with the dispersion given by our simulation.  \citet{2003A&A...407..177C} have shown that the minimum mass below which we expect stochastic effects, as well as the dispersion around the mean or median value, have a very weak dependence on the cluster age during the first 10 Myr.

\begin{figure*}
\centering
   \includegraphics[width=6cm,bb=70 360 450 770]{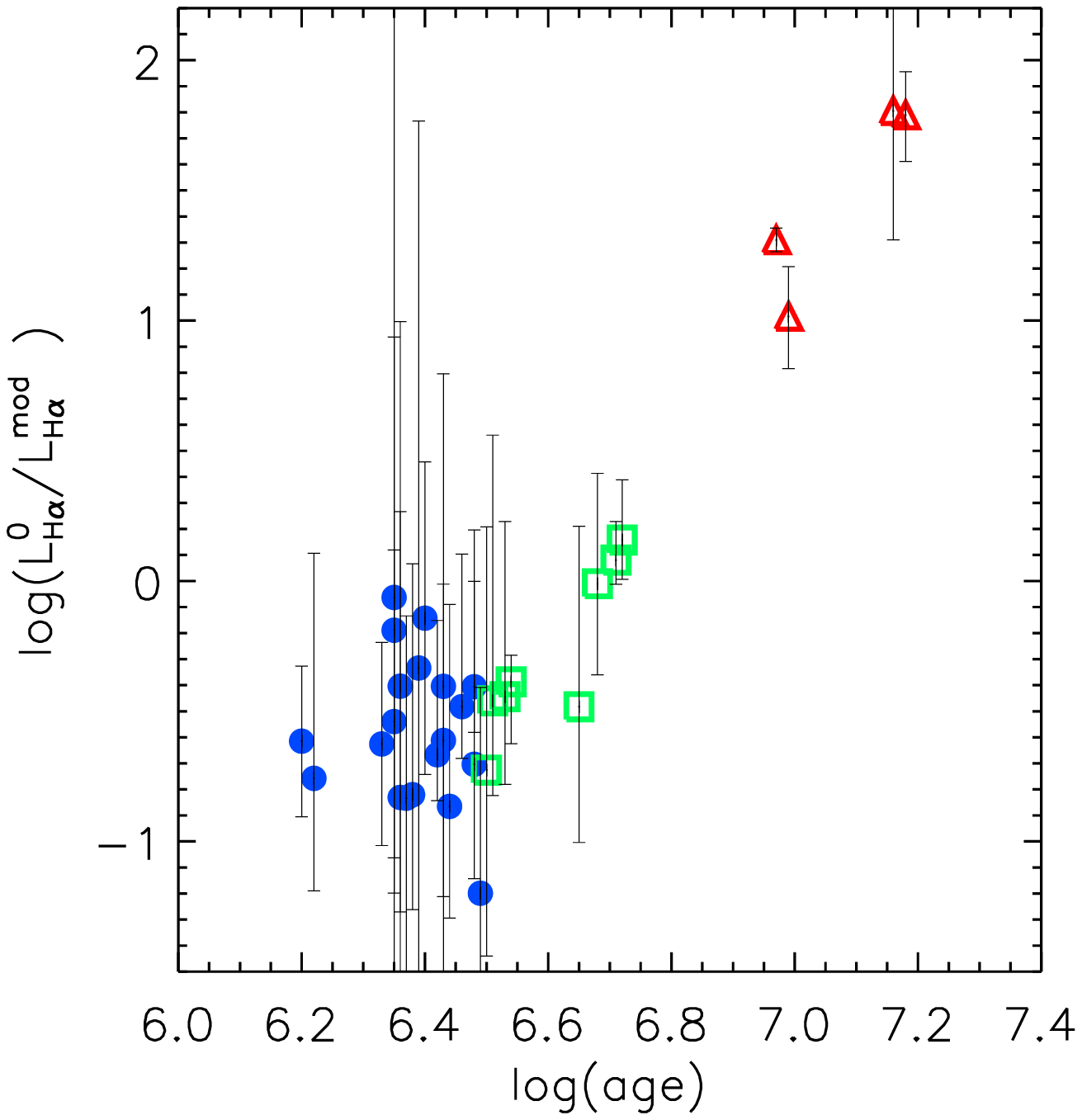}
   \includegraphics[width=6cm,bb=70 360 450 770]{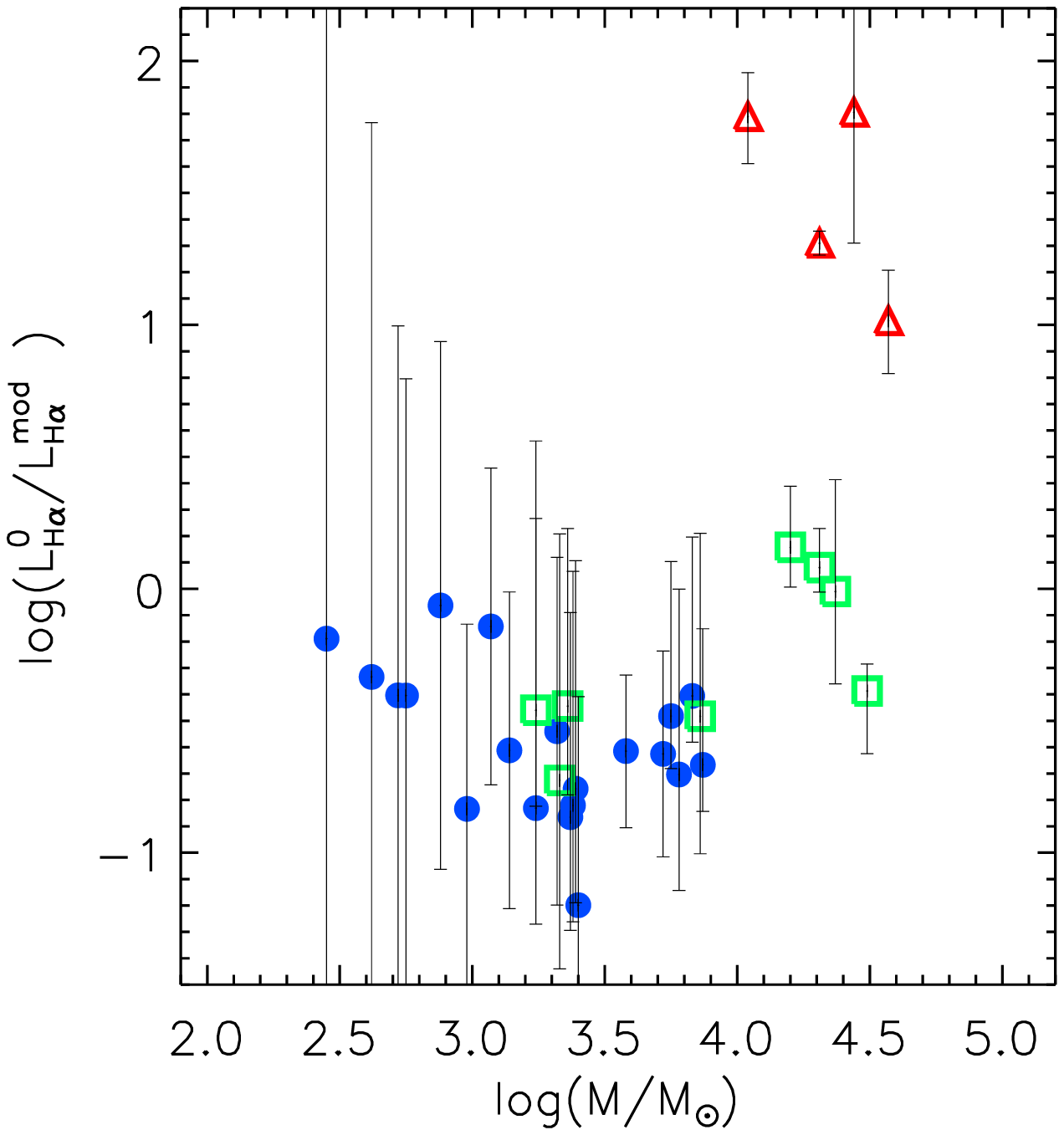}
   \includegraphics[width=6cm,bb=60 360 440 770]{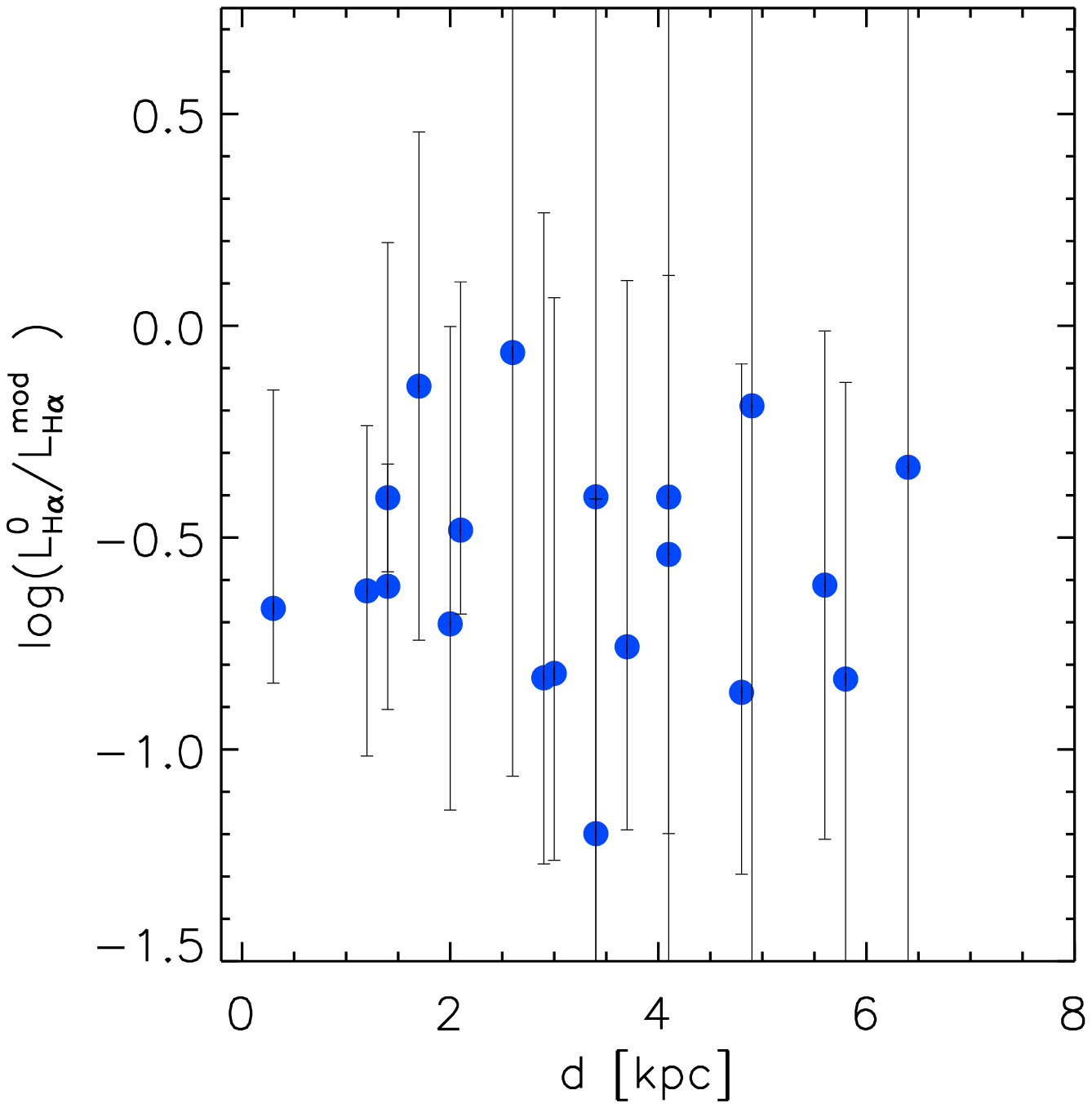}
\caption{Ratio of the observed (extinction corrected) to the model H$\alpha$ luminosity vs. the age of the clusters ({\em left-panel}), the mass of the clusters ({\em central-panel}), and  the galactocentric distance for clusters with log(age) $< 6.5$ ({\em right-panel}). In these plots, the model H$\alpha$ luminosities of clusters with masses
M $< 1450$ \msun, were determined with the simulations described in the text, taking into account the incomplete sampling of the IMF. The error bars include both the dispersion of the model H$\alpha$ luminosity and the uncertainty due to the error on the age of the clusters. Symbols are the same as for Fig. \ref{age_mass}.}
\label{samp_eff} %
\end{figure*}

The expected H$\alpha$ luminosities of low-mass clusters are lower when stochastic effects are considered, as shown in Fig. \ref{samp_eff}. The error bars in the figure include both the dispersion of the model H$\alpha$ luminosity and the uncertainty due to the error on the age of the clusters. 
The mean L$^0_{H\alpha}/$L$^{mod}_{H\alpha}$ ratio\footnote{Note that to derive the mean ratio we discarded the two lowest mass clusters with the largest error bars.} for the youngest clusters ($t <$ 3Myr) is 0.30$^{+0.8}_{-0.07}$.
Therefore, taking into account the stochastic effects reduces the discrepancy between the expected and observed values, but the large uncertainties prevent us to draw any firm conclusions on how the missing H$\alpha$ flux depends on the cluster mass, age, or galactocentric radius, as we show in Fig. \ref{samp_eff}.

\subsection{Dust absorption}

Dust within or around the cluster can absorb the UV photons emitted by the hot
massive stars, reducing the amount of ionising radiation, and contributing to the lack
of observed H$\alpha$ photons.

The good spatial correlation between 24 $\mu$m and H$\alpha$ emissions
shows that the dust is predominantly heated by OB stars within the \hii regions.
The {\em Spitzer} resolution at 70 and 160 $\mu$m is not sufficient to resolve our sources, and we estimated the total IR luminosity from the 24 $\mu$m emission following Verley et al. (2009). This agrees with what other studies found when the IR luminosity is dominated by emission at wavelengths longer than 24 $\mu$m.
Nevertheless, the total IR luminosities are lower than what is required by the modest extinctions obtained
from the SED-fitting procedure. Some additional contribution to $L_{TIR}$ might come from a hot dust component heated
by young stars in these sources. The SED indicates an excess above the expected stellar emission at 3.6 and 4.5 $\mu$m.
However, given the limited MIR emission detected around the clusters and, more generally, the low dust content of M33, it is difficult to explain the missing ionising radiation as an effect of dust absorption only  (see Sects. 6.4 and 6.5), even though this mechanism would explain the observed trend with cluster age.

On the other hand, the way dust is distributed within or around the \hii regions may produce different attenuation of the emitted light, and the usual assumption of a homogeneous dust screen may lead to an underestimation of the effects of dust extinction \citep{1997A&A...323..697P}.
For example, a clumpy dust distribution would explain the missing fraction of ionising radiation (40\%) observed in the super star clusters of SBS 0335-052 \citep{2008AJ....136.1415R}. A more detailed analysis of the dust properties for these clusters, taking into account more complex dust distributions, will be addressed elsewhere (Giovanardi et al. 2010, in preparation).

\subsection{Leakage of UV photons}

UV photons not absorbed by dust grains can escape from the star-forming region and explain
the origin of the extended diffuse UV emission observed by GALEX in the M33 disc
\citep{2005ApJ...619L..67T}.

The importance of taking into account the leakage of ionising photons from
\hii regions has been increasingly discussed in recent years, especially
after the discovery of the diffuse ionised  gas (DIG) in the
ISM of the Milky Way \citep{1974ApJ...192L..53R},
and in the discs of other spiral galaxies \citep{1998ApJ...506..135G}.
The DIG components of galaxies such as M31 \citep{1996ASPC..106....1W}, the LMC \citep{1997MNRAS.291..827O,1998PASA...15..141O}, the SMC \citep{2002ApJ...564..704R},
amount to the 30\% and 50\% of the total H$\alpha$ luminosity.
Studies of field OB stars in M33 \citep{2000ApJ...541..597H,2003ApJ...586..902H} have given evidence that at least 20\% - 30\% Lyman continuum photons must be able to escape from the \hii regions in this galaxy.
Nonetheless, there is still a debate about whether the energy budget of photons leaking out of \hii regions is sufficient to ionise the DIG or if additional mechanisms should be considered \citep{2008AJ....135.1291V}.

Theoretical models have shown that the amount of leakage is regulated by the structure
and clumpiness of the ISM \citep{1993ApJ...417..579M, 2004MNRAS.353.1126W}.
Inhomogeneities in the medium favour the leakage of radiation because the
presence of low-density regions may increase the mean free path of the photons
\citep{2008AJ....135.1291V, 2008AJ....136.1415R}.
The leakage also depends on the spectrum of the ionising radiation, and on
the evolutionary stage of the cluster, with the fraction of escaping photons being larger when
stars experience a hard Wolf-Rayet phase, which occurs, depending on the mass and metallicity of a star, between 3 and 6 Myr \citep{2002ApJ...565L..79C}.

Relating the leakage of ionising photons to the beginning of the W-R phase (Castellanos et al. 2002) may explain the age-dependence of the L$^0_{H\alpha}/$L$^{SB99}_{H\alpha}$ ratio for clusters older than 3 Myr, but the issue would still remain open for the youngest systems.

\subsection{Deficiency or delayed formation of massive stars}

The previous considerations cast some
doubts on leakage and dust absorption as the only processes causing the deficiency of ionising radiation.

If the star-formation process is not instantaneous, clusters might reach their expected H$\alpha$ luminosity only after all stars have formed, especially the most massive ones, which determine the cluster H$\alpha$ luminosity rather than the total mass.
Massive stars might form later during the cluster evolution either
because they need particular physical conditions and processes to take place
\citep{2005IAUS..227..196P} or because the time-scale for the gas to collapse is longer in a low-density environment such as the disc of M33 \citep{1999ASPC..187..145E}. Therefore, the delayed formation of the most massive stars appears as an interesting and promising way to explain why only very young clusters are underluminous in H$\alpha$, and this should be investigated in detail in future work. Particularly, it would be interesting to know whether the excess of H$\alpha$ emission in the oldest clusters could also be explained by this scenario without the assumption of multiple burst events.

A strong deficiency of massive stars compared to what is predicted by a stochastically sampled IMF should also be considered. For example, truncated models of the stellar IMF
\citep[and references therein]{2009ApJ...695..765M,2009MNRAS.395..394P} predict lower cluster luminosities than stochastically sampled models (Corbelli et al. 2009). This is because truncated models link the cluster mass to the most massive stars that can form in a cluster, without considering the occasional presence of outliers. The IMF over the whole galaxy in this case would deviate from the power law assumed in each cluster, and it will be steeper. Assessing the details of the IMF in M33 and its variance requires the availability of deep colour-magnitude diagrams for the stellar members of each cluster. This is difficult to attain in compact young clusters and is clearly beyond the scope of this study.

\section{The gaseous environments of young clusters}

Finally, we investigated whether the mass of young clusters is related to the surface density of the different phases of the surrounding ISM (ionised, atomic, molecular gas).

As described in Sect. 2.4, the spatial resolution of the CO and \hi maps is 13\arsec and
24\arsec $\times$ 48\arsec respectively. Thus, the size of the clusters (see Table \ref{phottab}) is smaller than the resolution of the CO images for roughly half of the objects in our sample, while none of the clusters can be resolved with the beam size of the 21-cm map.
For this reason we limit our analysis to areas of the ISM that are larger than the clusters.
Using an aperture equal to the lowest resolution gas map (H{\sc i}) i.e. of 24~arcsec,
centred on the optical position of each object, we measured the gas surface densities associated with it.

\begin{figure}[h]
   \centering
\includegraphics[width=10cm,bb=53 345 630 840]{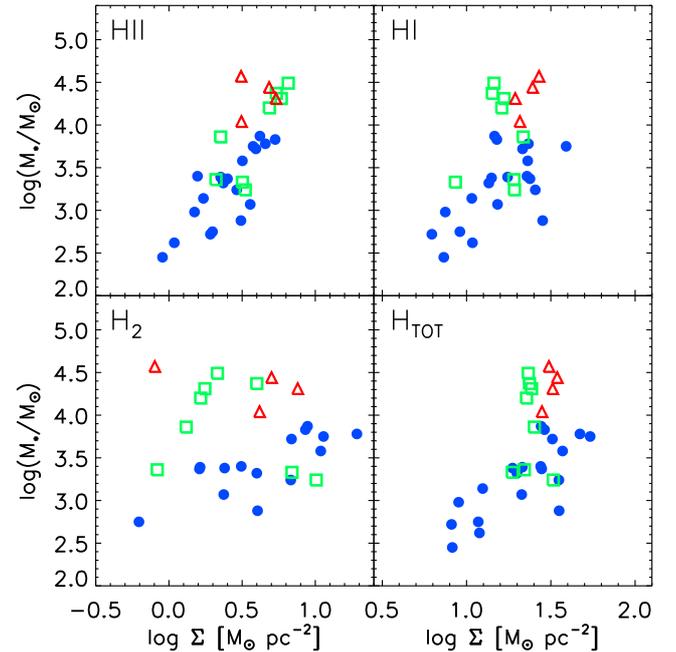}
   \caption{Surface density of ionised ({\em top-left}), atomic ({\em top-right}), molecular
   ({\em bottom-left}), and total ({\em bottom-right}) hydrogen as a function of the stellar mass of the cluster. Symbols and colour codes are the same as for Fig. \ref{age_mass}.}
   \label{gas} 
    \end{figure}

Figure \ref{gas} shows the measured surface density of the different gaseous components as well as the total gas surface density compared to the cluster stellar mass. There is no trend
with cluster age, the surface density of the surrounding ISM is not affected by
the early stages of cluster evolution. The ionised gas seems to better correlate with the cluster mass (as expected, since the most massive clusters have a higher number of ionising photons).
Concerning the \hi gas, one can only infer from the figure that the majority of the clusters are found in a narrow regions of surface density (N$_{HI}= 1 \div 3$ $\times$
$10^{21}$ cm$^{-2}$), which is no big surprise given the rather flat \hi distribution in
the M33 disc. Clusters are born in correspondence to \hi filaments, but there is no clear
connection between the stellar mass and the gas surface density \citep{2010A&A...510A..64V}.

The {\em bottom-right} panel of Fig. \ref{gas} shows the sum of all gaseous components (ionised, atomic, molecular) versus the stellar mass. It shows that the gaseous environment of
these clusters is dominated by the atomic component. For a change of two orders of magnitude in cluster mass there is only half an order of magnitude change in the total gas column density. Because the scatter of this correlation is large, as Verley et al. (2010) have already pointed out, the gas surface density is not the only parameter which determines the star-formation rate and the
mass of the newly born clusters.

\section{Summary and conclusions}

We analysed multi-wavelength observations of a sample of young star clusters and associations in M33 combining UV (GALEX), optical ($UBVRI$), and MIR ({\em Spitzer}/IRAC-MIPS), 3-mm
(BIMA-FCRAO) and 21-cm (WSRT) observations. The UV and optical data sets were used to derive their age and mass, and the line-of-sight extinction by fitting  the measured SEDs to a set of single-burst stellar population models. The model SEDs were reddened using various extinction laws (Milky Way, LMC, SMC), and the best-fit solutions were obtained with the LMC laws.

We find ages between 2 and 15 Myr and stellar masses between 3$\times 10^2$ and $4\times10^4$ \msun. The extinction towards these stellar systems, A$_V$ varies between 0.3 and 1 mag, indicating that they are not highly obscured even in an early evolutionary phase.
As expected, older clusters are less extinct than younger ones.
We also find a correlation between mass and radius with M$\propto$ R$^{1.5\pm0.2}$.
There is no clear trend between extinction and other parameters such as
metallicity or total gas surface density.
The fractional luminosities at 8 $\mu$m and 24 $\mu$m decrease with the galactocentric distance, suggesting a lower PAHs and small-grain abundance in the outer regions of the galaxy. A comparison between the model SEDs and the {\em Spitzer}/IRAC fluxes shows an excess emission at 3.6 $\mu$m and 4.5 $\mu$m, which may be due to a
hot dust component, heated by the ionising radiation of the cluster massive stars.

We find a discrepancy between the observed H$\alpha$ luminosity and that expected from population synthesis models, especially in young ($<$ 3 Myr) and low-mass (M $< 3 \times 10^3$ \msun) clusters. The ratio of the observed-to-model H$\alpha$ luminosity increases with cluster age and ranges between 20\% and 90\%. We showed that this difference can be reduced but not removed if one takes into account the uncertainties of the age from the SED-fitting procedure and the stochasticity of the IMF at the high mass-end.
Similarly, the gap between the best-fit bolometric luminosity (derived assuming a fully sampled IMF) and the total observed luminosity (UV and IR) cannot be fully explained if only a  stochastically sampled IMF is considered. We then discussed
other possible scenarios:

\begin{enumerate}

\item Absorption of Lyman continuum photons by dust
in \hii regions cannot fully explain this discrepancy because
the IR emissions inferred from the 24 $\mu$m luminosity is generally lower
than expected from the derived extinction. This suggests that UV continuum photons
are missing, but not because of dust absorption only.

\item The escape of UV ionising and non-ionising photons from the \hii region because of a
clumpy and porous ISM is a likely scenario that would be supported by the large UV
and H$\alpha$ diffuse fractions observed in this galaxy.

\item Finally we discussed the possibility of a delayed formation of the most massive stars in these clusters as a consequence of the increase in the observed-to-expected H$\alpha$ luminosity ratio with age and mass.
 Because the oldest clusters in our sample are also the most massive ones, a stronger suppression of massive
star formation in low-mass systems than what is predicted by a stochastically sampled
IMF would also alleviate the problem.

\end{enumerate}

Finally, we analysed the surface density of the gas (ionised, atomic, and molecular)
in a region about 200~pc wide around the cluster. A marginal correlation only is
found between the cluster mass and the total gas surface density, implying that
on these large scales this
is not the only parameter that determines the total mass of stars.

\begin{acknowledgements}

We thank the anonymous referee for the comments and suggestions that contributed to improve the paper.
This work is based on observations made with the NASA Galaxy Evolution Explorer and with the {\em Spitzer} Space Telescope. GALEX is operated for NASA by the California Institute of Technology under NASA contract
NAS5-98034.  The {\em Spitzer} Space Telescope is operated by the Jet Propulsion Laboratory, California Institute of Technology under a contract with NASA. This research draws upon data provided by {\em The
Resolved Stellar Content of Local Group Galaxies Currently Forming Stars}, PI: Dr. Philip
Massey, as distributed by the NOAO Science Archive. NOAO is operated by the Association
of Universities for Research in Astronomy (AURA), Inc. under a cooperative agreement with
the National Science Foundation.

\end{acknowledgements}

\bibliographystyle{aa} 
\bibliography{M33clust_bib} 

\end{document}